\documentclass[11pt]{article}
\usepackage[margin=1in]{geometry}
\usepackage{bm}
\usepackage{hyperref}
\usepackage{multirow}
\usepackage{amssymb,amsmath,amsthm,amsfonts}
\usepackage{cancel}
\usepackage{bm}
\usepackage{textgreek}
\usepackage{mathtools}
\usepackage{enumitem}
\usepackage[numbers,comma,sort&compress]{natbib}
\usepackage{authblk}
\usepackage{graphicx}
\usepackage[font=small]{caption}
\usepackage{tablefootnote} 
\usepackage{float}
\usepackage{physics}
\usepackage{footnote}
\usepackage{xcolor}
\usepackage{mathrsfs}
\usepackage{bbm}
\usepackage{makecell}
\usepackage{pbox}
\usepackage{graphicx}%
\usepackage{multirow}%
\usepackage{amsmath,amssymb,amsfonts}%
\usepackage{amsthm}%
\usepackage{mathrsfs}%
\usepackage[title]{appendix}%
\usepackage{xcolor}%
\usepackage{textcomp}%
\usepackage{manyfoot}%
\usepackage{booktabs}%
\usepackage{algpseudocode}%
\usepackage{listings}%
\usepackage{cleveref}
\usepackage{subfig}

\allowdisplaybreaks[4]

\renewcommand{\d}{\mathrm{d}}
\newcommand{\im}{\mathrm{i}}

\newcommand{\eq}[1]{Eq.~(\ref{#1})}
\newcommand{\Pra}[1]{{\color{black}{\vspace{2mm} \noindent\textbf{Parameter Setting: }#1}}}
\newcommand{\Results}[1]{{\color{black}{\vspace{2mm} \noindent\textbf{Experiment Result: }#1}}}

\renewcommand{\Tr}{\mathrm{Tr}}  
\def\lr#1{\left\langle {#1} \right\rangle}
\newcommand{\lrb}[1]{\left(#1\right)}

\newtheorem{theorem_}{Theorem}
\newtheorem{lemma_}{Lemma}
\newtheorem{corollary_}{Corollary}
\newtheorem{definition_}{Definition}
\newtheorem{proposition_}{Proposition}
\newtheorem{remark_}{Remark}

\crefname{figure}{Fig.}{Figs.}
\Crefname{figure}{Fig.}{Figs.}

\crefname{corollary_}{Corollary}{Corollaries}
\Crefname{corollary_}{Corollary}{Corollaries}

\newcommand{\note}[1]{#1}

\begin{document}

\title{Quantum Langevin Dynamics for Optimization}
\author{Zherui Chen\thanks{Equal contribution.}      \thanks{Xingjian College, Tsinghua University; Department of Mathematics, University of California, Berkeley. Email: \href{mailto:chenzherui007@berkeley.edu}{chenzherui007@berkeley.edu}}
\qquad
Yuchen Lu$^{*}$\thanks{Xingjian College, Tsinghua University; Ecole Polytechnique Fédéderale de Lausanne (EPFL). Email: \href{mailto:yuchen.lu@epfl.ch}{yuchen.lu@epfl.ch}}
\qquad
Hao Wang\thanks{School of Electronics Engineering and Computer Science, Peking University. Email: \href{mailto:2000013065@stu.pku.edu.cn}{2000013065@stu.pku.edu.cn}}
\qquad
Yizhou Liu\thanks{Corresponding author. Department of Engineering Mechanics, Tsinghua University. Email: \href{mailto:liuyz18@tsinghua.org.cn}{liuyz18@tsinghua.org.cn}}
\qquad
Tongyang Li\thanks{Corresponding author. Center on Frontiers of Computing Studies, School of Computer Science, Peking University.  Email: \href{mailto:tongyangli@pku.edu.cn}{tongyangli@pku.edu.cn}}}
\date{}

\maketitle

\begin{abstract}
    We initiate the study of utilizing Quantum Langevin Dynamics (QLD) to solve optimization problems, particularly those nonconvex objective functions that present substantial obstacles for traditional gradient descent algorithms. Specifically, we examine the dynamics of a system coupled with an infinite heat bath. This interaction induces both random quantum noise and a deterministic damping effect to the system, which nudge the system towards a steady state that hovers near the global minimum of objective functions. We theoretically prove the convergence of QLD in convex landscapes, demonstrating that the average energy of the system can \note{converge to} zero in the low temperature limit with \note{an exponential convergence rate}. Numerically, we first show the energy dissipation capability of QLD by retracing its origins to spontaneous emission. Furthermore, we conduct detailed discussion of the impact of each parameter. Finally, based on the observations when comparing QLD with the classical Fokker-Plank-Smoluchowski equation, we propose a time-dependent QLD by setting temperature and $\hbar$ as time-dependent parameters, which can be theoretically proven to converge better than the time-independent case and also outperforms a series of state-of-the-art quantum and classical optimization algorithms in many nonconvex landscapes.
\end{abstract}

\section{Introduction}\label{sec1}

Continuous optimization is a branch of optimization theory concerned with the optimization of continuous variables within a mathematical model. Within this domain, gradient descent and its improved versions emerge as a widely embraced subclass of continuous optimization algorithms, having achieved widespread utilization across diverse domains, including but not limited to machine learning and associated disciplines~\cite{nocedal1999numerical,lecun2015deep,ruder2016overview,jain2017non}.
More recently, there have been systematic studies of gradient descents via the analysis of their continuous-time limits as differential equations \cite{su2014differential,wibisono2016variational,jordan2018dynamical,shi2021understanding}. In particular, the continuous-time limit of stochastic gradient descents become the overdamped Langevin equation~\cite{shi2020learning}, and algorithms based on the classical Langevin equation with stochastic gradients were previously proposed and widely used in various machine learning and Bayesian modeling contexts~\cite{welling2011bayesian,ahn2012bayesian, chen2014stochastic}, such as the Stochastic Gradient Langevin Dynamics algorithm for nonconvex optimization~\cite{zhang2017hitting}, Langevin Monte Carlo for sampling~\cite{dwivedi2018log}, continuous-time Langevin diffusion for stochastic calculus~\cite{chewilog}, etc. Over the preceding decades, efficient quantum algorithms for various optimization problems have been proposed, including linear programming~\cite{li2019sublinear,van2019quantum,casares2020quantum,bouland2023quantum,gao2023logarithmic}, semi-definite programming~\cite{brandao2017quantum,van2017quantum,brandao_22017quantum,van2019improvements,kerenidis2020quantum,augustino2023quantum}, polynomial optimization~\cite{rebentrost2019quantum,li2021optimizing}, convex optimization~\cite{chakrabarti2020quantum,van2020convex}, escaping from saddle points for nonconvex optimization~\cite{zhang2021quantum}, etc. 
Despite this, leveraging the nature of quantum mechanics to propose powerful optimization algorithms~\cite{leng2023quantum,zhang2021quantum,liu2023quantum} is still a challenging task and solicits further developments.

The Bregman-Lagrangian framework was proposed in Ref.~\cite{wibisono2016variational}, which generated a large class of accelerated methods in the continuous-time limit. Ref.~\cite{leng2023quantum} derived Quantum Hamiltonian Descent (QHD) by using the propagator of the quantum dynamics to quantize the Bregman-Lagrangian framework. When taking stochastic noise into account, Ref.~\cite{shi2020learning} provided a general theoretical analysis of the effect of the learning rate in Stochastic Gradient Descent (SGD). Viewing SGD from the perspective of dynamic systems, it can be considered as a specific classical open dissipative system with stochastic noise, named as classical Langevin system \cite{ray1999notes,gardiner1985handbook,shi2020learning}. This observation inspired us to harness a distinct category of open dissipative quantum systems, known as quantum Langevin system \cite{gardiner2004quantum}, with the explicit purpose of constructing an optimization algorithm. 

Physically, the mutual influence of a system, with a few degrees of freedom, and a heat bath, with many degrees of freedom, on each other is the central concept in the physics of noise both in quantum and classical cases. The variables within the bath exert an influence on the system by introducing stochastic terms into the system's equations of motion. Langevin introduced a class of equations to describe this kinds of classical systems:
\begin{align}
  m\ddot{x} = -\nabla V(x) -\gamma \dot{x} + \sqrt{2\gamma kT}\xi(t).
\end{align}

The analogous equations for quantum systems have been formulated and have found applications in numerous instances of physical interest \cite{gardiner2004quantum}. The quantum case under consideration involves a model of a heat bath comprising harmonic oscillators characterized by distinct masses $m_n$ and spring constants $k_n$. 
The Hamiltonian governing the entire system is expressed as follows \cite{ford1965statistical,gardiner2004quantum}:
\begin{align}
  H_{\text{total}} = H + H_B \nonumber                    
            = H + \sum_{n}\left( \frac{p_{n}^{2} }{2m_{n}}+ \frac{k_{n}}{2}(q_{n}-X)^{2} \right),
            \label{eq:originH}
\end{align}
where $H=p^2/(2m)+V$ is the Hamiltonian of the system and $X$ represents a specific operator within the system that is coupled with the heat bath. Here, $p$ is the momentum operator of the system, $m$ is the mass of the system, $V$ represents the potential energy of the system, $q_n$ and $p_n$ are the position and momentum operator of the $n$-th oscillator, respectively.

In order to exploit quantum Langevin dynamics for solving optimization problems, we need a suitable framework for describing quantum Langevin systems. The quantum Langevin equation for an arbitrary system operator $Y$ can be readily derived from the Heisenberg equation of motion \cite{sakurai1995modern}, i.e.,
\begin{align}
  \dot{Y} = \frac{\im}{\hbar}[H_{\text{total}}, Y].
\end{align}
\note{Although the above equation intuitively illustrates the Langevin dynamics we consider, it involves the dynamics of each oscillator in the heat bath, making it difficult to capture the energy changes in the system of interest. Therefore, it is necessary to reconsider this problem from a statistical perspective.}

\note{We return to the Schr\"{o}dinger picture and follow the ideas in~\cite{lampo2019quantum,caldeira1983path,gardiner2004quantum}}, which adopt the Born-Markov approximation and use path integral approach. When the correlation time of the noise is significantly shorter than the typical time scale of the damped motion of the system, we obtain the quantum Brownian motion master equation:
\begin{align}
\hspace{-1mm}\frac{\d \rho}{\d t} = -\frac{\im}{\hbar}[H,\rho] -\frac{\eta}{\hbar}\left(\frac{2mkT}{\hbar}[x,[x,\rho]] + \im[x,[p,\rho]_+]\right)
\label{eq:CLeq}
\end{align}
where $H=p^2/(2m)+V$, $p$ is the momentum operator of the system, $x$ is the position operator, and $\eta = \gamma/2m$ represents the characteristic damping rate  with system mass $m$, $k$ is the Boltzmann constant. However, it is well known that this master equation is only applicable in high-temperature scenarios and fails to satisfy the positivity constraint of the density operator~\cite{gao1997dissipative,diosi1993high,karrlein1997exact}. Ultimately, we refer to Ref.~\cite{gao1997dissipative} and employ Lindblad functional (\eq{eq:Gao_Lindblad}) to describe the quantum Langevin dynamics, with a detailed introduction provided in Section \ref{sec:Lindblad_Approach}.

The \textbf{optimization problem} we consider is to find the global minimum of a (convex or nonconvex) function $V(x)$:
\begin{align}
    \min_{x} V(x),
\end{align}
where $V(x)$ is assumed to be continuously differentiable in the domain and is also the potential energy of the system in \eq{eq:Gao_Lindblad}. One can implement Quantum Langevin Dynamics (QLD) by starting from an easily prepared initial state and evolving the quantum system according to \eq{eq:Gao_Lindblad}. The global minimum of the objective function $V(x)$ can be obtained by measuring the position observable $x$ at the end of the evolution.\footnote{\note{Theoretically, the measurement operator is continuous, i.e., $\int\ket{x}\bra{x}\d x = I$. In practice, the problem is always defined on a finite domain. We discretize the domain of $x$ and the set of eigenstates of $x$ becomes $\{\ket{x_i}: i = 0,1,\ldots,n-1\}$, which can be represented with a set of computational basis. Thus, the measurement operator becomes $\sum_{i}\ket{x_i}\bra{x_i}$.}} Therefore, our objective is to establish a theoretical foundation guaranteeing the convergence to the global minimum of QLD. When the parameters of QLD is time-independent, namely, $\eta,\hbar$ and $T$ are constants during evolution, we establish the convergence of QLD within convex landscapes \note{(Theorem \ref{theorem:convex_converge} and \ref{thm:QLD_convex_highD})} and \note{quasar-convex landscape (Theorem \ref{thm:con_non_convex_qld})}. Informally speaking, with approximation of $\hbar \Omega \ll kT$ where $\Omega$ is the oscillator frequency of the ground state near the global minimum, we have the convergence rate of the average potential,
\begin{align}
  \lr{V}_t - V(x^*) \le \text{\note{$\zeta$}}kf(T) + O(e^{-\eta t}),
\end{align}
\note{where we denote $\zeta = 2 + O(\frac{\hbar\Omega}{kT})$ as a constant throughout the paper}, $\lr{V}_t$ denotes the expectation value of $V$ with respect to the density matrix of the system, $V(x^*)$ is the potential value at the global minimum $x^*$ and $f(T)$ is a function with respect to $T$ that satisfies $\lim_{T\to 0} f(T) = 0$. Consequently, the average energy of the system can approach zero in the low temperature limit. When $V$ is a quadratic form, the evolution of the system has an analytical solution. One interesting discovery related to the quadratic potential is that the system can reach the ground state of the quadratic potential when the evolution time $t\to +\infty$ and \note{the temperature} $T\to 0$.
Formal description can be found in Section \ref{sec:con_OLD}. However, it is always challenging to give a convergence in the general nonconvex case. Despite our best efforts, we are constrained to confine our theoretical discussions to \note{quasar-convex} case. Thus, for other nonconvex landscapes, we resort to meticulous numerical experiments to compensate for the absence of theoretical assurances.

In the realm of numerical experiments, we initially substantiate our intuition regarding the fundamental cause of energy dissipation in QLD. Noteworthy is the observation that despite the finite nature of the heat bath, the system retains the capacity to escape local minima, transitioning from excited states to ground states due to spontaneous emission.  \note{We conduct experiments within quadratic frameworks, comparing results with analytic solutions to demonstrate strong agreement with our theoretical results.} To offer optimal strategies for harnessing thermal and tunneling effects when confronting intricate nonconvex problems, we engage in a detailed discussion delineating the role of each parameter. Broadly speaking, our results indicate that $\eta$ determines the convergence rate, temperature $T$ reflects the thermal effect to some extent and the Plank constant $\hbar$ reflects the tunneling effect.

\begin{figure}[htbp]
  \centering
  \includegraphics[width=0.6\textwidth]{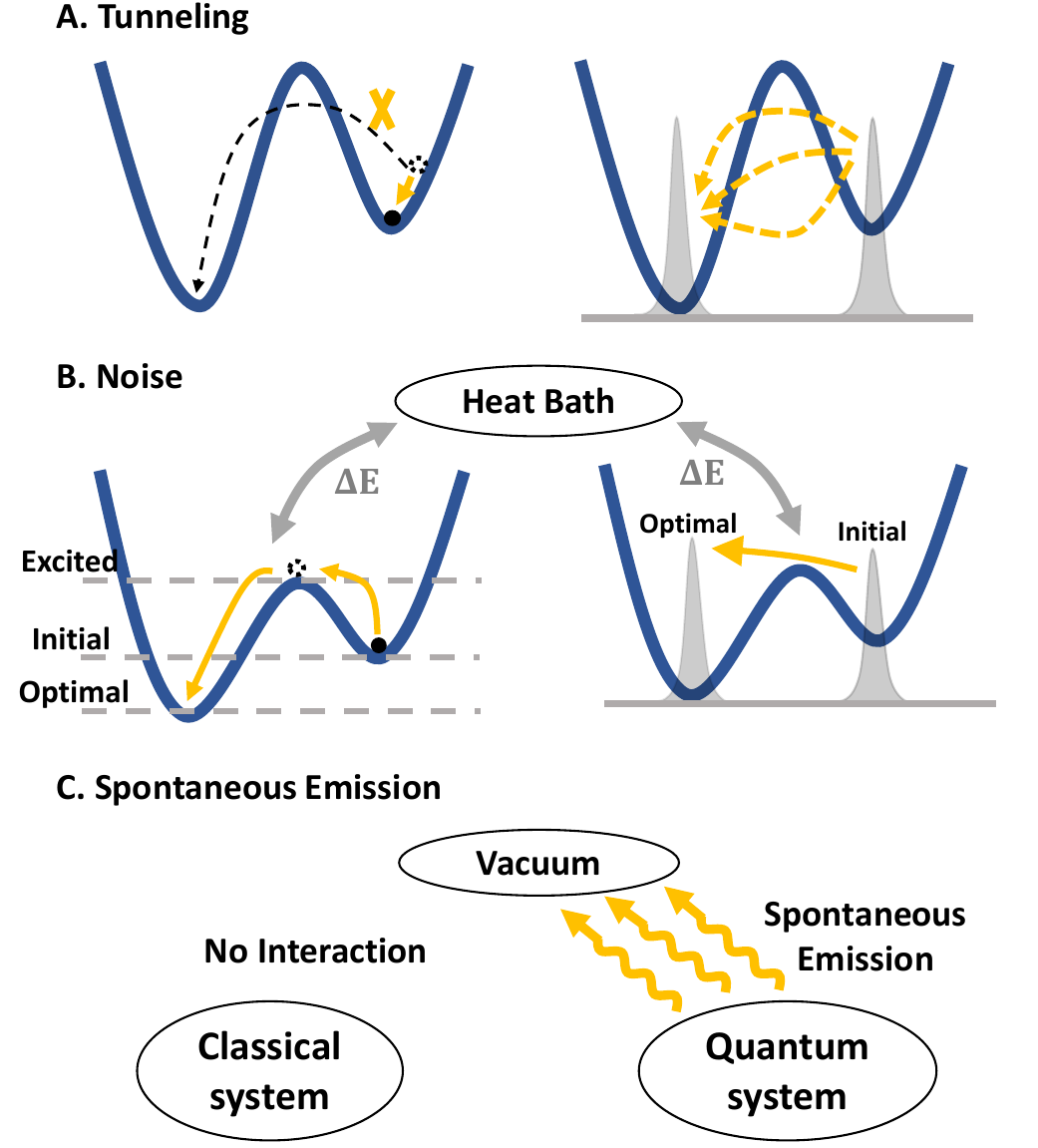}
  \caption{Differences between classical dynamics (left) and quantum dynamics (right). \textbf{A.} Tunneling effect: An illustrative example where classical Gradient Descent algorithms will be trapped because of bad initialization, while quantum algorithms can escape easily. \textbf{B.} Noise: The effect of noise can be described by the system coupling with a heat bath.  \textbf{C.} Special thing of quantum noise: The crucial difference between classical noise and quantum noise emerges when interacting with vacuum. Owing to the principles of quantum mechanics, spontaneous emission exists even in a vacuum heat bath, which indicates the fundamental origin of energy dissipation in the quantum Langevin system.}
  \label{fig:randomness}
\end{figure}

We conduct a comparative analysis between time-independent QLD and overdamped classical Langevin dynamics. In the majority of landscapes that classical dynamics can adeptly resolve, time-independent QLD does not demonstrate any discernible advantages. However, a notable advantage becomes evident in highly nonconvex landscapes. At elevated temperatures, classical dynamics exhibits superior performance by supplying the requisite energy to escape local minima. In contrast, time-independent QLD excels in low-temperature environments. This observation substantiates our intuition that quantum noise differs fundamentally from its classical counterpart, as depicted in Fig.~\ref{fig:randomness}. Quantum Hamiltonian Descent (QHD)~\cite{leng2023quantum} 
exclusively leverages the tunneling effect and surpasses a curated selection of state-of-the-art gradient-based classical solvers, as well as the conventional Quantum Adiabatic Algorithm (QAA) \cite{leng2023quantum}. It has been demonstrated that noise confers benefits upon classical algorithms in nonconvex landscapes, whereas its impact on quantum algorithms remains unexplored.

Building upon these observations, we find it imperative to explore the realm of time-dependent QLD. Specifically, we recognize the need for pronounced thermal and tunneling effects initially to facilitate escape from local minima. As the system converges to its final stage, these effects need to be gradually deactivated. Consequently, we present a theoretical analysis for time-dependent QLD and propose a cooling strategy for implementation. The convergence of time-dependent QLD is similar to the time-independent one when $\hbar = \hbar(t)$ decreasing monotonically with respect to time and $\eta = \eta(t)$, which says $\lr{V}_t - V(x^*)\le \text{\note{$\zeta$}}kT + O(e^{-\int_0^{t}\eta(t')\d t'})$. For the case of time-dependent temperature, there exists a cooling strategy $T = T(t)$ that guarantees convergence. In numerical experiments, time-dependent QLD significantly outperforms its time-independent counterpart and surpasses other algorithms, including QHD, QAA, and SGD, across a diverse array of landscapes. In theory, the query complexity of QLD scales as \note{$\tilde{{O}}(t)$, where $\tilde{{O}}(\cdot)$ denotes the asymptotic complexity up to poly-logarithmic factors}. This complexity is among the most efficient of all comparable algorithms."

The rest of this paper is laid out as follows. In Section \ref{sec:pre}, we introduce our notations and provide details for explicit description of QLD in \eq{eq:Gao_Lindblad}. Theoretically, we prove the convergence of QLD \note{in the convex case (Theorem \ref{theorem:convex_converge}) and the quasar-convex case (Theorem \ref{the:Nonconvex_convergence}), referring to} Section \ref{sec:con_OLD}. Numerically, we first trace back to spontaneous emission to show the essential cause of energy dissipation in QLD, which verifies our intuition, see Section \ref{sec:spon}. Then we verify our theoretical results of quadratic case in Section \ref{sec:verification} and discuss the roles of each parameter and their relations with thermal and tunneling effects in Section \ref{sec:roles_of_parameters}. In Section \ref{sec:comparison} we compare QLD with classical dynamics, which informs the development of time-dependent QLD to fully leverage the advantages inherent in quantum dynamics. In Section \ref{sec:convergence_time-dependent-qld}, we prove the convergence of time-dependent QLD in convex case. The performance of time-dependent QLD is evaluated in eight representative landscapes, demonstrating superior performance through comparisons with QHD, QAA, and classical algorithms, shown in Section \ref{sec:three_phases_theorem}. Finally, we evaluate and compare the query complexity of QLD with other algorithms in Section \ref{app:query_com}, demonstrating that QLD is superior to other algorithms in terms of query complexity.

\section{\label{sec:pre}Preliminaries}
\noindent
\textbf{Notations. }We denote the Hamiltonian of system as $H$, the Hamiltonian of heat bath as $H_B$, the Hamiltonian of heat bath and system as $H_{\text{total}} = H + H_B$. Let $E_k$ be the kinetic energy of the system and $V$ be the potential energy of the system, so $H=E_k+V$. Given two quantum operators $A$ and $B$, we use $[A,B] = AB-BA$ for denoting the commutation, while $[A,B]_{+} = AB + BA$ for the anti-commutation. We denote the characteristic damping rate of the system as $\eta$ and the temperature of heat bath as $T$. Given a quantum operator $O(t)$, its time-dependent expectation value with respect to density matrix $\rho(t)$ is calculated by $\langle O(t)\rangle_t=\mathrm{Tr}(\rho(t) O(t))$\footnote{\note{We note here that in our setup, the operator $O(t)$ is defined in the Schr\"{o}dinger picture, which might explicitly depend on time.}}.

\label{sec:Lindblad_Approach}

Density matrix approach is often used to describe the dynamics of an open quantum system. Based on this point, Caldeira and Leggett proposed a well-known master equation named after themselves as the Caldeira-Leggett equation (\eq{eq:CLeq}) \cite{caldeira1983path,gardiner2004quantum} to describe the dynamics of quantum Langevin system. However, the Caldeira-Leggett equation faces some serious probelms, such as the non-positivity of the density matrix and only available for high temperature region \cite{gao1997dissipative,diosi1993high,karrlein1997exact}. Refs.~\cite{gao1997dissipative, lindblad1976generators} fixed these problems by introducing a Lindblad equation, which is a more general form of the Caldeira-Leggett equation.

A Lindblad equation can be presented in the standard form:
\begin{align}
  \frac{\d \rho}{\d t}=  \mathcal{L}(\rho)  
  =  -\frac{\im}{\hbar}\left[H, \rho\right]+\sum_{j=1}^{m}\left(2L_{j} \rho L_{j}^{\dagger}- \left[L_{j}^{\dagger} L_{j}, \rho\right]_+\right),
  \label{eq:Lindblad_standard}
\end{align}
where $L_j$ for $j\in[m]$ are linear \note{operators} known as jump operators. 
Ref.~\cite{gao1997dissipative} studied the case where the Hamiltonian $H$ is given by $H = \frac{p^{2}}{2m} + V$ and the unique Lindblad operator $L$ is a linear combination of the position operator $x$ and the momentum operator $p$,
\begin{align}
  L= \mu x+\im\nu p,~~L^{\dagger}=\mu x-\im\nu p,
  \label{eq:Lindblad_operator_L}
\end{align}
where
\begin{align}
  \mu^{2}(T) & =\frac{\eta m \Omega}{2 \hbar} \operatorname{coth}\left(\frac{\hbar \Omega}{4 k T}\right),\nonumber \\
  \nu^{2}(T) & =\frac{\eta}{2 \hbar m \Omega} \tanh \left(\frac{\hbar \Omega}{4 k T}\right)
  \label{eq:coefficient_mu_nu}
\end{align}
with the accompanying relation $2\mu\nu\hbar = \eta$, and $k$ is the Boltzmann constant, $T$ is the temperature, $\eta>0$ is the characteristic damping rate.

This Lindblad equation can be written out explicitly according to \eq{eq:Lindblad_standard} and \eq{eq:Lindblad_operator_L},
\begin{align}
  \frac{\d \rho}{\d t}+\frac{\im}{\hbar}\left[H^{\prime}, \rho\right] =  -\mu^{2}[x,[x, \rho]]-2 \im \mu \nu\left[x,[p, \rho]_{+}\right] -\nu^{2}[p,[p, \rho]],
  \label{eq:Gao_Lindblad}
\end{align}
where $H^{\prime}=H-2 \mu \nu \hbar x p$. We also call this equation the Lindblad equation of \textbf{Quantum Langevin Dynamics}, abbreviated as \textbf{QLD}, which is also the name of our algorithm.  The coordinate space representation of \eq{eq:Gao_Lindblad}~\cite{gao1997dissipative} is also used in the numerical experiments of QLD, stated here for completeness:
\begin{align}
 \frac{\partial \rho\left(x, y, t\right)}{\partial t}&+\frac{\im}{\hbar}\left[\widetilde{H}(x)-\widetilde{H}^{*}\left(y\right)\right] \rho\left(x, y, t\right)
    =  -\left(\mu^{2}(T)\left(x-y\right)^{2}\right. \nonumber \\
      & \left. + \eta\left(x-y\right)\left(\frac{\partial}{\partial x}-\frac{\partial}{\partial y}\right)  -\nu^{2}(T) \hbar^{2}\left(\frac{\partial}{\partial x}+\frac{\partial}{\partial y}\right)^{2}\right) \rho\left(x, y, t\right),
    \label{eq:Gao_Lindblad_xy}
\end{align}
where $\widetilde{H}(x)= H(x)+\im \hbar\eta x \frac{\partial}{\partial x}+\im \frac{\hbar \eta}{2}$.

Although Ref.~\cite{gao1997dissipative} does not elucidate the description of Quantum Langevin Dynamics in high dimension explicitly, its generalization is quite natural. For completeness, we write down the expression of  high-dimensional QLD here. Suppose that the system is in the $\mathbb{R}^{d  }$ space with ${d  }$-dimensional position operator $x=(x_1,x_2,\ldots, x_{d  })^\top $ and momentum operator $p=(p_1,p_2,\ldots, p_{d  })^\top $, intriguing the following commutation relations 
\begin{align}
    [x_i,p_j]=\im \delta _{ij} \hbar,~[x_i,x_j]=0,~[p_i,p_j]=0, ~\text{for }i,j=1,2,\ldots, {d  }.
    \label{eq:commutation_highD}
\end{align}
The Hamiltonian $H$ of high-dimensional QLD is 
\begin{align}
    H=\frac{p_1^2+\cdots+p_{d  }^2}{2m}+V(x_1,\ldots, x_{d  }),
    \label{eq:hamiltonian_highD}
\end{align}
where \note{$V(x_1,\ldots, x_d)\colon \mathbb{R}^{d  } \to \mathbb{R}$} is the potential energy (and also the objective function we want to optimize). The Lindblad operator of the $j$-th dimension in space $\mathbb{R}^{d  }$ is  
\begin{align}
    L_j=\mu x_j + \im \nu p_j,\quad \text{for } j=1,2,\ldots, {d  },
    \label{eq:lindblad}
\end{align}
where  the parameters $\mu$ and $\nu$ are the coefficients of the Lindblad operators. Here we suppose that each dimension holds the same coefficients $\mu$ and $\nu$ given by \eq{eq:coefficient_mu_nu}.
Thus, the Quantum Langevin Dynamics in $\mathbb{R}^ {d  }$ space is
\begin{align}
    \frac{\d \rho}{\d t} = -\frac{\im}{\hbar}[H,\rho] + \sum_{j=1}^{d  } \left( 2L_j\rho L_j^\dagger - \left[ L_j^\dagger L_j,\rho \right]_+ \right),
    \label{eq:highDQLD}
\end{align}
where $\rho$ is the density matrix, $H$ is the Hamiltonian given by \eq{eq:hamiltonian_highD}, and $L_j$ is the Lindblad operator in \eq{eq:lindblad}.

\section{Convergence of QLD}\label{sec:con_OLD}

In this section, our primary focus is the time-independent scenario of QLD, implying that all parameters $\eta,\hbar,T,m,\Omega,k$ remain constant throughout evolution. We first discuss the convergence of QLD for the convex functions, which are defined as follows:
\begin{definition_}
    Let $x^*$ be the global minimum of the differentiable function $V\colon\mathbb{R}^d\to \mathbb{R}$. The function $V$ is convex if for all $x\in \mathbb{R}^d$,
    \begin{align}
        V(x^*) - V(x) \ge \nabla V(x)\cdot (x^* - x).
    \end{align}
    \label{def:convex}
\end{definition_}
In the convex case, the convergence rate of QLD is $O(e^{-\eta t})$ as demonstrated in Theorem \ref{thm:QLD_convergence}. The generalization of QLD to high-dimensional convex case is quite natural while keeping the same convergent speed $O(e^{-\eta t})$, see Section \ref{sec:highD_convex}. Subsequently, we compute the analytical solution of the quadratic potential, the simplest convex function, in Section~\ref{sec:analytical_solution}. The most remarkable discovery related to the quadratic potential is that the system will reach the ground state of the quadratic potential when the evolution time $t\to +\infty$ and the temperature $T\to 0$.
Lastly, we prove that QLD continues to converge to the global minimum even in a specific nonconvex scenario, where the objective function satisfies $x\cdot \nabla V(x)\ge rV(x)\ge 0\text{ for }r\in(0,1)$, at a rate of $O(e^{-\eta rt})$. More details can be referred to Theorem \ref{thm:con_non_convex_qld} of Section~\ref{sec:non_convex_case}. 

\subsection{\label{sec:convex_case}Convergence of QLD in the convex case}

In this subsection, we mainly focus on the one-dimensional case. Our goal is to demonstrate that the expectation value of the one-dimensional objective function $V(x)$ converges to the global minimum $V(x^*)$ at a defined rate under the evolution of QLD. This necessitates the consideration of the expectation value of the Hamiltonian $H$, as if we can establish that the expectation value of $H$ diminishes, it follows that the expectation value of $V$ will also decrease. With this understanding, we initially present Lemma \ref{lem:Ehrenfest}, which enables us to calculate the time evolution of the expectation value of any observable evolving by standard Lindblad equation. After that, we show some commutation relations used in the calculation referring to Lemma \ref{lem:commutation_relations}. Subsequently, we verify the convergence of the expectation value of $H$ in Proposition \ref{prop:non_increasing}, and with the assistance of this proposition, we validate the convergence of the expectation value of $V$ in Theorem \ref{thm:QLD_convergence}. 

\begin{lemma_}[Generalized Ehrenfest theorem]  \label{lem:Ehrenfest}
  Given a time-dependent quantum observable $O(t)$ and let $\rho(t)$ be the solution of the standard Lindblad equation \eq{eq:Lindblad_standard}. Then the time derivative of the expectation value of $O(t)$ is given by
  \begin{small}
    \begin{align}
      \frac{\d}{\d t}\langle O(t)\rangle_t= \lr{ \frac{\d}{\d t} O(t)}_t +\frac{\im}{\hbar}\langle [H,O(t)]\rangle_t \nonumber                                            
      +                                     \sum_{j=1}^{m}\left\langle\left(-L_j^{\dagger}\left[L_j,O(t)\right]+\left[L_j^{\dagger},O(t)\right]L_j \right)\right\rangle_t.
    \end{align}
  \end{small}
\end{lemma_}
\begin{proof}
  \begin{align}
    \frac{\d}{\d t}\langle O(t)\rangle_t & = \frac{\d}{\d t}\Tr(O(t)\rho(t))\nonumber                                                        \\
                                         & =\Tr\left(\frac{\d O(t)}{\d t}\rho(t)\right)+\Tr\left(O(t)\frac{\d \rho(t)}{\d t}\right)\nonumber \\
                                         & =\lr{ \frac{\d}{\d t} O(t)}_t+\Tr\left(O(t)\frac{\d \rho(t)}{\d t}\right).
    \label{eq:generalized_Ehrenfest_1}
  \end{align}
  Let us focus on the second term $\Tr\left(O(t)\frac{\d \rho(t)}{\d t}\right)$. Inserting \eq{eq:Lindblad_standard} into the second term, we get
  \begin{align}
      & \Tr\left(O(t)\frac{\d \rho(t)}{\d t}\right) \nonumber                                                                                              \\
    = & \Tr\left( -\frac{\im}{\hbar}O(t)[H, \rho(t)]+\sum_{j=1}^{m}O(t)\left(2L_{j} \rho(t) L_{j}^{\dagger}- L_{j}^{\dagger} L_{j} \rho(t)- \rho(t) L_{j}^{\dagger} L_{j}\right)\right)\nonumber \\
    = & -\frac{\im}{\hbar}\Tr \left(O(t)[H, \rho(t)]\right) +\Tr \left(\sum_{j=1}^{m}O(t)\left(2L_{j} \rho(t) L_{j}^{\dagger}- L_{j}^{\dagger} L_{j} \rho(t)- \rho(t) L_{j}^{\dagger} L_{j}\right)\right),
    \label{eq:generalized_Ehrenfest_2}
  \end{align}
  where the first term $\Tr \left(O(t)[H, \rho(t)]\right)$ could be simplified as follows:
  \begin{align}
    \Tr \left(O(t)[H, \rho(t)]\right) & =\Tr\left(OH\rho\right)-\Tr\left(O\rho H\right) \nonumber  \\
                                      & =\Tr\left(OH\rho\right)-\Tr\left(HO \rho \right) \nonumber \\
                                      & =\Tr\left([O,H]\rho \right) \nonumber                      \\
                                      & =-\langle[H,O]\rangle_t.
    \label{eq:generalized_Ehrenfest_3}
  \end{align}
  The second term of \eq{eq:generalized_Ehrenfest_2} becomes
  \begin{align}
      & \sum_{j=1}^{m}\Tr \left(O(t)\left(2L_{j} \rho(t) L_{j}^{\dagger}- L_{j}^{\dagger} L_{j} \rho(t)- \rho(t) L_{j}^{\dagger} L_{j}\right)\right)\nonumber \\
    = & \sum_{j=1}^{m}\Tr \left(O(t)2L_{j} \rho(t) L_{j}^{\dagger}\right)-\sum_{j=1}^{m}\Tr \left(O(t)L_{j}^{\dagger} L_{j} \rho(t)\right) -\sum_{j=1}^{m}\Tr \left(O(t)\rho(t) L_{j}^{\dagger} L_{j}\right)\nonumber                                                                            \\
    = & \sum_{j=1}^{m}\Tr \left(2 L_{j}^{\dagger}O(t)L_{j} \rho(t)\right)-\sum_{j=1}^{m}\Tr \left(O(t)L_{j}^{\dagger} L_{j} \rho(t)\right) -\sum_{j=1}^{m}\Tr \left(L_{j}^{\dagger} L_{j} O(t)\rho(t)\right)\nonumber                                                                            \\
    = & \sum_{j=1}^{m}\Tr\left(\left(2 L_{j}^{\dagger}O(t)L_{j} - O(t)L_{j}^{\dagger} L_{j}-L_{j}^{\dagger} L_{j} O(t)\right)\rho(t)\right)\nonumber          \\
    = & \sum_{j=1}^{m}\Tr\left(\left(-L_{j}^{\dagger}[L_{j},O(t)]+\left[L_{j}^{\dagger},O(t)\right]L_{j} \right)\rho(t)\right)\nonumber                       \\
    = & \sum_{j=1}^{m} \left\langle\left(-L_j^{\dagger}[L_j,O(t)]+\left[L_j^{\dagger},O(t)\right]L_j \right)\right\rangle_t.
    \label{eq:generalized_Ehrenfest_4}
  \end{align}
  Inserting \eq{eq:generalized_Ehrenfest_2}, \eq{eq:generalized_Ehrenfest_3} and \eq{eq:generalized_Ehrenfest_4} into \eq{eq:generalized_Ehrenfest_1}, we complete the proof.
\end{proof}

\begin{lemma_}[Commutation relations]\label{lem:commutation_relations}
  Let $x$ and $p$ be the one-dimensional position operator and momentum operator, respectively. $f$ is any continuously differentiable function. Then
  \begin{subequations}
    \begin{align}
       & [x,p^2]=2\im \hbar p, \label{eq:com_1}                 \\
       & [f(x),p]=\im \hbar \nabla f, \label{eq:com_2}          \\
       & [\nabla f(x),p]=\im \hbar \nabla^2 f, \label{eq:com_4} \\
       & [f(x),p^2]=\im \hbar[\nabla f,p]_+, \label{eq:com_3}   \\
       & [p,xp]=[p,px]=-\im\hbar p, \label{eq:com_5}            \\
       & [x,px]=[x,xp]=\im \hbar x, \label{eq:com_7}            \\
       & [p^2,xp]=[p^2,px]=-2\im \hbar p^2, \label{eq:com_6}    \\
       & [x^2,xp]=[x^2,px]=2\im \hbar x^2.
    \end{align}
  \end{subequations}
\end{lemma_}

\begin{proposition_} \label{prop:1_initial_state}
  If $\lim_{x\to \infty} V(x) = +\infty$, there always exists an initial density matrix  $\rho(0)$ that satisfies $\lr{H}_{0}>R $, for any $R\in \mathbb{R}, R>0$.
\end{proposition_}
\begin{proof}
  This proof is quite straightforward. We only need to ensure the initial density matrix $\rho(0)$  is heavily weighted in the region where $V$ is large to realize $\lr{V}_{0}>R$. Because $\lr{V}_{0}\le \lr{H}_0$, we have $\lr{H}_{0}>R $.
\end{proof}

\begin{proposition_}\label{prop:non_increasing}
  Assume that $V$ is a continuously differentiable convex function and $\text{\note{$\lr{\nabla^{2}V}_t$}} < +\infty$.\footnote{\note{This assumption depends on the property of the potential function $V(x)$ specifically. Suppose that the domain of $x$ is $\mathcal{K}$. In practice, $\mathcal{K}$ is always finite, which implies that $\nabla^2 V(x)\le \max_{x\in\mathcal{K}}\nabla^2 V(x)$.}} If $\hbar \Omega \ll kT$ and the initial condition satisfies $\lr{H}_{0} -\text{\note{$\zeta$}}kT> 0$, the inequality $\lr{H}_{t}-\text{\note{$\zeta$}}kT\le (\lr{H}_{0}-\text{\note{$\zeta$}}kT)e^{-\eta t} = O(e^{-\eta t})$ holds under the evolution of \eq{eq:Gao_Lindblad}.
\end{proposition_}

\begin{proof}
  Without loss of generality, we still set $x^* = 0$ and $V(x^*) = 0$.     By Lemma~\ref{lem:Ehrenfest}, we have
  \begin{align}
    \frac{\d}{\d t}\langle H\rangle_t = & \left\langle \frac{\d}{\d t} H\right\rangle_t +\frac{\im}{\hbar}\langle [H,H]\rangle_t  + \left\langle - L^{\dagger}[L,H] + \left[L^{\dagger},H\right]L \right\rangle_t.
    \label{eq:energy_derivative_HHH}
  \end{align}
  The first and second terms are zero, so we only need to calculate the third term. Inserting \eq{eq:Lindblad_operator_L} into the third term, we get
  \begin{align}
      & \left\langle-L^{\dagger}[L,H]+\left[L^{\dagger},H\right]L \right\rangle_t \nonumber                                                          \\
    = & \left\langle-L^{\dagger}[\mu x+\im\nu p,p^2/(2m)+V]+ \left[\mu x-\im\nu p,p^2/(2m)+V\right]L \right\rangle_t \nonumber                                                                     \\
    = & \left\langle-L^{\dagger}\left(\frac{\mu}{2m}[x,p^2]+\cancel{\mu[x,V]}+\cancel{\im\frac{\nu}{2m}[p,p^2]}+\im \nu[p,V] \right)\right.\nonumber \\
      & +\left.\left( \frac{\mu}{2m}[x,p^2]+\cancel{\mu[x,V]}-\cancel{\im \frac{\nu}{2m}[p,p^2]}-\im \nu[p,V]\right)L \right\rangle_t\nonumber       \\
    = & \left\langle(-\mu x+\im  \nu p)\left(\im \frac{\mu}{m} \hbar p+\nu \hbar\nabla V\right)+\left( \im \frac{\mu}{m}\hbar p-\nu\hbar\nabla V\right)(\mu x+\im \nu p) \right\rangle_t\nonumber                                     \\
    = & \left\langle -\im \frac{\mu^2}{m}\hbar (xp-px)-2\mu\nu\hbar( x\cdot \nabla V+p^2/m)\right \rangle_t + \left\langle\im\nu^2\hbar(p\cdot\nabla V-\nabla V \cdot p)\right\rangle_t\nonumber                                                         \\
    = & \left\langle -\im \frac{\mu^2}{m}\hbar [x,p]-2\mu\nu\hbar( x\cdot\nabla V+p^2/m)+\im \nu^2\hbar[p,\nabla V]\right\rangle_t\nonumber          \\
    = & \left\langle \frac{\mu^2}{m}\hbar^2  -2\mu\nu\hbar( x\cdot\nabla V+p^2/m)+\nu^2\hbar^2 \nabla^2 V\right\rangle_t\nonumber                    \\
    = & \hbar^2 \left(\frac{\mu^2}{m}+\nu^2\langle\nabla^2 V\rangle_t\right) -2\mu\nu\hbar\langle x\cdot\nabla V+p^2/m\rangle_t.
  \end{align}
  In the above derivations, Lemma~\ref{lem:commutation_relations} is also used to simplify.

  Known that $2\hbar\mu\nu=\eta$, put this into the above equation, then we get
  \begin{subequations}
    \begin{align}
      \frac{\d}{\d t}\langle H\rangle_t= & \hbar^2 \left(\frac{\mu^2}{m}+\nu^2\langle\nabla^2 V\rangle_t\right) -2\mu\nu\hbar\langle x\cdot\nabla V+p^2/m\rangle_t  \label{eq:energy_derivative_H_1}         \\
      =                                  & \left(\frac{\hbar^2\mu^2}{m}+\frac{\eta^2}{4\mu^2}\langle\nabla^2 V\rangle_t\right) -\eta\langle x \cdot\nabla V+p^2/m\rangle_t.   \label{eq:energy_derivative_H}
    \end{align}
  \end{subequations}
  According to the approximation $\hbar \Omega \ll kT$, \eq{eq:coefficient_mu_nu} becomes
  \begin{align}
    \mu^2=           & \frac{\eta m \Omega}{2 \hbar} \left(\frac{4kT}{\hbar \Omega} + \frac{\hbar\Omega }{12kT} + O\left(\left(\frac{\hbar \Omega}{kT}\right)^3\right)\right), \label{eq:mu_approx_1}\\
    \frac{1}{\mu^2}= & \frac{2 \hbar}{\eta m \Omega} \left(\frac{\hbar \Omega}{4kT} + O\left(\left(\frac{\hbar \Omega}{kT}\right)^3\right)\right).
    \label{eq:mu_approx}
  \end{align}
  Inserting the above equations into \eq{eq:energy_derivative_H}, \eq{eq:energy_derivative_H} becomes
  \begin{align}
    \frac{\d}{\d t}\langle H\rangle_t= & \eta\cdot  2kT+\eta\cdot\frac{\hbar\Omega}{kT}\left(\frac{\hbar \Omega}{24}+\frac{\hbar\langle\nabla^2 V\rangle_t}{8m\Omega} \right)  + O\left(\left(\frac{\hbar \Omega}{kT}\right)^3\right)-\eta\left\langle x \cdot\nabla V+\frac{p^2}{m}\right\rangle_t.
    \label{eq:energy_derivative_H_approx}
  \end{align}
  The assumption $\text{\note{$\lr{\nabla^{2}V}_t$}} < +\infty$ and approximation $\frac{\hbar\Omega}{kT}\ll 1$ guarantee that $\eta\cdot\frac{\hbar\Omega}{kT}\cdot\frac{\hbar}{8m\Omega}\langle\nabla^2 V\rangle_t$ is negligible compared to $\eta\cdot  2kT$, which also tells us how small $\frac{\hbar\Omega}{kT}$ must be at least to ensure the approximation work. \note{Putting all the negligible terms into a constant $O\left(\eta\frac{\hbar\Omega}{kT}\right)$}, we have
\note{\begin{align}
    \frac{\d}{\d t}\langle H\rangle_t= & \eta\cdot  \left(kT\left(2+ O\left(\frac{\hbar\Omega}{kT}\right) \right)-\langle x \cdot\nabla V+p^2/m\rangle_t  \right) \label{eq:energy_derivative_H_approx_final} \\
    \leq                               & \eta\cdot  \left(kT\left(2+ O\left(\frac{\hbar\Omega}{kT}\right) \right) -\langle x \cdot\nabla V+p^2/(2m)\rangle_t \right) \nonumber                                  \\
    \leq                               & \eta\cdot  \left(kT\left(2+ O\left(\frac{\hbar\Omega}{kT}\right) \right)-\langle V+p^2/(2m)\rangle_t \right),   \label{eq:approximation_H_leq}
  \end{align}}
where the last inequality comes from the property of convex functions $x\cdot \nabla V(x) \ge V(x)\geq 0$. Finally, \note{denote $\zeta = 2+ O\left(\frac{\hbar\Omega}{kT}\right)$,} we have
  \begin{align}
    \frac{\d}{\d t} \lr{H}_{t} \le \eta \cdot (\text{\note{$\zeta$}} kT - \lr{H}_{t}).
    \label{eq:energy_derivative_H_final_1111}
  \end{align}
  Construct a function $h(t) = (\lr{H}_{t}-\text{\note{$\zeta$}} kT)e^{\eta t}$, and it is direct to verify that $\frac{\d h(t)}{\d t} \le 0$ for $\forall t > 0$ by \eq{eq:energy_derivative_H_final_1111}. Consequently, we get $h(t)\leq h(0)$,
  \begin{align}
    \lr{H}_{t}-\text{\note{$\zeta$}} kT\le (\lr{H}_{0}-\text{\note{$\zeta$}} kT)e^{-\eta t} = O(e^{-\eta t}).
  \end{align}
\end{proof}

\begin{theorem_}[Convergence of QLD]\label{thm:QLD_convergence}
  Assume that $V$ is a continuously differentiable convex function and $\text{\note{$\lr{\nabla^{2}V}_t$}} <+ \infty$. Let $x^*$ be the unique global minimum of $V(x)$. Then the solution $\rho(t)$ to \eq{eq:Gao_Lindblad}, with approximation $\hbar \Omega \ll kT$, satisfies that
\begin{align}
    \lr{V}_{t} - V(x^*) \le \text{\note{$\zeta$}} kT + O(e^{-\eta t}),
    \label{eq:V_converge_convex}
\end{align}
  if the initial density matrix $\rho(0)$ satisfies $\lr{H}_{0}-2kT \ge 0$, where \note{$\zeta$ denotes the constant $2+O\left(\frac{\hbar\Omega}{kT}\right)$, and} $\langle \cdot \rangle_t$ denotes the expectation value with respect to the density matrix $\rho(t)$, i.e., $\langle \cdot\rangle_t = \Tr (\cdot\rho(t))$.
  \label{theorem:convex_converge}
\end{theorem_}
\begin{proof}
  Without loss of generality, we can position the global minimum of $V$ at the origin by shifting the function vertically and horizontally, which derives $x^*=0$ and $V(x^*)=0$. Then we only need to prove
  \begin{align}
    \lr{V}_{t} \le \text{\note{$\zeta$}} kT + O(e^{-\eta t}).
  \end{align}
  Because $\lr{\frac{p^2}{2m}}_t+\lr{V}_t=\lr{H}_t$ and all these three terms are greater than or equal to  zero, we have
  \begin{align}
    \lr{V}_{t}\le \lr{H}_t.
    \label{eq:VleqH}
  \end{align}
  By Proposition~\ref{prop:non_increasing}, we get $ \lr{V}_{t}\le \lr{H}_t\leq \text{\note{$\zeta$}} kT+O(e^{-\eta t})$.
\end{proof}

\note{In the theorem above, we prove the convergence of QLD with the assumption $\hbar \Omega \ll kT$. According to the physical significance of these parameters, $\hbar \Omega$ can be seen as a characteristic energy of quantum effect, while $kT$ can be seen as a characteristic energy of thermal effect. Thus, this theorem implies that the convergence in convex cases of QLD is guaranteed by the thermal effect of the heat bath, which provides dissipation. Intuitively, the benefits of quantum effect is reflected in the non-convex situations. Due to the difficulty of theoretical analysis in non-convex case, we verify our intuition with numerical experiments in Section~\ref{sec:roles_of_parameters}.}

\begin{remark_}\label{remark:QLD_convergence}
  If the desired accuracy is $\varepsilon$, which means we want $\lr{V}_{t} - V(x^*) \le \varepsilon$ when $t\to +\infty$. To achieve this, we only need to set $kT=\varepsilon/\text{\note{$\zeta$}}$ and ensure the approximation $\hbar \Omega\ll kT$ still holds.
\end{remark_}

\subsection{Generalization to the high-dimensional case} \label{sec:highD_convex}
This subsection mainly generalizes the one-dimensional case, discussed in Section \ref{sec:convex_case}, to the high-dimensional case following the high-dimensional model put forward in \eq{eq:highDQLD}. The technical roadmap used in this part is to calculate the linear combination for each dimension. The same convergent speed $O(e^{-\eta t})$ still holds here, as demonstrated in Theorem \ref{thm:QLD_convex_highD}.
\begin{theorem_}\label{thm:QLD_convex_highD}
      Assume that $V(x)\colon \mathbb{R}^d  \to  \mathbb{R}$ is an $d  $-dimensional continuously differentiable convex function and $\text{\note{$\lr{\nabla^{2}V}_t$}} <+ \infty$, where $\nabla^2 V=\sum _{j=1}^d   \frac{\partial ^2 V}{\partial x_j^2}$ is the Laplacian of $V$. Let $x^*$ be the unique global minimum of $V(x)$. Then the solution $\rho(t)$ to \eq{eq:highDQLD}, with approximation $\hbar \Omega \ll kT$, satisfies that
\begin{align}
    \lr{V}_{t} - V(x^*) \le \text{\note{$\zeta$}}k dT + O(e^{-\eta t}),
\end{align}
  if the initial density matrix $\rho(0)$ satisfies $\lr{H}_{0}-\text{\note{$\zeta$}}k d T \ge 0$, where \note{$\zeta$ denotes the constant $2+O\left(\frac{\hbar\Omega}{kT}\right)$, and} $\langle \cdot \rangle_t$ denotes the expectation value with respect to the density matrix $\rho(t)$, i.e., $\langle \cdot\rangle_t = \Tr (\cdot\rho(t))$.
\end{theorem_}

\begin{proof}
    This proof is mainly intrigued by the proof of Theorem  \ref{thm:QLD_convergence} with the same logic as Section \ref{sec:convex_case}. Following the same procedure as in \Cref{prop:non_increasing} and referring to the high-dimensional commutation relations in \eq{eq:commutation_highD}, we can obtain  the time derivative of  expectation value of the Hamiltonian in $\mathbb{R}^d  $ space, i.e., \eq{eq:energy_derivative_H_1} changing to
\begin{align}
    \frac{\d }{\d t}\lr{H}_t= \hbar ^2 \lrb{\frac{\mu ^2 d  }{m}+\nu ^2 \lr{\nabla^2 V}_t} -2\mu \nu \hbar \lr{x\cdot \nabla V+\frac{p^2}{m}}_t,
\end{align}
where $\nabla^2 V=\sum _{j=1}^d   \frac{\partial ^2 V}{\partial x_j^2}$ is the Laplacian of $V$, and $x\cdot \nabla V = \sum_{j=1}^d   {x_j} \frac{\partial V}{\partial x_j}$ is the inner product of $x$ and the gradient of $V$, and $p^2=\sum _{j=1}^d   p_i^2$ is the square of the momentum operator. With the same approximation $\hbar \Omega \ll kT$ (\eq{eq:mu_approx_1} and \eq{eq:mu_approx}), \eq{eq:energy_derivative_H_approx_final} becomes 
  \note{\begin{align}
    \frac{\d}{\d t}\langle H\rangle_t= & \eta\cdot  \left(kdT\left(2+ O\left(\frac{\hbar\Omega}{kT}\right) \right)-\langle x \cdot\nabla V+p^2/m\rangle_t  \right)   \nonumber\\
    \leq&  \eta \cdot (\zeta kd T - \lr{H}_{t}),
  \end{align}}
  where the last inequality also uses the property of convex functions $x\cdot \nabla V(x) \ge V(x)\geq 0$. The subsequent logic is exactly same as the procedure of Section \ref{sec:convex_case}, neglected for simplicity. Finally, we arrive at 
    \begin{align}
        \lr{V}_{t} - V(x^*) \le \text{\note{$\zeta$}}kdT + O(e^{-\eta t}).
    \end{align}
\end{proof}

Similar discussion to Theorem \ref{thm:QLD_convex_highD} can also be applied in following \Cref{thm:con_non_convex_qld}, \Cref{cor:QLD_convergence_time-dependent_eta,cor:QLD_convergence_time-dependent_hbar,cor:QLD_convergence_time-dependent_T}, and the generalization of high-dimension is quite obvious. We only need to multiply the dimensionality $d  $  in front of  the temperature related term $T$ or $f(T)$, while the convergent speed is not influenced at all. 

\subsection{\label{sec:analytical_solution}Analytical solution for the quadratic potentials}
It is intractable to solve the Lindblad equation of QLD \eq{eq:Gao_Lindblad} analytically in most cases. However, we can solve it analytically for quadratic potentials. Here, we consider the simplest one-dimensional case with quadratic potential $V(x)=\frac{1}{2}m\Omega^2 x^2$, which also implies that the system is a harmonic oscillator with energy level $\mathcal{H}_k=(\frac{1}{2}+k)\hbar \Omega$ ($k=0,1,\ldots$). For the multidimensional scenario where $V(\mathbf{x})=\frac{1}{2}x^T \mathbf{A}x$ with a positive semidefinite matrix $\mathbf{A}$, it can be solved by the normal mode analysis method (regularization)~\cite{goldstein2002classical}. 
In the following part, we mainly calculate the time evolutions of expectation value of kinetic energy $\lr{\frac{p^2}{2m}}_t$ and potential energy $\lr{V}_t$. An interesting thing we find is that the system always reaches the ground state of harmonic oscillator with ground state energy $\mathcal{H}_0=\frac{1}{2}\hbar \Omega$ if $T\to0$ when $t\to+\infty$, no matter what initial density matrix $\rho(0)$ we set.

At first, we need to calculate $\frac{\d}{\d t}\langle V\rangle_t$ similar to the previous calculation of $\frac{\d}{\d t}\langle H\rangle_t$ in Proposition \ref{prop:non_increasing}. By Lemma~\ref{lem:Ehrenfest} and Lemma~\ref{lem:commutation_relations}, we have
\begin{align}
     \frac{\d}{\d t}\langle V\rangle_t \nonumber                                                                                                                               
  = & \cancel{\left\langle \frac{\d}{\d t} V\right\rangle_t} +\frac{\im}{\hbar}\langle [H,V]\rangle_t + \left\langle-L^{\dagger}[L,V]+\left[L^{\dagger},V\right]L \right\rangle_t\nonumber \\
  = & \frac{\im}{\hbar}\langle [p^2/(2m)+V,V]\rangle_t  + \left\langle-L^{\dagger}[\mu x+\im\nu p,V]+\left[\mu x-\im\nu p,V\right]L \right\rangle_t\nonumber                                                                      \\
  = & \frac{\im}{\hbar}\langle [p^2/(2m),V] +\cancel{[V,V]}\rangle_t+\langle-L^{\dagger}(\mu \cancel{[x,V]}+\im\nu [p,V]) +\left(\mu \cancel{[x,V]}-\im\nu [p,V]\right)L \rangle_t\nonumber                                                                                                         \\
  = & \frac{\im}{2m\hbar}\langle [p^2,V]\rangle_t+\langle-L^{\dagger}\im\nu [p,V]-\im\nu [p,V]L \rangle_t\nonumber                                                              \\
  = & \frac{\im}{2m\hbar}\langle-\im\hbar[\nabla V,p]_+\rangle_t +\langle(-\mu x+\im\nu p)\nu \hbar\nabla V-\nu \hbar \nabla V(\mu x+\im\nu p) \rangle_t\nonumber                                                                          \\
  = & \frac{1}{2m}\langle[\nabla V,p]_+\rangle_t-2\mu\nu \hbar\langle x \cdot\nabla V \rangle_t+\hbar^2\nu^2 \langle \nabla^2 V \rangle_t.
\end{align}
Then, inserting $V(x)=\frac{1}{2}m\Omega^2 x^2$ into above equation and implementing Lemma~\ref{lem:commutation_relations} again, we get
\begin{align}
  \frac{\d}{\d t}\langle x^2\rangle_t & =\frac{1}{m}\langle[x,p]_+\rangle_t-4\mu\nu \hbar\langle x^2 \rangle_t+2\hbar^2\nu^2 \nonumber \\
                                      & =\frac{1}{m}\langle[x,p]_+\rangle_t-2\eta\langle x^2 \rangle_t+2\hbar^2\nu^2.
  \label{eq:energy_derivative_V}
\end{align}
From \eq{eq:energy_derivative_H}, we have
\begin{align}
     \frac{\d}{\d t}\left \langle \frac{p^2}{2m}+\frac{1}{2}m\Omega ^2x^2\right\rangle_t =\hbar^2 \left(\frac{\mu^2}{m}+\nu^2m\Omega^2\right) -\eta\langle m\Omega^2x^2+p^2/m\rangle_t.
  \label{eq:energy_derivative_H_3}
\end{align}
\eq{eq:energy_derivative_H_3} minus \eq{eq:energy_derivative_V}:
\begin{align}
  \frac{\d}{\d t}\langle {p^2}\rangle_t & =-m\Omega^2\langle[x,p]_+\rangle_t-2\eta\langle p^2 \rangle_t+2\hbar^2\mu^2.
  \label{eq:energy_derivative_p2_2}
\end{align}

Observing that the time differential equations of  $\langle x^2\rangle_t$ and  $\langle {p^2}\rangle_t$ have been collected in \eq{eq:energy_derivative_V} and \eq{eq:energy_derivative_p2_2}, if we want to formulate a complete set of differential equations, time differential equation of $\langle[x,p]_+\rangle_t$ must be obtained. By Lemma~\ref{lem:Ehrenfest} and Lemma~\ref{lem:commutation_relations}, we can calculate
\begin{align}
  \frac{\d}{\d t}\left \langle[x,p]_+\right \rangle_t= \cancel{\left\langle \frac{\d}{\d t} [x,p]_+\right\rangle_t}+\frac{\im}{\hbar}\left \langle\left[H,[x,p]_+\right]\right \rangle_t 
  +                                                     \left\langle-L^{\dagger}[L,[x,p]_+]+\left[L^{\dagger},[x,p]_+\right]L \right\rangle_t,
  \label{eq:de_x_p}
\end{align}
where
\begin{align}
     \left \langle\left[H, [x,p]_+ \right]\right \rangle_t  \nonumber                                                                                          
  = & \left \langle\left[\frac{p^2}{2m}+\frac{1}{2}m\Omega^2x^2, [x,p]_+ \right]\right \rangle_t \nonumber                                                      \\
  = & \frac{1}{2m}\left \langle\left[p^2, [x,p]_+ \right]\right \rangle_t+\frac{1}{2}m\Omega^2\left \langle\left[x^2, [x,p]_+ \right]\right \rangle_t \nonumber \\
  = & \frac{1}{2m}\left \langle-4\im \hbar p^2\right \rangle_t+\frac{1}{2}m\Omega^2\left \langle4\im \hbar x^2\right \rangle_t \nonumber                        \\
  = & \frac{-2\im \hbar }{m}\left \langle p^2\right \rangle_t+2 \im \hbar m\Omega^2\left \langle x^2\right \rangle_t
  \label{eq:commutator_H_x_p}
\end{align}
and
\begin{align}
    & \left\langle-L^{\dagger}[L, [x,p]_+ ]+\left[L^{\dagger}, [x,p]_+ \right]L \right\rangle_t\nonumber                               \\
  = & \left\langle-L^{\dagger}[\mu x+i\nu p, [x,p]_+ ]+\left[\mu x-i\nu p, [x,p]_+ \right]L \right\rangle_t\nonumber                   \\
  = & \left\langle -L^{\dagger} \left( \mu\left[ x, [x,p]_+ \right]+ \im \nu\left[ p, [x,p]_+ \right]\right)  \right\rangle_t +\left\langle \left( \mu\left[ x, [x,p]_+ \right]- \im \nu\left[ p, [x,p]_+ \right]\right) L\right\rangle_t\nonumber             \\
  = & \left\langle -L^{\dagger} \left( 2\im\mu \hbar x+ 2 \nu \hbar p\right)  \right\rangle_t
  +\left\langle \left( 2\im\mu \hbar x- 2 \nu \hbar p \right) L\right\rangle_t\nonumber                                                \\
  = & \left\langle -(\mu x-\im \nu p) \left( 2\im\mu \hbar x+ 2 \nu \hbar p\right)  \right\rangle_t +\left\langle \left( 2\im\mu \hbar x- 2 \nu \hbar p \right) (\mu x+\im \nu p)\right\rangle_t\nonumber                            \\
  = & -4\hbar \mu\nu \left\langle [x,p]_+  \right\rangle_t\nonumber                                                                    \\
  = & -2\eta\left\langle [x,p]_+  \right\rangle_t.
  \label{eq:commutator_x_p}
\end{align}
Inserting \eq{eq:commutator_H_x_p} and \eq{eq:commutator_x_p} into \eq{eq:de_x_p}:
\begin{align}
    \frac{\d}{\d t}\left \langle [x,p]_+ \right \rangle_t=\frac{2 }{m}\left \langle p^2\right \rangle_t-2  m\Omega^2\left \langle x^2\right \rangle_t-2\eta\left\langle [x,p]_+  \right\rangle_t.
    \label{eq:de_x_p_2}
\end{align}

For convenience, we restate all three equations together here
\begin{subequations}
    \begin{align}
       & \frac{\d}{\d t}\langle x^2\rangle_t=\frac{1}{m}\langle[x,p]_+\rangle_t-2\eta \langle x^2 \rangle_t+2\hbar^2\nu^2, \label{eqsystem1}                                                                          \\
       & \frac{\d}{\d t}\left \langle[x,p]_+\right \rangle_t=\frac{2 }{m}\left \langle p^2\right \rangle_t-2  m\Omega^2\left \langle x^2\right \rangle_t-2\eta \left\langle[x,p]_+ \right\rangle_t, \label{eqsystem2} \\
       & \frac{\d}{\d t}\langle p^2\rangle_t=-m\Omega^2\langle[x,p]_+\rangle_t-2\eta \langle p^2 \rangle_t+2\hbar^2\mu^2,\label{eqsystem3}
    \end{align}
\end{subequations}
which form a complete group of differential equations.
\eq{eqsystem1} gives
\begin{align}
  \frac{1}{m}\langle [x,p]_+ \rangle_t=\frac{\d}{\d t} \langle x^2\rangle_t+2 \eta \langle x^2 \rangle_t -2\hbar^2\nu^2.
  \label{eq:xpmoment_xp}
\end{align}
Inserting it into \eq{eqsystem2}, then $\left\langle p^2\right \rangle_t$ can be presented as a function of $\langle x^2\rangle_t$ and its time derivatives
\begin{align}
  \left \langle p^2\right \rangle_t= & \frac{m^2}{2} \frac{\d^2}{\d t^2}\left \langle x^2\right \rangle_t+2m^2 \eta  \frac{\d}{\d t} \left \langle x^2\right \rangle_t  +m^2\left(2 \eta ^2+\Omega^2\right)\left \langle x^2\right \rangle_t-2\hbar^2 \nu ^2 m^2  \eta . \label{eq:xpmoment_p2}
\end{align}

Finally, we insert \eq{eq:xpmoment_xp} and the time derivative of \eq{eq:xpmoment_p2} into \eq{eqsystem3}, which gives
\begin{align}
  \frac{\d^{3}\langle x^{2} \rangle_{t}}{\d t^{3}} +6 \eta  \frac{\d^{2}\langle x^{2} \rangle_{t}}{\d t^{2}} + (4\Omega^{2}+12 \eta ^{2})\frac{\d\langle x^{2} \rangle_{t}}{\d t} 
  +8 \eta  (\Omega^{2}+ \eta ^{2})\langle x^{2} \rangle_{t} - D=0,
  \label{x2moment}
\end{align}
where $D$ is a constant
\begin{align}
  D = \frac{4}{m^{2}}\mu^{2}\hbar^{2} + 4(\Omega^{2} + 2 \eta ^{2})\nu^{2}\hbar^{2}.
\end{align}
The analytical solution of \eq{x2moment} is
\begin{align}
  \lr{x^{2}}_{t} = & e^{-2 \eta  t}\left (A + Be^{2\im \Omega t}+Ce^{-2\im\Omega t}\right)+\frac{D}{8 \eta  (\Omega^{2}+ \eta ^{2})}\nonumber \\
  =                & e^{-2 \eta  t}\left(A + \tilde{B} \cos(2\Omega t)+\tilde{C} \sin(2\Omega t)\right)+\frac{D}{8 \eta (\Omega^{2}+ \eta ^{2})},
  \label{eq:x2moment_solution}
\end{align}
where $A$,$B$,$C$ are three constants determined by the initial condition of the system and $A$,$\tilde{B}$,$\tilde{C}\in \mathbb{R}$.
This also means that the convergence rate of $\lr{V(x)}_{t} = \frac{1}{2}m\Omega^{2}\lr{x^{2}}_{t}$ is $O(e^{-2 \eta  t})$.

Calculating the time derivatives of \eq{eq:x2moment_solution}, we have
\begin{align}
  \frac{\d}{\d t} \lr{x^{2}}_{t} = & e^{-2 \eta  t} \left(-2 \eta A + 2\cos(2\Omega t)(- \eta \tilde{B}+\Omega \tilde{C}) +2\sin(2\Omega t) (-\Omega \tilde{B} - \eta  \tilde{C})  \right),
\end{align}
and
\begin{align}
     \frac{\d^2}{\d t^2}\lr{x^{2}}_{t}
  =  e^{-2 \eta  t} \left(4 \eta ^2A + 4\cos(2\Omega t)( \eta ^2\tilde{B} -\Omega^2\tilde{B}-2  \eta \Omega \tilde{C}) \right.\nonumber \\\left.  +4\sin(2\Omega t) ( \eta ^2\tilde{C}-\Omega^2 \tilde{C} +2 \eta \Omega \tilde{B})  \right).
\end{align}
Inserting the above two equations into  \eq{eq:xpmoment_p2},
we get
\begin{align}
  \left \langle \frac{p^2}{m^2}\right \rangle_t= & e^{-2 \eta  t}  \left(A\Omega^2 -\Omega^2 \tilde{B} \cos (2\Omega t) -\Omega^2 \tilde{C}   \sin(2\Omega t) \right)+\frac{(2 \eta ^2 +\Omega^2)D }{8 \eta  (\Omega^{2}+ \eta ^{2})}-2\hbar^2 \nu ^2   \eta .
  \label{eq:p2moment_solution}
\end{align}

When $t\to +\infty$, the convergence  value of average potential is
\begin{align}
  \lr{V}_f=\frac{1}{2}m\Omega^2\lr{x^{2}}_{t\to +\infty}\to \frac{1}{2}m\Omega^2\frac{D}{8 \eta (\Omega^{2}+ \eta ^{2})}.
  \label{eq:V_f}
\end{align}
When $T\to 0$ and $t\to +\infty$, $ \lr{V}_f=\lr{\frac{1}{2}m\Omega^{2}x^{2}}_{t}\to \frac{1}{4}\hbar \Omega$.

When $t\to+\infty$, the convergence value of average kinetic energy is
\begin{align}
  \lr{E_k}_f= \lr{\frac{p^2}{2m}}_{t\to +\infty}\to \frac{m(2 \eta ^2 +\Omega^2)D }{16 \eta  (\Omega^{2}+ \eta ^{2})}-m\hbar^2 \nu ^2   \eta .
  \label{eq:E_k_f}
\end{align}
When $T\to 0$ and $t\to +\infty$, $ \lr{E_k}_f=\lr{\frac{p^2}{2m}}_{t\to+\infty}\to \frac{1}{4}\hbar \Omega$.
The above evidences suggest the system reaches the ground state of energy level $\mathcal{H}_0=\frac{1}{2}\hbar \Omega$.

We can also view this fact from another point of view. Inserting \eq{eq:coefficient_mu_nu} into \eq{eq:energy_derivative_H_3}, the time derivative of $\lr{H}_{t}$ is
\begin{align}
  \frac{\d \lr{H}_{t}}{\d t} = & -\eta \left\langle m\Omega^{2}x^{2}+\frac{p^{2}}{m}\right\rangle_{t} + \frac{1}{2}\eta \Omega \hbar \coth\left(\frac{\hbar \Omega}{4kT}\right) + \frac{1}{2}\eta \Omega \hbar \tanh\left(\frac{\hbar \Omega}{4kT}\right)\nonumber                                                                       \\
  =                            & -2\eta\lr{H}_{t} + \eta \hbar \Omega \coth \left( \frac{\hbar \Omega}{2kT} \right).\label{eq:energy_l_HH}
\end{align}
Trying to present the energy level of the system by a time-dependent energy level number $n_t$ ($n_t\in \mathbb{R} $), we have $\lr{H}_{t} = (n_{t} + 1/2)\hbar \Omega$. The evolution of $n_{t}$ indicates the energy level transition of the system. Then \eq{eq:energy_l_HH} becomes
\begin{align}
  \frac{\d \lr{H}_{t}}{\d t} = \left(-2n_{t} -1 +\coth \left( \frac{\hbar \Omega}{2kT}\right)\right)\eta \hbar \Omega .
  \label{eq:converge}
\end{align}
When $n_{t}>\frac{1}{2}\left(\coth \left( \frac{\hbar \Omega}{2kT} \right)-1\right)$, $\frac{\d  \lr{H}_{t}}{\d  t}<0$ makes $n_{t}$ keep decreasing until $n_{t}\to \frac{1}{2}\left(\coth \left( \frac{\hbar \Omega}{2kT} \right)-1\right)$. Unless the temperature is extremely small ($T\to0\Rightarrow\coth\left( \frac{\hbar \Omega}{2kT}\right)\to1^+ $), the system cannot reach the ground state.

\subsection{\label{sec:non_convex_case}Convergence of QLD in the quasar-convex case}
For general nonconvex potential, we cannot prove the convergence of QLD. However, when facing some special nonconvex ones, QLD can still converge to the global minimum within desired precision. The set of quasar-convex functions are defined as follows~\cite{hardt2018gradient,hinder2020near}:
\begin{definition_}[Quasar-Convexity]
    Let $r\in (0,1]$ and let $x^*$ be a minimizer of the differentiable function $V\colon\mathbb{R}^d\to \mathbb{R}$. The function $V$ is $r$-quasar-convex with respect to $x^*$ if for all $x\in \mathbb{R}^d$,
    \begin{align}
        V(x^*) - V(x) \ge \frac{1}{r}\nabla V(x)\cdot (x^* - x).
    \end{align}
    \label{eq:quasar-convex}
\end{definition_}
In the case $r=1$, Definition~\ref{eq:quasar-convex} is simply star-convexity~\cite{nesterov2006cubic,lee2016optimizing}. We find QLD can still converge to the global minimum within desired precision in this case. Here, we still assume this kind of potential is continuously differentiable with unique global minimum $x^*=0$ and $V(x^*)=0$, which satisfies $x\cdot \nabla V(x) \ge rV(x)$. In this case, we establish the following convergence rate of QLD by applying similar proof techniques to Theorem \ref{thm:QLD_convergence}.
\begin{theorem_}[Convergence of QLD in quasar-convex case]\label{thm:con_non_convex_qld}
  Assume that $V$ is a continuously differentiable function, which satisfies $\text{\note{$\lr{\nabla^{2}V}_t$}}< +\infty$ and $x\cdot \nabla V(x)\ge rV(x)\ge 0$, where $r\in (0,1)$ is a constant. Let $x^*$ be the unique global minimum of $V(x)$. Then the solution $\rho(t)$ to  \eq{eq:Gao_Lindblad}, with approximation $\hbar \Omega \ll kT$, satisfies that
  \begin{align}
    \lr{V(x)}_{t} - V(x^*) \le \frac{\text{\note{$\zeta$}}}{r}kT + O(e^{-\eta rt}),
  \end{align}
  if the initial density matrix $\rho(0)$ satisfies $\lr{H}_{0}-\frac{\text{\note{$\zeta$}}}{r}kT \ge 0$\note{, where $\zeta$ denotes the constant $2+O\left(\frac{\hbar\Omega}{kT}\right)$}.
  \label{the:Nonconvex_convergence}
\end{theorem_}
\begin{proof}
  Without loss of generality, we assume $x^*=0$ and $V(x^*)=0$. Then we only need to prove
  \begin{align}
    \lr{V(x)}_{t} \le \frac{\text{\note{$\zeta$}}}{r}kT + O(e^{-\eta rt}).
  \end{align}
  In this case, \eq{eq:energy_derivative_H_approx_final} still holds with approximation $\hbar \Omega \ll kT$ and $\text{\note{$\lr{\nabla^{2}V}_t$}}< +\infty$. With the constraint $x\cdot\nabla V(x)\ge rV(x)\ge 0$, \eq{eq:energy_derivative_H_approx_final} becomes
  \begin{align}
    \frac{\d}{\d t}\langle H\rangle_t= & \eta\cdot  \left(\text{\note{$\zeta$}}kT-\langle x \cdot\nabla V+p^2/m\rangle_t \right) \nonumber                              \\
    \leq                               & \eta\cdot  \left(\text{\note{$\zeta$}}kT-\langle rV+p^2/m\rangle_t \right) \nonumber                                           \\
    =                                  & \eta\cdot \left( \text{\note{$\zeta$}} kT-\lr{r(V(x)+\frac{p^{2}}{2m}) + (1-\frac{r}{2})\frac{p^{2}}{m}}_{t}\right)  \nonumber \\
    \le                                & -\eta r\lr{H(x)}_{t} + \text{\note{$\zeta$}}\eta kT.
    \label{eq:energy_derivative_H_approx_final_2}
  \end{align}
  Let $h(t) = (\lr{H}_t-\frac{\text{\note{$\zeta$}}}{r}kT) e^{\eta r t}$, then verify that $\frac{\d h(t)}{\d t} \le 0$ for $\forall t > 0$ by \eq{eq:energy_derivative_H_approx_final_2}. Consequently, we get $h(t)\leq h(0)$,
  \begin{align}
    \lr{H(x)}_{t} - \frac{\text{\note{$\zeta$}}}{r}kT \le \left(\lr{H}_{0}-\frac{\text{\note{$\zeta$}}}{r}kT\right)e^{-\eta r t} = O(e^{-\eta r t}).
  \end{align}
Hence we have
  \begin{align}
    \lr{V(x)}_{t}\le \lr{H}_t \le \frac{\text{\note{$\zeta$}}}{r}kT + O(e^{-\eta rt}).
  \end{align}
\end{proof}
\begin{remark_}
To achieve a desired accuracy $\varepsilon$, i.e., $\lr{V}_{t} - V(x^*) \le \varepsilon$ when $t\to +\infty$, we only need to set $kT=r\varepsilon/\text{\note{$\zeta$}}$ and ensure that the approximation $\hbar \Omega\ll kT$ still holds.
\end{remark_}

\section{Numerical results of time-independent QLD}
In this section, we conduct numerical experiments for time-independent QLD, i.e., all parameters remain constant throughout the evolution.
In Section \ref{sec:spon}, we initially delve into the origin of dissipation, grounding our understanding in the intuitive concept of spontaneous emission. Section \ref{sec:spon} also demonstrates that spontaneous emission can endow the system with the ability to transit to the ground state and achieve stability as the size of the heat bath increases.
Section \ref{sec:Implement_QLD} primarily discusses the numerical methods employed to solving QLD. These technical details are essential for maintaining numerical stability during numerical computation.
We then employ numerical experiments to validate the theoretical results concerning the quadratic potential from Section~\ref{sec:analytical_solution}, as detailed in Section~\ref{sec:verification}. In the experiments of the quadratic potential, we found that the numerical results are in alignment with the theoretical predictions. This consistency validates the precision of our numerical experiments.
Subsequently, roles of each parameter in QLD are individually discussed in Section \ref{sec:roles_of_parameters}. We discover that the parameter $\eta$ influences the speed of convergence in QLD, while the magnitudes of $\hbar$ and $T$ impact the system's ability to escape from local minima. Finally, Section \ref{sec:compareFPE} presents a comparison of the  performance between the time-independent QLD and its classical counterpart, the Fokker-Planck-Smoluchowski equation. This comparison demonstrates the consistency of the classical and quantum thermal effects.

All results and plots are obtained by simulations on classical computers (Intel$^\circledR$ Core$^\text{TM}$ i9-13900KS Processor with 128GB memory) via MATLAB 2023a, also including numerical experiments in Section \ref{sec:comparison}.

\subsection{\label{sec:spon}Verification of the intuition: spontaneous emission}
QLD describes the dynamics when the number of oscillators in heat bath is infinite. The essential cause of energy dissipation in QLD can trace back to spontaneous emission in quantum mechanics \cite{kudlis2022dissipation}. Intuitively, we have an assumption: the system has the ability to escape from local minimum as long as the number of oscillators in heat bath is finite.

\begin{figure}[ht]
    \centering
    \includegraphics[width = 0.4\textwidth]{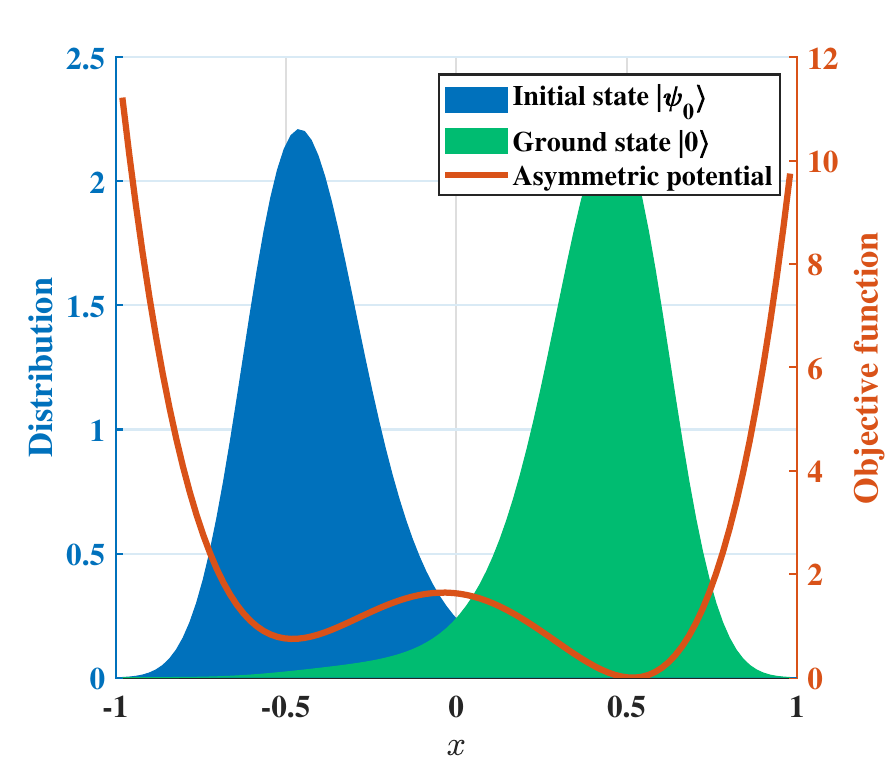}
    \caption{Landscape of spontaneous emission: asymmetric potential (orange line), the distribution of initial state $\ket{\psi_0}$ (blue area), and the distribution of  ground state $\ket{0}$ (green area).}
    \label{fig:spon_landscape}
\end{figure}

We utilize a specific numerical case to support our assumption. We choose a new landscape, an asymmetric quadratic potential
\begin{align}
  V(x) = V_0\left( \left( \left( \frac{x}{a} \right)^2 -1 \right)^2 - 1 - 0.3\left( \frac{x}{a} \right)\right),
  \label{fig:asymmetric_potential}
\end{align}
where $V_0= mw^2a^2/8$, $a = 0.5$, $m = 10$, $w = 2$. The location of local minimum is $x_{local} = -0.5$, while the location of global minimum is $x^* = 0.5$. Denote the ground state to be $\ket{0} = \ket{\psi_g(x)}$, and let the
initial state $\ket{\psi_0} = \ket{\psi_0(x)}$ and ground state  be symmetric about $x= 0$, i.e., $\psi_{0}(x)= \psi_{g}(-x)$, shown in Fig.~\ref{fig:spon_landscape}.

\begin{figure}[ht]
    \centering
    \subfloat[A single two-energy-level oscillator in heat bath.]
    {\label{fig:spon-single}\includegraphics[width=0.56\textwidth]{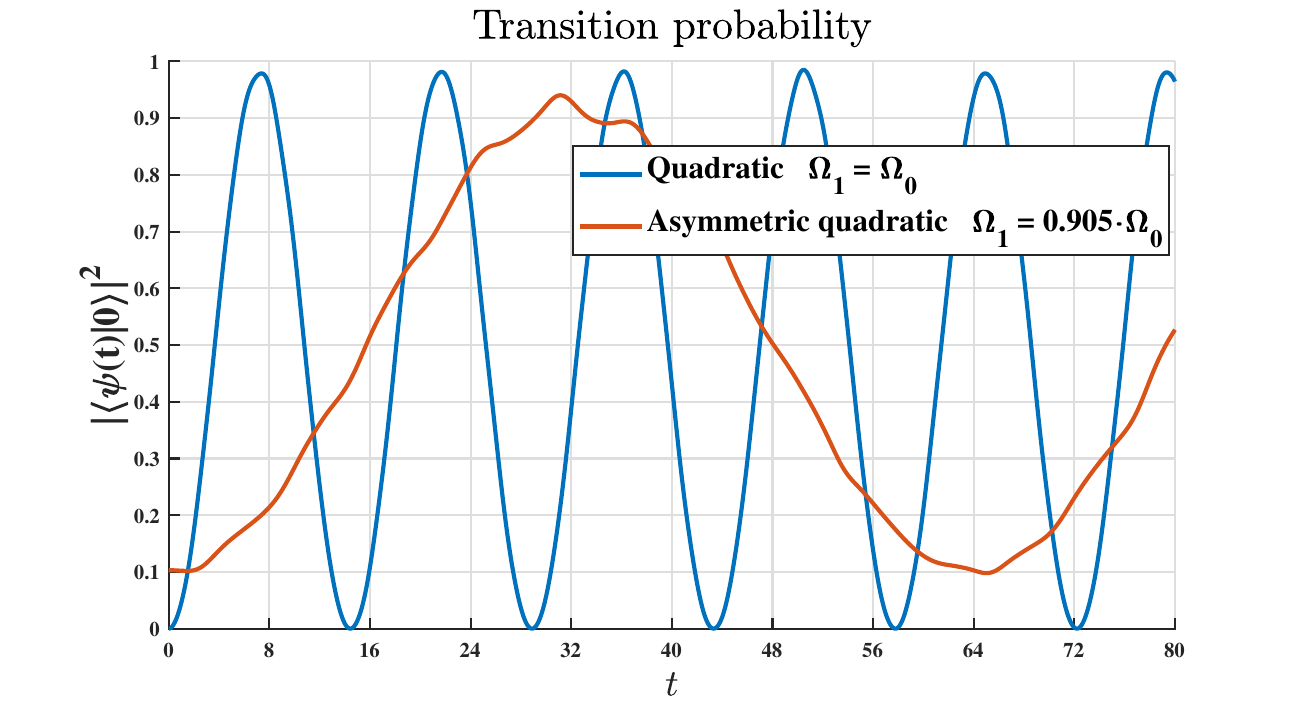}}
    \subfloat[Increasing the number of oscillators in heat bath.]{\label{fig:Pt_Multiparticle}\includegraphics[width = 0.435\textwidth]{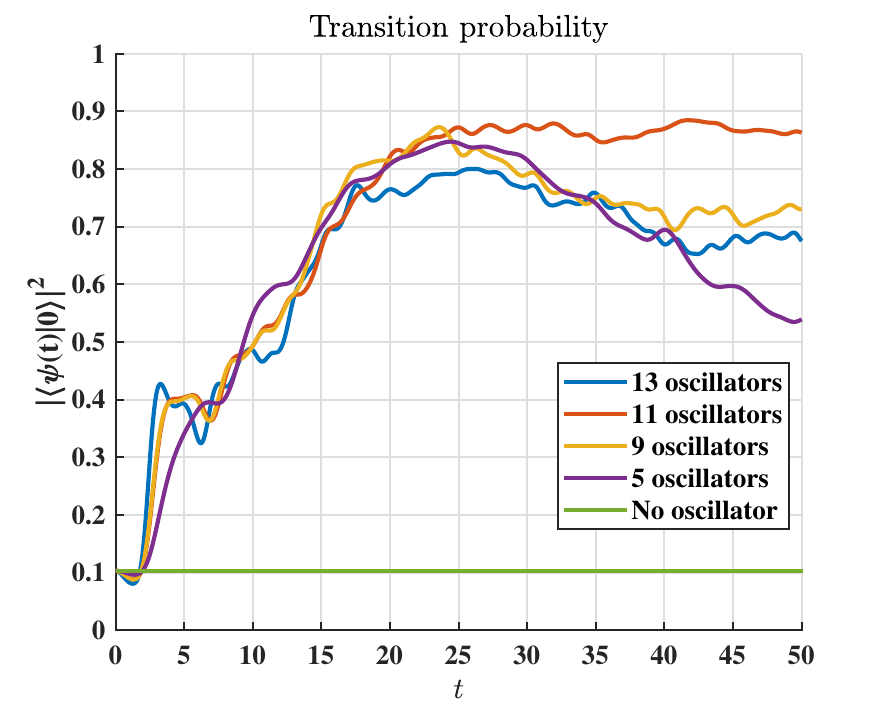}}
    \caption{Spontaneous emission. (a) Transition probability to the ground state, i.e., $|\langle \psi(t)| 0 \rangle |^{2}$. The quadratic case is a benchmark ($V(x) = \frac{1}{2}m\Omega_0^2 x^2, \Omega_0= w$), which exhibits characteristics of spontaneous emission (blue line). When $\Omega_1 = 0.905\Omega_0$, the system maximum probability of transition is larger than $0.9$ (orange line). (b) Transition probability when increasing the number of harmonic oscillators in heat bath. Since the frequencies of oscillators are randomly chosen, the results of this experiment are not unique. We choose the best one to verify our statements.}
\end{figure}

The finite heat bath is simulated by numerically solving the Schr\"{o}dinger equation using symplectic leapfrog scheme \cite{Mauger}, where the Hamiltonian is given by \eq{eq:originH}. Since the simulation cost increases exponentially with respect to the number of oscillators, the energy level of oscillators is restricted to two, which means the Hamiltonian of the harmonic oscillator can be written as $H = \mathrm{diag}(1/2,3/2)\hbar \Omega$.

We first verify that the system can still reach the ground state with high probability, when there is only a single two-energy-level oscillator and the Hamiltonian of heat bath is $H_{B} =\mathrm{diag}(1/2,3/2)\hbar \Omega_1$. The parameter setting and results are shown in Fig.~\ref{fig:spon-single}, which implicates that there exists a frequency $\Omega_1 = 0.905\Omega_0$ such that the system can reach the ground state with high probability $|\langle \psi(t)| 0\rangle|^2 > 0.9$.

We increase the number of oscillators in heat bath, where $n = 0,5,9,11,13$. The Hamiltonian of heat bath is $H_{B} = \sum_{n}\mathrm{diag}(1/2,3/2)\hbar \Omega_n$. For a physical system in the real world, the number of modes of frequency $\Omega_n$ should satisfy a specific distribution. For simplification, we choose the uniform distribution, namely the frequencies of oscillators being randomly chosen in the range $\Omega_n \in[0.7\Omega_0,1.3\Omega_0]$. The results are shown in Fig.~\ref{fig:Pt_Multiparticle}, which implicate that the probability of transition gradually stabilizes and stays in a high level with the increasing number of oscillators.

\subsection{\label{sec:Implement_QLD}Implementation of numerical experiments for QLD}

The simulation of QLD is achieved by numerically solving \eq{eq:Gao_Lindblad_xy}. We employ the implicit Euler method and an approximately uniform distribution $p(x)$ to ensure the stability of numerical computation. All parameters are configured in accordance with an actual physical system after nondimensionalization. Comprehensive details are provided in Appendix \ref{app:Implement_QLD}.

\subsection{\label{sec:verification}Quadratic potentials}
First, we verify whether the convergence values satisfy \eq{eq:V_f} and  \eq{eq:E_k_f} by the following parameter setting.\footnote{This parameter setting is also consistent with Refs. \cite{gao1995femtosecond,gao1997dissipative,gao1998lindblad} after nondimensionalizing.}

\Pra{Resolution $n=128$, domain $x\in [-1,1]$, space step $\Delta x=\frac{2}{n}$, time step $\Delta t=0.05\Delta x^2$, total evolution time $t_f=10$, $\hbar=2.1108$,  $kT=165.8080~(T=200)$, characteristic damping rate $\eta=5$, $\Omega=223.2728$, $m=1$, initial state $\psi_0(x)=\sqrt{p(x)}$. \footnote{\note{We note here that the condition $\hbar\Omega \ll kT$ is not required in quadratic cases, because according to Section~\ref{sec:analytical_solution}, the convergence of QLD in quadratic potentials is derived analytically without any assumptions.}}}

\begin{figure*}
  \centering
  \includegraphics[width=1\textwidth]{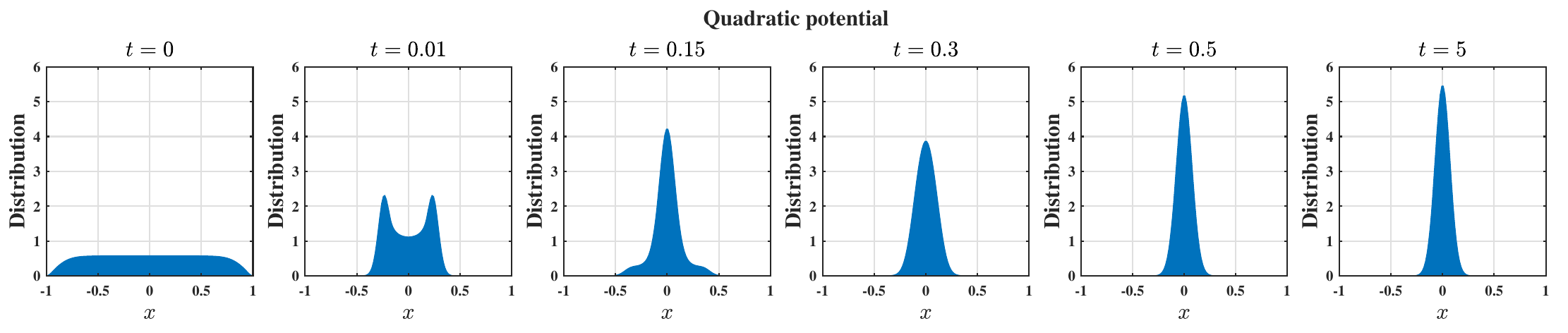}
  \caption{The time evolution of probability distributions  for quadratic potential at mild temperature. x-coordinate represents the position $x$, and y-coordinate represents the diagonal elements of density matrix. Six different moments $t=0,~0.01,~0.15,~0.3,~0.5,~5$ with notable features are chosen. At the beginning of evolution, we can observe the oscillation of the probability distribution. After $t>0.5$, the probability distribution converges to a stable distribution gradually because of dissipation.}
  \label{fig:Quadratic_high_T}
\end{figure*}

\begin{figure*}
  \centering
  \includegraphics[width=0.9\textwidth]{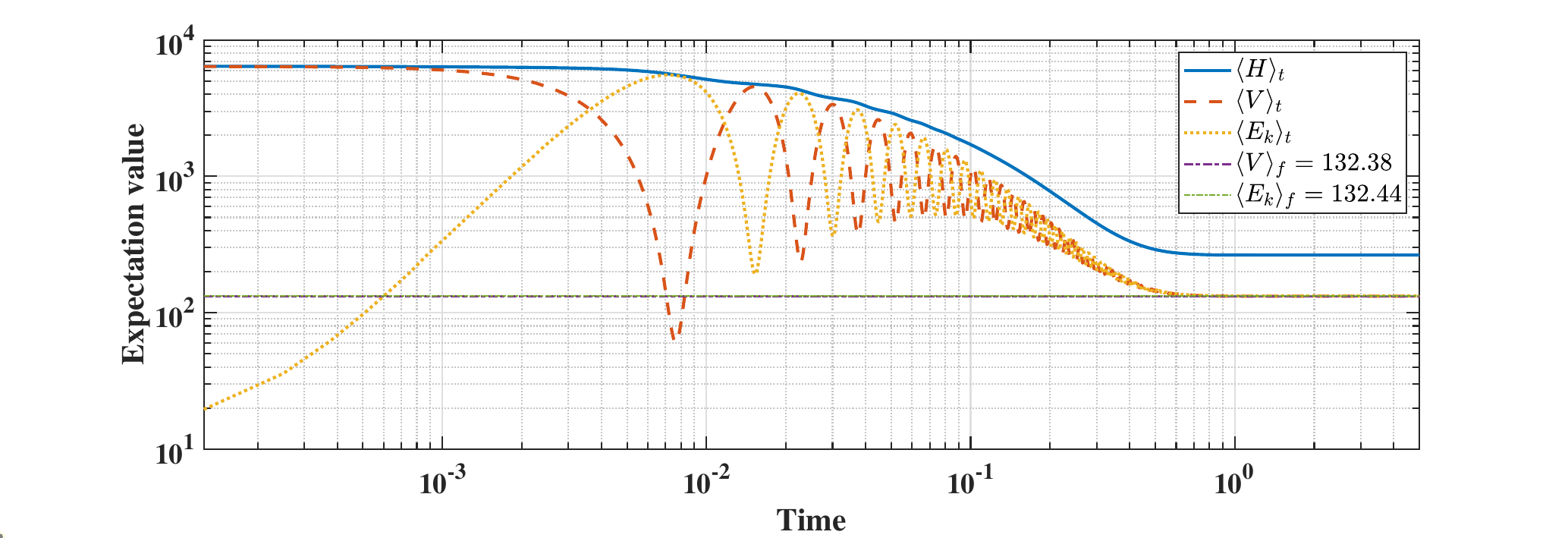}
  \caption{The time evolution of expectation values for quadratic potential at mild temperature. x-coordinate represents the time $t$, and y-coordinate represents the expectation value of Hamiltonian $\lr{H}_t$, potential energy $\lr{V}_t$ and kinetic energy $\lr{E_k}_t$. The analytical convergence values $\lr{V}_f=132.3762,~\lr{E_k}_f=132.4368$ are also shown in the figure as convergence lines. We can easily observe the oscillation of the expectation values at the beginning consistent with analytical solutions \eq{eq:x2moment_solution} and \eq{eq:p2moment_solution}, which originates from $\cos(2\Omega t)$ and $\sin(2\Omega t)$ factors in these equations. During the whole process, the oscillation decays gradually, which results from the damping factor $\exp(-2\eta t)$.}
  \label{fig:Quadratic_high_T_evolution_time}
\end{figure*}

\Results{The evolution of the probability distributions is shown in Fig.~\ref{fig:Quadratic_high_T}. The evolutions of the expectation values of Hamiltonian $\lr{H}_t$, potential energy $\lr{V}_t$ and kinetic energy $\lr{E_k}_t$ are shown in Fig.~\ref{fig:Quadratic_high_T_evolution_time}. The analytical solutions of convergence value $\lr{V}_f$ and $\lr{E_k}_f$ can be calculated by \eq{eq:V_f} and  \eq{eq:E_k_f}, which are $\lr{V}_f=132.3762,~\lr{E_k}_f=132.4368$. Numerical experiments also give the convergence value $\lr{V}_f^\prime=132.1086,~\lr{E_k}_f^\prime=132.6604$ with respect to density matrix at the final moment $t_f=5$. As a result, numerical results are very close to theoretical results, while differences between them might be caused by numerical errors.}


Then, we try to verify whether the system reaches the ground state at the low temperature regime, stated at the end of Section \ref{sec:analytical_solution}. The following parameter setting is used.

\Pra{Resolution $n=128$, domain $x\in [-1,1]$, space step $\Delta x=\frac{2}{n}$, time step $\Delta t=0.05\Delta x^2$, total evolution time $t_f=10$, $\hbar=2.1108$,  $kT=0.0083~(T=0.01)$, characteristic damping rate $\eta=5$, $\Omega=223.2728$, $m=1$, initial state $\psi_0(x)=\sqrt{p(x)}$.}

\begin{figure*}
  \centering
  \includegraphics[width=1\textwidth]{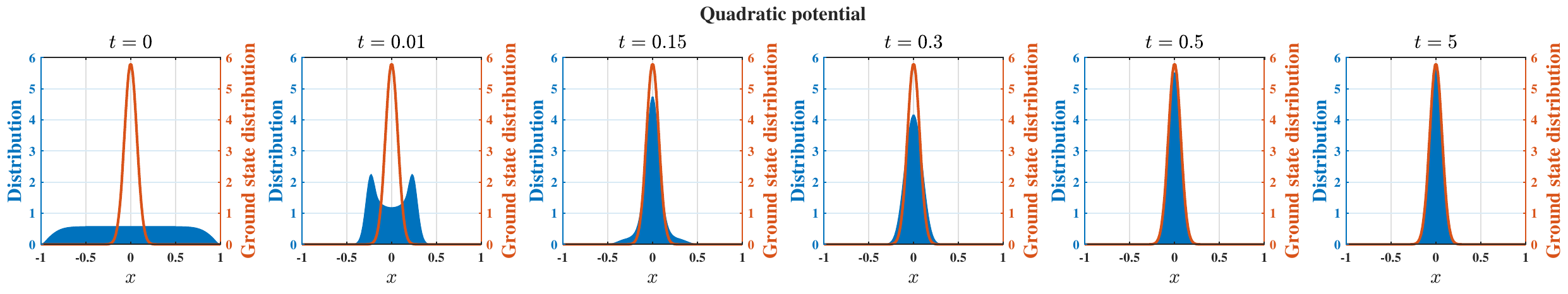}
  \caption{The time evolution of probability distributions for quadratic potential at low temperature. x-coordinate represents the position $x$, and y-coordinate represents the diagonal elements of density matrix. Six different moments $t=0,~0.01,~0.15,~0.3,~0.5,~5$ with notable features are chosen. The orange line is the analytical solution of ground state probability distribution. In the end, the system reaches the ground state.}
  \label{fig:dis_qua_ground}
\end{figure*}

\begin{figure*}
  \centering
  \includegraphics[width=0.9\textwidth]{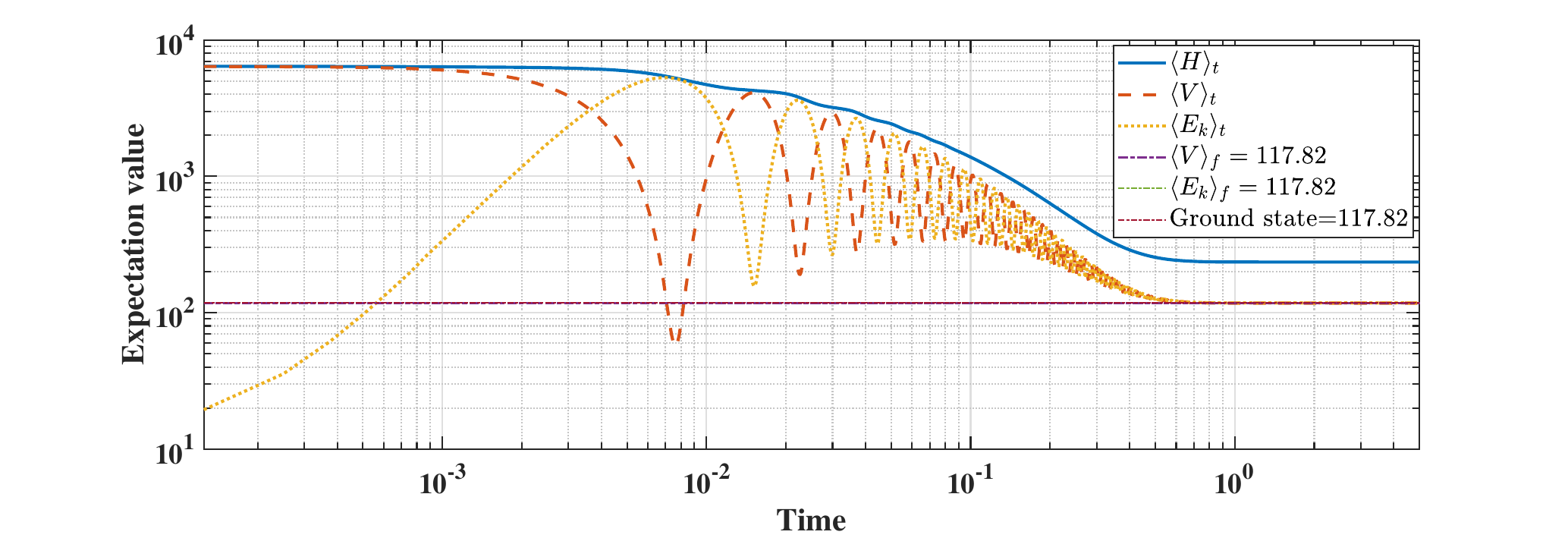}
  \caption{The time evolution of expectation values for quadratic potential at low temperature. x-coordinate represents the time $t$, and y-coordinate represents the expectation value of Hamiltonian $\lr{H}_t$, potential energy $\lr{V}_t$ and kinetic energy $\lr{E_k}_t$. The analytical convergence values $\lr{V}_f=117.8211,~\lr{E_k}_f=117.8211$ and $\frac{1}{4}\hbar \Omega=117.8211$ are also shown in the figure as convergence lines. It could be observed that the system reaches ground state when $t=5$.}
  \label{fig:qua_ground}
\end{figure*}

\Results{The evolutions are shown in Fig.~\ref{fig:dis_qua_ground} and Fig.~\ref{fig:qua_ground}. Still using same notations as previous, we have $\lr{V}_f=117.8211,~\lr{E_k}_f=117.8211$, $\lr{V}_f^\prime=117.4373,~\lr{E_k}_f^\prime=117.8249$, analytical solution of ground state potential energy and kinetic energy is $\frac{1}{4}\hbar \Omega =117.8211$. This consistency suggests that the system has reached the ground state. Denote $\rho_f$ to be the numerical solution of density matrix at the final moment $t=5$, and $\rho_g$ to be the analytical solution of the ground-state density matrix. We use fidelity and quantum relative entropy to evaluate the closeness between the numerical result and analytical ground state, which derives fidelity $F(\rho_f,\rho_{g})=\left(\Tr \sqrt{\sqrt{\rho_f}\rho_g\sqrt{\rho_f}}\right)^2=0.9999902$ and  quantum relative entropy (in Kullback-Leibler divergence) $D_{KL}(\rho_f\|\rho_g)=\Tr (\rho_f(\ln(\rho_f)-\ln(\rho_g)))=6.3466\times 10^{-4}$, showing that the numerical result and analytical ground state are very close to each other.}

\subsection{\label{sec:roles_of_parameters}Roles of each parameter}
In this part, we discuss the role of each parameter in QLD separately. In \eq{eq:Gao_Lindblad}, parameters $\eta$, $T(kT)$, $\hbar$ play important roles in the behavior of the system. In order to analyze the roles of these parameters, we fix other parameters and change one of $\eta$, $T(kT)$, $\hbar$ each time.

In all experiments of this section, we use a double well potential because it can test the ability of QLD in escaping from the local minimum by quantum thermal effect and dissipative quantum tunneling effect while it keeps the simplicity of the potential. The double well potential is set in domain $x\in [-1,1]$ as
\begin{align}
  V(x)=83.9808x^4-64.8x^2+7.2x+17.0680,
  \label{eq:double_well_before}
\end{align}
whose global minimum appears at $x^*=-0.6472$.

In order to evaluate the performance of algorithms, we introduce ``success probability'' \cite{leng2023quantum} to measure the probability of finding the global minimum. The success probability is defined as the probability of finding the global minimum $x^*$ within a certain tolerance $\epsilon=0.15$, which is given by
\begin{align}
  \Pr(\text{success})=\Pr(|x(t)-x^*|\leq \epsilon).
  \label{eq:success_proba}
\end{align}
The introduction of the concept ``success probability" is necessitated by the presence of numerous objective functions with local minima near zero. Traditional loss functions struggle in these scenarios, failing to show the superiority of QLD.

\subsubsection{Roles of characteristic damping rate $\eta$}
No matter in quantum case or classical case, $\eta$ influences the convergence speed of the system \cite{shi2020learning,gao1998lindblad}. $\eta$ is related to the damping and dissipative effect in quantum Langevin system. When $\eta$ increases, the system will reach the stationary distribution faster, and vice versa. We also guess that if $\eta$ is too large, the system will reach the stationary solution so quickly that the system cannot escape from the local minimum, weakening the performance of QLD. This guess is also verified in our following experiments.
In order to evaluate the role of $\eta$ individually, we set appropriate temperature and $\hbar$, and then use different $\eta$ to see the speed of convergence.

\begin{figure*}[htbp]
  \centering
  \includegraphics[width=1\textwidth]{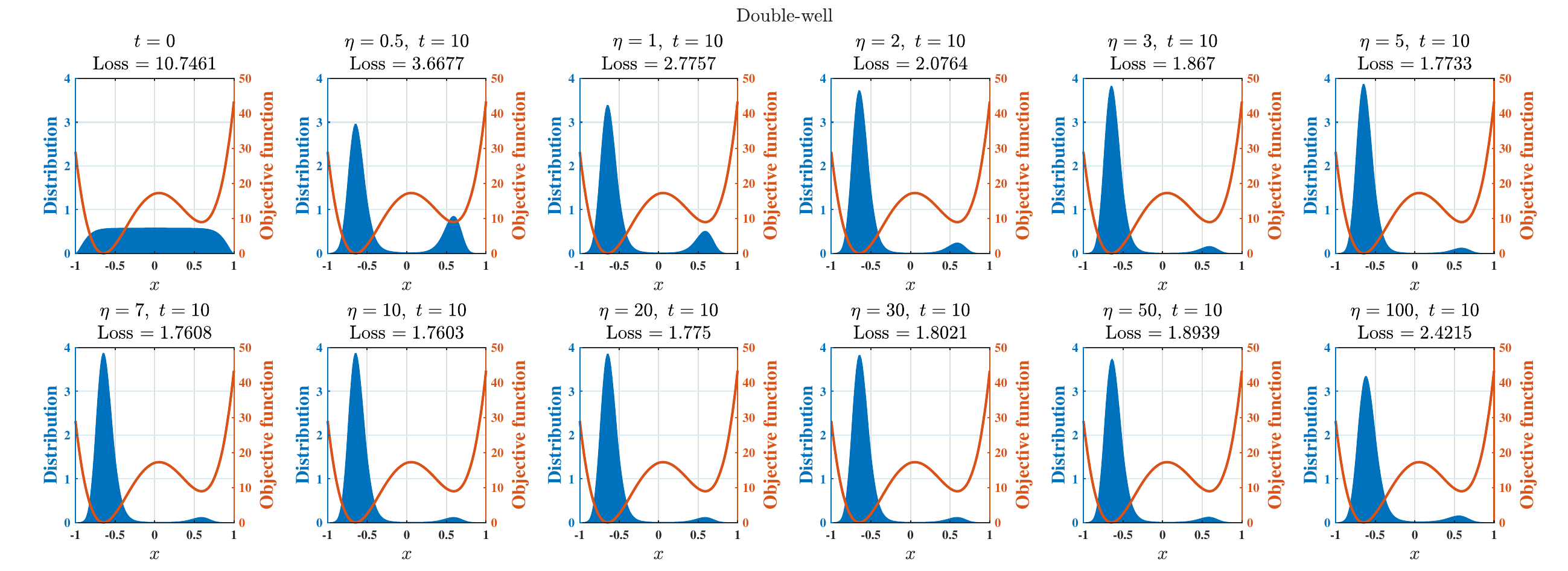}
  \caption{The probability distributions with different $\eta$ at final moment $t = 10$. x-coordinate represents the position $x$, and y-coordinate represents the diagonal elements of density matrix (blue area). Orange line is the double-well objective function. The first subplot shows the approximately uniform distribution at $t=0$, and the following subplots are distributions at final moment $t=10$ with different $\eta$. Losses at final moment are written in the subtitles of subplots. We can observe that $\eta$ affects the evolution speed of QLD.}
  \label{fig:double-well_diff_gamma}
\end{figure*}
\begin{figure*}[htbp]
  \centering
  \includegraphics[width=0.8\textwidth]{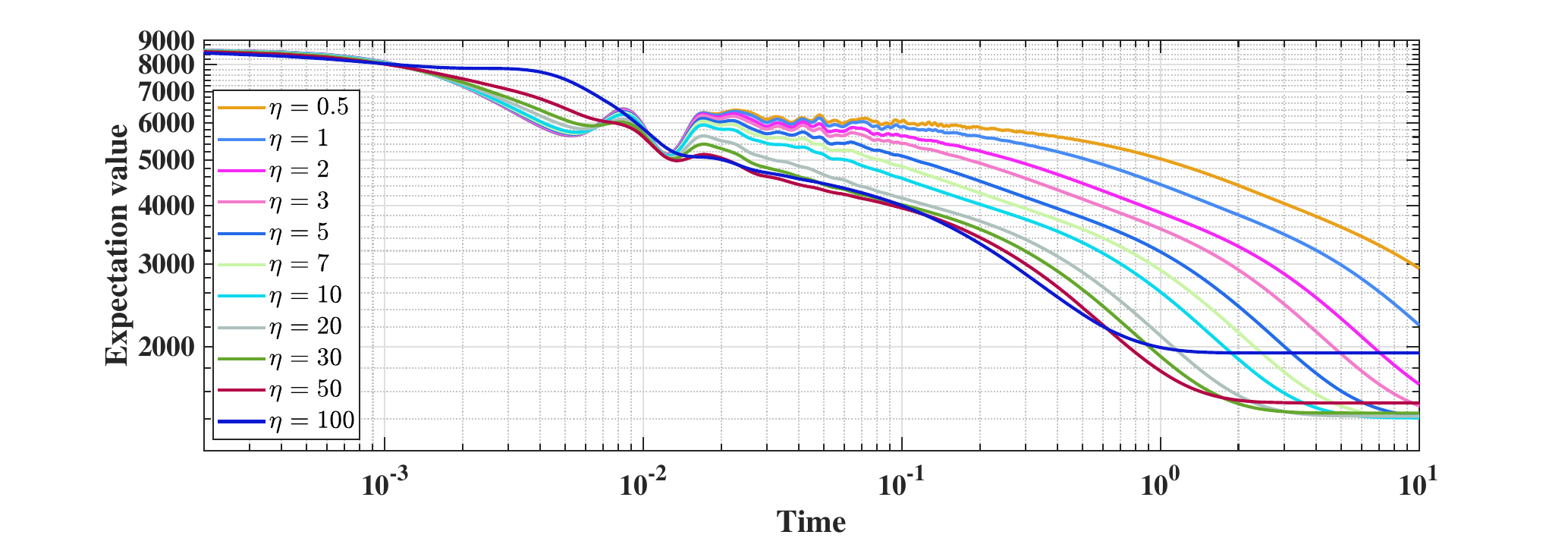}
  \caption{The time evolution of expectation values of potential. x-coordinate represents the time $t$, and y-coordinate represents the expectation value. From this figure, we can observe when $\eta$ is too small, the system has not reached the stationary distribution at $t=10$ still needing extra time to evolve. When $\eta$ is too large, the system reaches the stationary distribution so fast that system cannot escape from local minimum completely and the convergence value of potential energy increases. Within a suitable range of $\eta\in [5,30]$, $\eta$ does not have a significant impact on the expectation value at $t=10$.}
  \label{fig:double-well_diff_gamma_evolution_time}
\end{figure*}

\Pra{Use different damping rates $\eta=0.5, ~1, ~2, ~3,~5, ~7, ~10,$ $~20, ~30,~50, ~100$, while other parameters are invariant. Resolution $n=128$, domain $x\in [-1,1]$, space step $\Delta x=\frac{2}{n}$, time step $\Delta t=0.05\Delta x^2$, total evolution time $t_f=10$, $\hbar=2.1108$,  $kT=2072.6~(T=2500)$, $\Omega=300$, $m=1$, initial state $\psi_0(x)=\sqrt{p(x)}$.}

\Results{Fig.~\ref{fig:double-well_diff_gamma} shows the probability distributions of different $\eta$ at final moment $t_f = 10$. Fig.~\ref{fig:double-well_diff_gamma_evolution_time} shows the evolution of expectation values. From Fig.~\ref{fig:double-well_diff_gamma} and Fig.~\ref{fig:double-well_diff_gamma_evolution_time}, we can observe when $\eta$ is too small, the system has not reached the stationary distribution at $t=10$ still needing extra time to evolve. When $\eta$ is too large, the system reaches the stationary distribution so fast that system cannot escape from local minimum completely and the convergence value of potential energy increases. Within a suitable range of $\eta\in [5,30]$, $\eta$ does not have a significant impact on the expectation value at $t=10$. Nevertheless, we have to admit that all observations of $\eta$ are only available at this double well potential with form \eq{eq:double_well_before}, and changing the objective function may influence the behavior of $\eta$.}

\subsubsection{Roles of $\hbar$ and $T$}

The performance of QLD mainly depends on $\hbar$ and $T$. We first consider the roles of them separately, and then combine them to find the best strategy for optimization.

$\hbar$ is the unique parameter that only exists in the quantum case, which determines the quantum tunneling effect in QLD. There are plenty of previous results for dissipative quantum tunneling. The most famous one is Ref.~\cite{caldeira1983quantum} by Calderia and Leggett, with subsequent works \cite{harris1993quantum,chakravarty1984dynamics,leggett1987dynamics,fisher1985dissipative,hanggi1986escape,kelkar2017time,dolgitzer2021dynamical}. To study the tunneling effect, we need to turn off the thermal term, which means setting an extremely low temperature in the experiments. In order to show the power of quantum tunneling, the initial wave function $\psi_0$ is set near the local minimum $x_{local}\approx0.6$, and it follows a Gaussian distribution:
\begin{align}
  \psi_0=\frac1{\sqrt{2\pi\sigma^2}}\exp\left(-\frac{(x-x_1)^2}{2\sigma^2}\right), ~~x_1=0.6,~\sigma=0.08.
  \label{eq:initial_wave_function_Gaussian}
\end{align}

\Pra{Change the reduced Planck constant $\hbar=1,~2.1108,~4,~8,~12,~16,~20,~24,~28,$ $~32,~36$, while other parameters are invariant in each case. Resolution $n=128$, domain $x\in[-1,1]$, space step $\Delta x=\frac{2}{n}$, time step $\Delta t=0.05 \Delta x^2$, total evolution time $t_f=10$, $kT=0.0083~(T=0.01)$, $\eta=5$, $\Omega=300$, $m=1$, initial state: Gaussian distribution \eq{eq:initial_wave_function_Gaussian}.}

\begin{figure*}[htbp]
  \centering
  \includegraphics[width=1\textwidth]{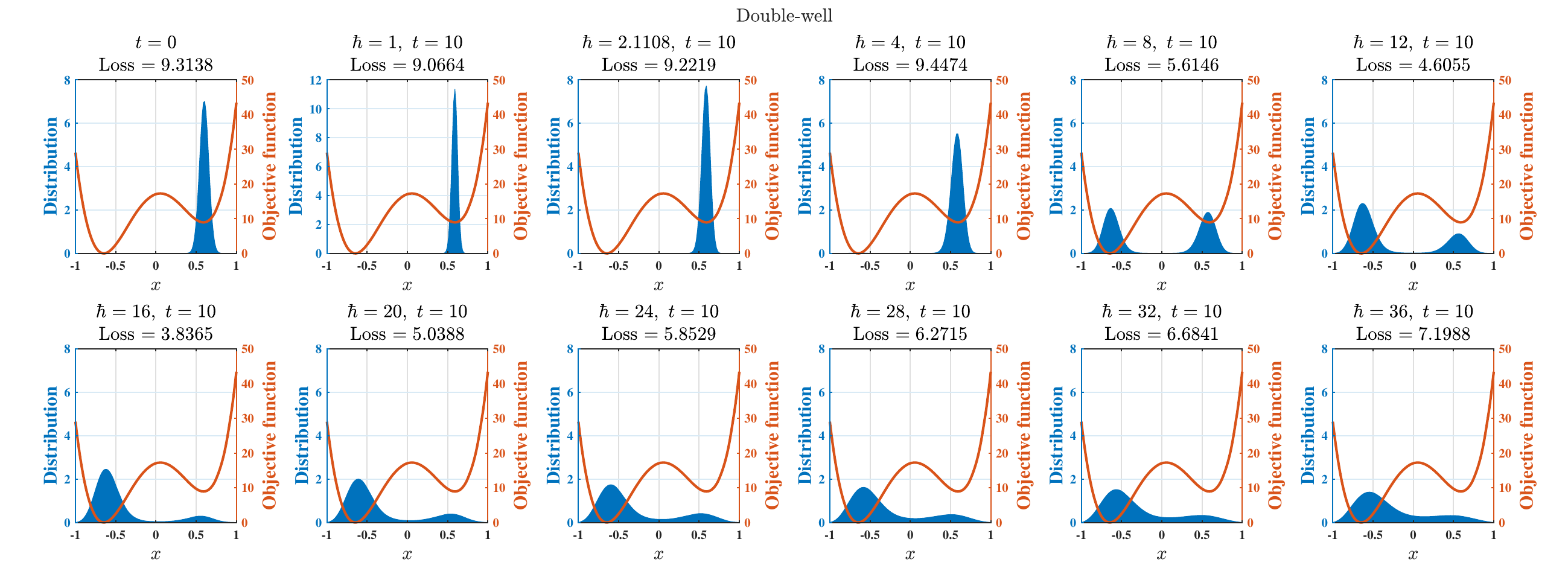}
  \caption{The distributions with different $\hbar$ at $t=10$. x-coordinate represents the position $x$, and y-coordinate represents the diagonal elements of density matrix (blue area). The first subplot shows the initial distribution \eq{eq:initial_wave_function_Gaussian} at $t=0$. Losses at final moment are written in the subtitles of subplots. We observe that quantum tunneling effect is stronger when increasing the value of $\hbar$, but when $\hbar$ is so large that it might cause significant fluctuation which makes distribution not concentrate on the global minimum.}
  \label{fig:double-well_diff_hbar}
\end{figure*}

\Results{The distributions with different $\hbar$ at $t=10$ are shown in the Fig.~\ref{fig:double-well_diff_hbar}. We observe that the quantum tunneling effect is stronger when increasing the value of $\hbar$, but when $\hbar$ is so large that it might cause significant fluctuation, making the system incapable of concentrating on the global minimum.}

Temperature determines the thermal effect in QLD. As shown in the above experiments of $\hbar$, in the double well landscape, the tunneling effect is negligible when $\hbar<4$ (not applicable in all landscapes). In order to evaluate the roles of temperature $T$ individually, we set $\hbar=2.1108$.
\begin{figure*}[ht]
  \centering
  \includegraphics[width=1\textwidth]{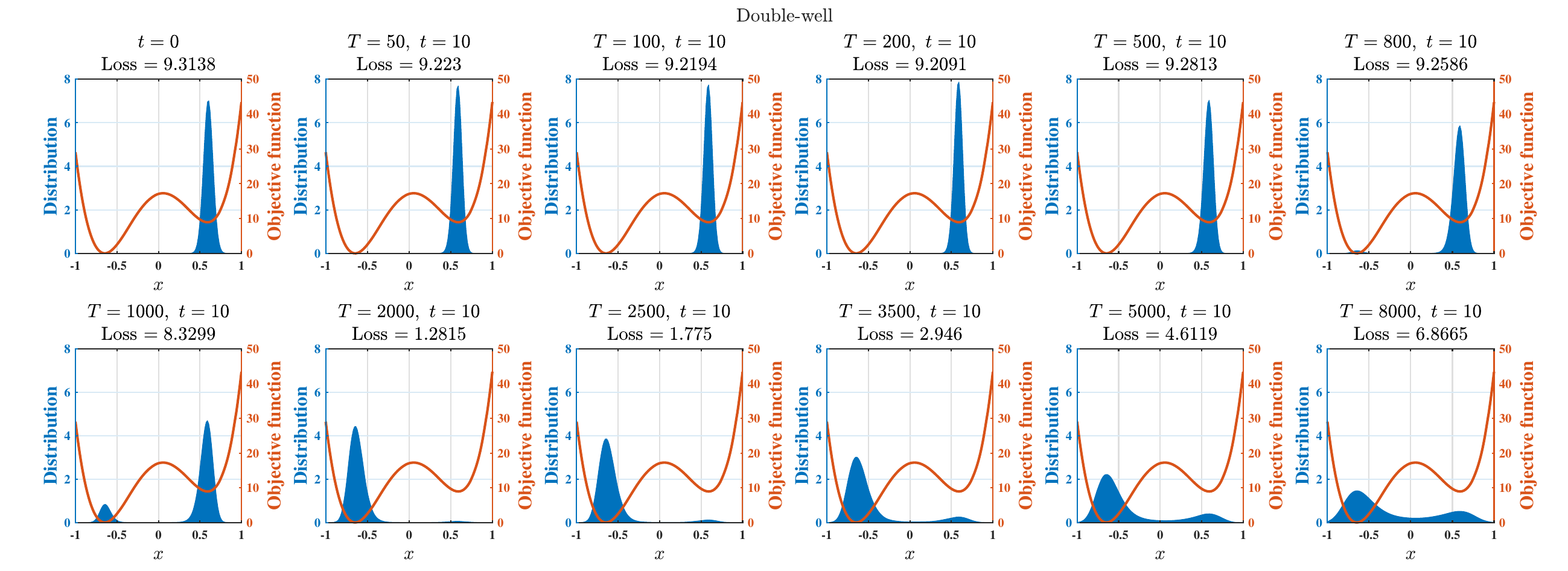}
  \caption{The probability distributions with different temperature $T$ at final moment $t = 10$. x-coordinate represents the position $x$, and y-coordinate represents the diagonal elements of density matrix (blue area). The first subplot shows the initial distribution \eq{eq:initial_wave_function_Gaussian} at $t=0$. We observe that the thermal effect is very subtle at the low temperature region, manifesting as the system being stuck at a local minimum and unable to escape, while at high temperature region, the thermal effect is significant, manifesting as the system being able to escape from the local minimum and reach the global minimum. However, if the $T$ is too high (e.g., $T=8000$), the thermal effect is so strong that the system is unable to concentrate on the minimum because of fluctuation.}
  \label{fig:T_different}
\end{figure*}

\Pra{Change the temperature $T=50, ~100, ~200, ~500, ~800,~1000, ~2000,~2500, ~3500,$ $~5000, ~8000$ ($k=0.82904$), while other parameters are invariant. Resolution $n=128$, domain $x\in[-1,1]$, space step $\Delta x=\frac{2}{n}$, time step $\Delta t=0.05 \Delta x^2$, total evolution time $t_f=10$, $\hbar=2.1108$, $\eta=20$, $\Omega=300$, $m=1$, initial distribution: Gaussian distribution \eq{eq:initial_wave_function_Gaussian}.}

\Results{The distributions with different $T$ at final moment $t = 10$ are shown in the Fig.~\ref{fig:T_different}. We observe that the thermal effect is very subtle at the low temperature region, manifesting as the system being stuck at a local minimum and unable to escape, while at high temperature region, the thermal effect is significant, manifesting as the system being able to escape from the local minimum and reach the global minimum. However, if the $T$ is too high (e.g., $T=8000$), the thermal effect is so strong that the system is unable to concentrate on the minimum because of fluctuation.}

When considering $T$ and $\hbar$ together, tunneling and thermal effect both exist. The performance and success probability of QLD not only depend on $T$ and $\hbar$, but also its evolution time.  We set a range of $T$ and $\hbar$ for $T\in [1000,2800]$ and $\hbar\in [6,24]$, then choose the double well landscape to discuss the combined effect.

\Pra{Conduct 100 cases, where $T = ~1000,~1200,~1400,~1600,~1800,~2000,~2200,$ $~2400,~2600,~2800$ and $\hbar = 6,~8,~10,~12,~14,~16,~18,~20,~22,~24$ respectively, while $\eta=20$ is fixed. We set an upper bound of the total evolution time $t_{f} \le 100$, since it takes too long for some cases to reach the steady state. $\Omega = 300$, $m = 1$, initial state $\psi_0(x)=\sqrt{p(x)}$. Establish the criterion of steady state: let the $k$-th iteration of success probability be $\Pr^{(k)}$, defined by \eq{eq:success_proba}. Cease iteration when $\left|\Pr^{(k)} - \Pr^{(k-1)}
  \right|<10^{-7}$.}

\begin{figure}[ht]
\centering
\subfloat{\includegraphics[width=0.44\linewidth]{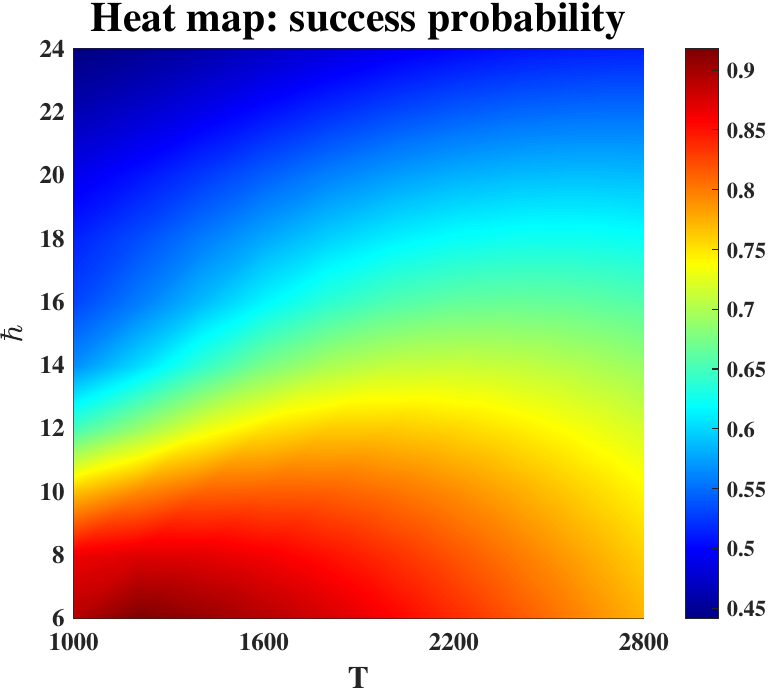} \label{fig:subfigureA}}
\subfloat{\includegraphics[width=0.44\linewidth]{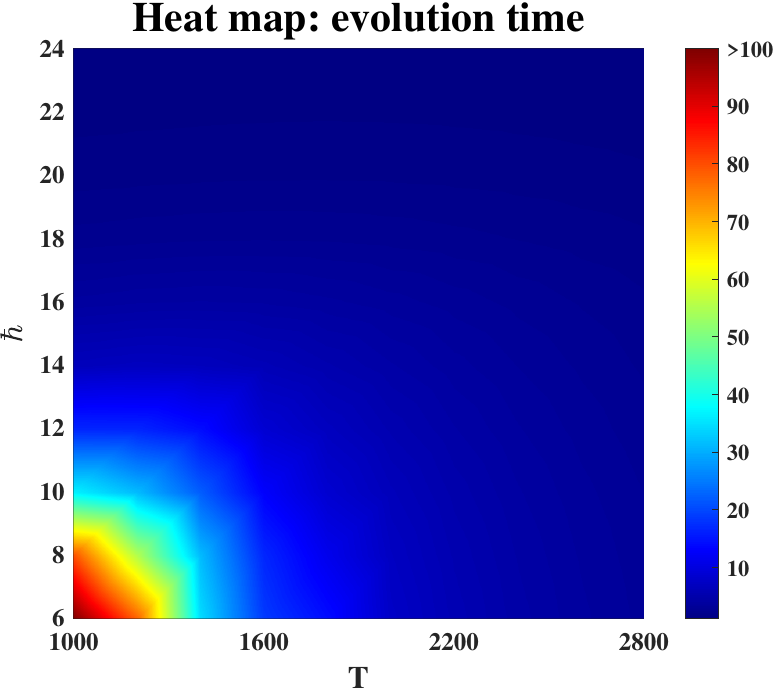}
\label{fig:subfigureB}}
\caption{In the double well landscape, there exists a best temperature when $\hbar$ is fixed. Furthermore, although smaller $\hbar$ leads to higher success probability, the evolution time increases dramatically with respect to the decrease of $\hbar$.}
\label{fig:heat_map_Succ}
\end{figure}

\Results{As shown in Fig.~\ref{fig:heat_map_Succ}, we observe that there exists a best temperature when $\hbar$ is fixed. Furthermore, although smaller $\hbar$ leads to higher success probability, the evolution time increases dramatically with respect to the decrease of $\hbar$. Thus, it is necessary to balance the cost of evolution time and the success probability of the algorithm.}

\subsection{\label{sec:compareFPE}Comparison with classical dynamics}

It is well known that the stochastic differential equation approximates the discrete-time SGD algorithm and the Fokker-Plank-Smoluchowski equation (FPE) can be derived from the lr-dependent SDE by It\^{o}'s formula \cite{shi2020learning}. In the overdamped classical Langevin dynamics, the classical probability distribution $\rho$ of the particle evolves according to the Smoluchowski equation \cite{ray1999notes}:
\begin{align}
  \frac{\partial \rho}{\partial t} = -\frac{1}{\gamma}\nabla \cdot (\rho \nabla V) + \frac{kT}{\gamma} \nabla^2 \rho,
  \label{eq:FPE}
\end{align}
where $\gamma$ is the friction coefficient that satisfies $\gamma = 2m\eta$. It is the counterpart of FPE derived from SDE.

\begin{figure}[htbp]
    \centering
    \includegraphics[width = 0.3\textwidth]{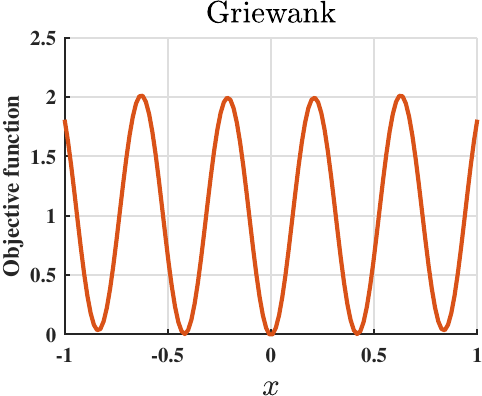}
    \caption{Objective function: Griewank function $y = \frac{(15x)^2}{4000} - \cos(15x) + 1$. The global minimum appears at $x^*=0$.}
    \label{fig:Griewank}
\end{figure}

We compare QLD with FPE. In some landscapes, classical algorithms have demonstrated efficacy, and the advantages of quantum algorithms are not manifested. Hence, we choose a landscape that poses a comparatively challenging task for classical algorithms for comparison, shown in Fig.~\ref{fig:Griewank}. This function is called the Griewank function with form \eq{eq:Griewank}, which is a nonconvex function with many local minima approximating zero.

\Pra{We fix the temperature of QLD and FPE while changing $\hbar$ of QLD. Parameters of QLD: Resolution $n = 128$, domain $x\in [-1,1]$, $\Delta x = \frac{2}{n}$, time step $\Delta t = 0.5\Delta x^2$, $\Omega = 300$, $m=1$. Change  $T =200, ~1200, ~2200$, $\hbar =4, ~10, ~16$. Parameters of FPE: all the parameters are the same as QLD, except $\gamma = 2m\eta = 40$.}

\begin{figure}[ht]
  \centering
  {\includegraphics[width = 0.255\textwidth]{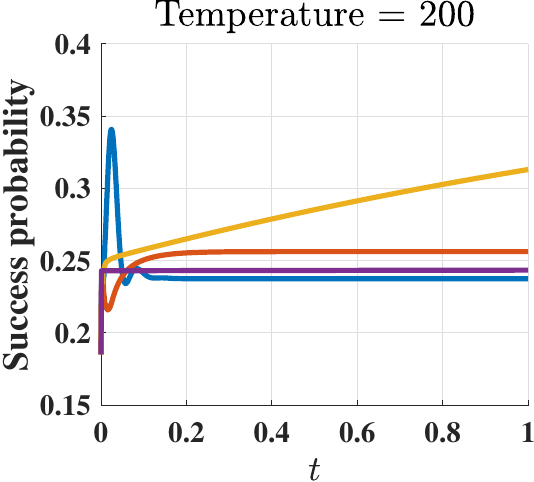}}
  {\includegraphics[width= 0.255\textwidth]{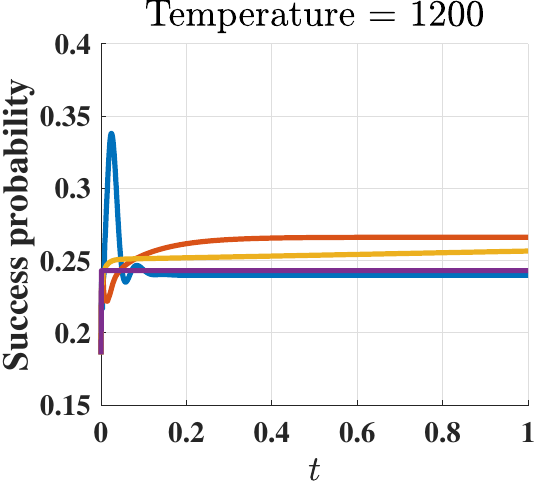}}
  {\includegraphics[width= 0.255\textwidth]{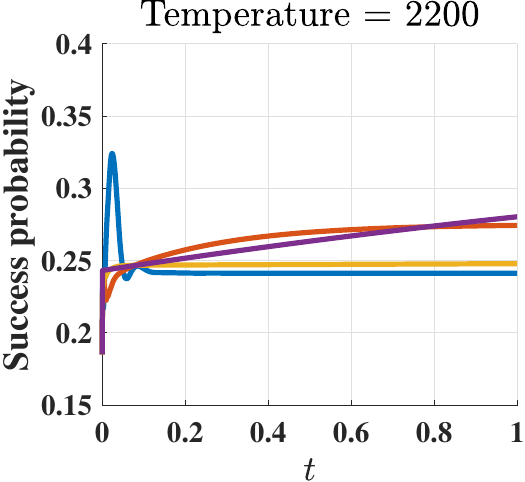}}
  {\includegraphics[width= 0.2\textwidth]{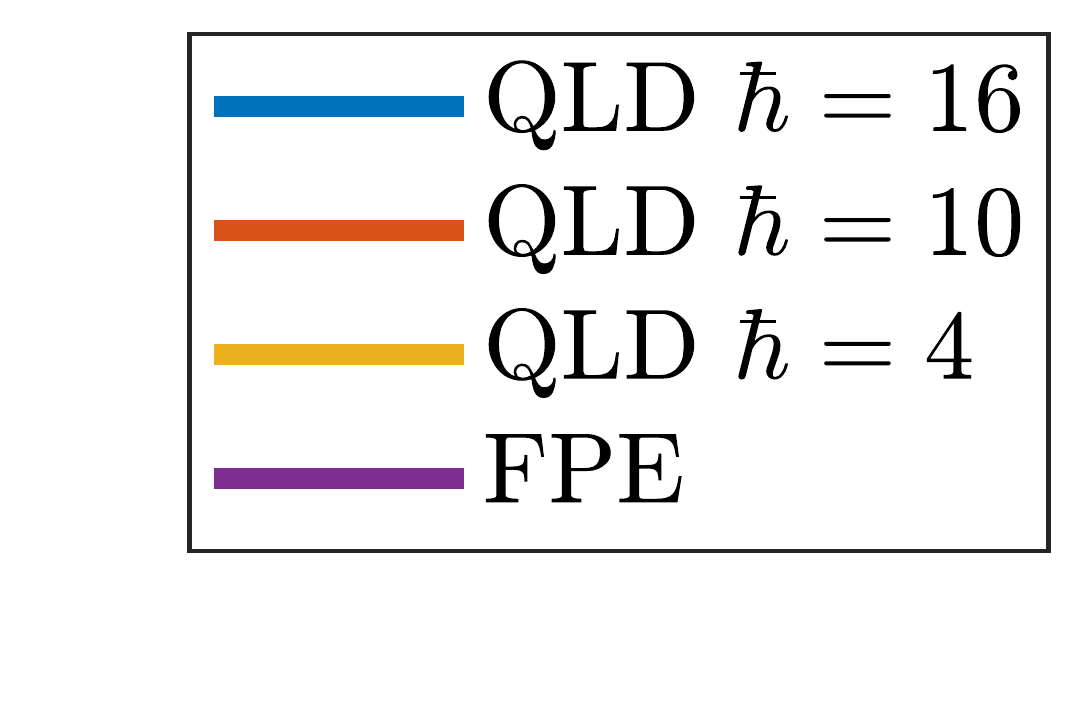}}
  \caption{Success probabilities of QLD (with different $\hbar$) and FPE at different temperatures. FPE demonstrates superior performance in high-temperature conditions, while QLD thrives in lower temperature environments. }
  \label{fig:compareFPE}
\end{figure}

\Results{As shown in Fig.~\ref{fig:compareFPE}, QLD and FPE exhibit entirely distinct characteristics. FPE performs better at high temperature, whereas QLD excels at low temperature. This observation aligns intuitively with classical dynamics, which necessitates higher temperatures for escaping local minima, consequently influencing the overall result quality. In contrast, QLD benefits from lower temperatures for effective energy dissipation, facilitating its transition to lower-energy subspace.}

\section{\label{sec:comparison}Time-dependent QLD}

Based on the conclusions drawn in Section \ref{sec:compareFPE}, QLD does not demonstrate a significant advantage in situations where classical algorithms already perform well. Conversely, in landscapes where classical algorithms exhibit shortcomings, QLD shows only marginal superiority. This implies that we have not fully leveraged the advantages inherent in quantum physics. Consequently, we propose a time-dependent QLD\note{, where we manipulate the quantum and thermal effect more precisely in order to deal with non-convexity in objective functions.}

In this section, we extend Theorem \ref{thm:QLD_convergence} to accommodate time-dependent scenarios. Specifically, we allow the parameters
$\eta,\hbar,T$ in \eq{eq:Gao_Lindblad} to be time-dependent, denoted as $\eta(t),\hbar(t),T(t)$. Notably, the convergence of the time-dependent QLD is not only maintained but also exhibits improved performance in convex and quasar-convex cases, see Section \ref{sec:convergence_time-dependent-qld} for the details. By introducing the parameter $\eta(t)$, the convergence speed of QLD becomes $O(e^{-\int_{0}^t\eta(t^\prime) \, dt^\prime})$. When carefully selecting an appropriate function $\eta(t)$, QLD can even achieve faster convergence. 
The time-dependent factors $\hbar(t)$ and $T(t)$ allow us to gradually decrease the strength of quantum tunneling and thermal fluctuation, respectively. This gradual decrease helps concentrate the optimization process towards the global minimum in the nonconvex case.

Section \ref{sec:objective_functions} introduces process for selecting nonconvex objective functions. These functions are categorized into five groups, each with distinct features. 
In Section \ref{sec:time-dependent-qld}, we provide an example of a specific time-dependent QLD for the nonconvex case by specifying the functions $\hbar(t)$ and $T(t)$. 

We compare QLD with other state-of-the-art algorithms, including Quantum Hamiltonian Descent (QHD) \cite{leng2023quantum}, Quantum Adiabatic Algorithm (QAA), Stochastic Gradient Descent (SGD) and Nesterov's accelerated gradient descent (NAGD) for different nonconvex objective functions, in Section \ref{sec:quantum_al_lists} and Section \ref{sec:classical_al_lists}. The first three algorithms are quantum algorithms, and the last two are classical algorithms. All quantum algorithms are simulated on classical computers to visualize the evolution of probability distributions. Finally, we summarize the results of comparison and put forward a three-phase diagram to explain the superiority of QLD in the nonconvex case, see Section \ref{sec:three_phases_theorem}.

\subsection{Convergence of time-dependent QLD}\label{sec:convergence_time-dependent-qld}
In this section, we prove the convergence of time-dependent QLD in the convex case. We first evolve the time-dependent  parameters $\eta(t),\hbar(t),T(t)$ to QLD respectively in \cref{cor:QLD_convergence_time-dependent_eta,cor:QLD_convergence_time-dependent_hbar,cor:QLD_convergence_time-dependent_T}. Ultimately, we briefly discuss how to evolve them simultaneously.

\subsubsection{Time-dependent $\eta(t)$}
Note that the damping rate $\eta>0$ in Theorem \ref{thm:QLD_convergence} is a constant during the evolution. We generalize $\eta=\eta(t)>0$ for all $t\geq0$, with the understanding that we can incrementally increase $\eta$ to enhance the convergence speed of QLD, without compromising its effectiveness in escaping from local minima. Consequently, we can prove the following corollary of Theorem \ref{thm:QLD_convergence}.
\begin{corollary_}\label{cor:QLD_convergence_time-dependent_eta}
  Only let $\eta$ in Theorem \ref{thm:QLD_convergence} be a time-dependent parameter $\eta(t)$, and keep other parameters and requirements consistent with Theorem \ref{thm:QLD_convergence}. In this case, \eq{eq:V_converge_convex} becomes $$\lr{V}_{t} - V(x^*) \le \text{\note{$\zeta$}}kT + O(e^{-\int_0^t \eta(t^\prime)dt^\prime}).$$
\end{corollary_}
\begin{proof}
  The validity of Lemma \ref{lem:Ehrenfest} holds even when $L_j=L_j(t)$. Consequently,  \eq{eq:energy_derivative_HHH} remains applicable when $\eta=\eta(t)$. This, in turn, establishes that \eq{eq:energy_derivative_H} also continues to hold. Under the approximation $\hbar \Omega \ll kT$, \eq{eq:energy_derivative_H_final_1111} becomes
  \begin{align}
    \frac{\d}{\d t}\langle H\rangle_t\leq \eta(t)\cdot  \left(\text{\note{$\zeta$}}kT-\langle H\rangle_t \right),
    \label{eq:dH_t_time}
  \end{align}
  in which the only difference is $\eta=\eta(t)$. Then, defining a function $h(t)=(\lr{H}_{t}-\text{\note{$\zeta$}}kT)e^{\int_0^t\eta(t^\prime) dt^\prime}$, it can be directly verified that $\frac{\d h(t)}{\d t}\leq 0$ for $\forall t>0$ by \eq{eq:dH_t_time}. Therefore, $h(t)\leq h(0)$, and
  \begin{align}
    \lr{H}_{t}-\text{\note{$\zeta$}}kT&\le (\lr{H}_{0}-\text{\note{$\zeta$}}kT)e^{-\int_0^t\eta(t^\prime) dt^\prime} = O(e^{-\int_0^t\eta(t^\prime) dt^\prime}).
  \end{align}
  Using \eq{eq:VleqH}, we get $\lr{V}_{t} \leq \lr{H}_{t} \le \text{\note{$\zeta$}}kT + O(e^{-\int_0^t \eta(t^\prime)dt^\prime})$.
\end{proof}
By dynamically modulating the value of  $\eta(t)$ at various stages of the evolution process, we can effectively control the convergence rate of the QLD, thereby enabling acceleration or deceleration of its convergence speed as needed.

\subsubsection{Time-dependent $\hbar(t)$}
In the numerical experiments of time-independent $\hbar$, we observe that larger $\hbar$ can reinforce the effect of quantum tunneling to escape from local minima, however, it also makes the system incapable of concentrating on the minimum. In order to solve this problem, we prove the following corollary of Theorem \ref{thm:QLD_convergence} with
monotonically decreasing $\hbar(t)$  with respect to $t$.
\begin{corollary_}\label{cor:QLD_convergence_time-dependent_hbar}
  Only let $\hbar$ in Theorem \ref{thm:QLD_convergence} be a time-dependent parameter $\hbar(t)$, which satisfies $\frac{\d \hbar(t)}{\d t}\leq 0$ and $\hbar(t)>0$ for $\forall t\geq0$, while still keeping other parameters and requirements consistent with Theorem \ref{thm:QLD_convergence}. In this case, \eq{eq:V_converge_convex} still holds
  $$\lr{V}_{t} - V(x^*) \le \text{\note{$\zeta$}}kT + O(e^{- \eta t}).$$
\end{corollary_}
\begin{proof}
  The validity of Lemma \ref{lem:Ehrenfest} holds even when $\hbar=\hbar(t)$. Then the first term on the right-hand side of \eq{eq:energy_derivative_HHH} is non-zero now, becoming
  \begin{align}
    \lr{\frac{\d}{\d t}H }_t = \lr{\frac{\d}{\d t}\left( \frac{p^2}{2m}+V\right)}=\lr{\frac{p}{m}\frac{\d p}{\d t} }_t.
    \label{eq:dH_t_hbar}
  \end{align}
  Using $p = -\im \hbar(t) \nabla$, we get
  \begin{align}
    \frac{\d p}{\d t} = -\im \frac{\d \hbar}{\d t}\nabla = \frac{p}{\hbar}\frac{\d \hbar}{\d t} = p \frac{\d \ln \hbar}{\d t}.
  \end{align}
  Inserting the above equation into \eq{eq:dH_t_hbar}, \eq{eq:dH_t_hbar} becomes
  \begin{align}
    \lr{\frac{\d}{\d t}H }_t = \lr{\frac{p^2}{m}\frac{\d \ln \hbar}{\d t}}_t=\frac{\d \ln \hbar}{\d t} \cdot \lr{\frac{p^2}{m}}_t.
  \end{align}
  At the same time, it is not difficult to verify that the second term and the third term on the right-hand side of \eq{eq:energy_derivative_HHH} are still same here. Therefore, \eq{eq:energy_derivative_H} becomes
  \begin{align}
    \frac{\d}{\d t}\langle H\rangle_t = & \frac{\d \ln \hbar}{\d t} \cdot \lr{p^2/m}_t+\left(\frac{\hbar^2\mu^2}{m}+\frac{\eta^2}{4\mu^2}\langle\nabla^2 V\rangle_t\right) -\eta\langle x \cdot\nabla V+p^2/m\rangle_t .
    \label{eq:dH_t_hbar2}
  \end{align}
  Under the approximation $\hbar(t) \Omega \ll kT$, \eq{eq:approximation_H_leq} becomes
  \begin{align}
    \frac{\d}{\d t}\langle H\rangle_t\leq & \frac{\d \ln \hbar}{\d t} \cdot \lr{p^2/m}_t  + \eta\cdot \left(\text{\note{$\zeta$}}kT-\langle V+p^2/(2m)\rangle_t \right).
  \end{align}

  Because $\frac{\d \hbar (t)}{\d t}<0$, we get  $\frac{\d \ln \hbar (t)}{\d t}<0$, and $\lr{p^2/m}_t\geq 0$, inducing the following inequality
  \begin{align}
    \frac{\d}{\d t}\langle H\rangle_t\leq \eta \cdot\left(\text{\note{$\zeta$}}kT-\langle V+p^2/(2m)\rangle_t \right),
  \end{align}
  which is the same as \eq{eq:energy_derivative_H_final_1111}. Then, the rest of the proof is similar to that of Theorem \ref{thm:QLD_convergence} and Proposition \ref{prop:non_increasing}.
\end{proof}

\subsubsection{Time-dependent $T(t)$}

We also investigate the impact of a time-dependent temperature, $T(t)>0$, on the convergence of QLD. Given that $2kT$ influences the precision of QLD (refer to Remark \ref{remark:QLD_convergence}), possessing a progressively diminishing $2kT$ can enhance the precision of QLD. Additionally, it can intensify the thermal effect to escape from the local minimum when $T$ is initially high at the onset of the evolution process. Under the aforementioned perspective, we have the following corollary of Theorem \ref{thm:QLD_convergence}.

\begin{corollary_}\label{cor:QLD_convergence_time-dependent_T}
  There exists a function $f$ in relation to $T(\cdot)$, with properties: \\
  1. $\lim _{T\to 0^+}f(T)= 0^+$,  \\
  2. $f(T)\geq T$ for $\forall T>0$, \\
  3. $\d f(T)/\d T >0$ for $\forall T>0$. \\
  There also exists another function $G(\cdot)$ which depends on $f$ and is a monotonically decreasing function. If temperature $T$ is a time-dependent parameter $T(t)$ which satisfies $\d T/\d t\leq0$ and $T(t)\leq G(t)$ for $\forall t\geq0$, \eq{eq:V_converge_convex} becomes
  $$\lr{V}_{t} - V(x^*) \le \text{\note{$\zeta$}}k f(T(t)) + O(e^{- \eta t}),$$
  if we still keep other parameters and requirements consistent with Theorem \ref{thm:QLD_convergence}.
\end{corollary_}
\begin{proof}
Similar to the proof of Corollary \ref{cor:QLD_convergence_time-dependent_eta}, we can find that \eq{eq:energy_derivative_H_final_1111} holds here:
  \begin{align}
    \frac{\d}{\d t}\langle H\rangle_t+\eta \cdot\left(\lr{H}_t-\text{\note{$\zeta$}}kT(t)\right)\leq 0.
    \label{eq:dH_t_T_123}
  \end{align}
Defining a function $h(t)=(\lr{H}_t-2kf(T(t)))e^{\eta t}$, its time derivative can be calculated as
  \begin{align}
    \frac{\d h(t)}{\d t}= & \left(\frac{\d }{\d t}\lr{H}_t-\text{\note{$\zeta$}}k \frac{\d f}{\d T} \frac{\d T}{\d t}\right)e^{\eta t} +(\lr{H}_t-\text{\note{$\zeta$}}kf(T(t)))\eta e^{\eta t}.
  \end{align}
  If there exists functions $f(T)$ and $T(t)$ that satisfy (we will prove their existence later)
  \begin{align}
    \frac{\d f}{\d T} \frac{\d T}{\d t}+\eta f(T)\geq \eta T>0,
    \label{eq:inequality_T}
  \end{align}
  it is direct to verify that
  \begin{align}
         & \left(\frac{\d}{\d t} \lr{H}_t-\text{\note{$\zeta$}} k \frac{\d f}{\d T} \frac{\d T}{\d t}\right)+\eta(\lr{H}_t-\text{\note{$\zeta$}} kf(T(t))) \label{eq:inequality_H} \\
    \leq & \frac{\d}{\d t}\langle H\rangle_t+\eta \cdot\left(\lr{H}_t-\text{\note{$\zeta$}} kT(t)\right)\leq0,
  \end{align}
  in which the second inequality comes from \eq{eq:dH_t_T_123}. Multiplying $e^{\eta t}$ on \eq{eq:inequality_H}, we get
  \begin{align}
    \frac{\d h(t)}{\d t}\leq 0.
  \end{align}
  Then $h(t)\leq h(0)$ implies $\lr{H}_t-\text{\note{$\zeta$}}kf(T(t))\leq O(e^{-\eta t})$. Using \eq{eq:VleqH}, we get $\lr{V}_{t} \leq \lr{H}_{t} \le \text{\note{$\zeta$}}k f(T(t)) + O(e^{- \eta t})$.

  Therefore, we only need to prove the existence of functions $f(T)$ and $T(t)\leq G(t)$ satisfying all the properties in this corollary.

  Define $f(T)=T_c (e^{T/T_c}-1)$, which satisfies all properties of $f(\cdot)$ in this corollary, where $T_c$ is any constant temperature to ensure dimensional consistency.  Inserting $f(T)=T_c (e^{T/T_c}-1)$ into \eq{eq:inequality_T}, we get
  \begin{align}
    e^{{T/T_c}} \frac{\d T}{\d t}+\eta T_c (e^{T/T_c}-1)\geq \eta T,
    \label{eq:inequality_T2}
  \end{align}
  which can be rewritten as
  \begin{align}
    0\geq\frac{\d T}{\d t}\geq\eta  \frac{T-T_c (e^{T/T_c}-1)}{e^{T/T_c}}.
    \label{eq:inequality_T3}
  \end{align}
  Here, we mainly focus on $\frac{\d T}{\d t}\leq0$ consistent with the corollary, because \eq{eq:inequality_T2} is obviously true when $\frac{\d T}{\d t}>0$.
  \eq{eq:inequality_T3} can be transformed into (consider $\d t>0,\d T\leq0$)
  \begin{align}
    0\leq\frac{e^{T/T_c}}{T-T_c (e^{T/T_c}-1)}{\d T}\leq \eta \d t.
    \label{eq:inequality_T4}
  \end{align}
  Integrate both sides of the above equation
  \begin{align}
    \int_{T(0)}^{T(t)}\frac{e^{T/T_c}}{T-T_c (e^{T/T_c}-1)}{\d T}\leq \eta\int_0^t\d t^\prime= \eta t,
    \label{eq:inequality_T5}
  \end{align}
  where $T(t)\leq T(0)$ for $t>0$. Denote the integration $\int\frac{e^{T/T_c}}{T-T_c (e^{T/T_c}-1)}{\d T}=\mathcal{G}(T)$ \footnote{This integral does not have an analytical solution, and it is also non-integrable when the interval of integration includes 0. However, in the situations we are considering (such as numerical experiments), we can give $T$ a finite lower bound, thus making this integral integrable.}. Obviously, $\mathcal{G}(T)$ is a monotonically decreasing function with respect to $T$. As a result, we have
  \begin{align}
    \mathcal{G}(T(t))-\mathcal{G}(T(0))\leq \eta t,
  \end{align}
  which implies
  \begin{align}
    T(t)\leq \mathcal{G}^{-1} \big(\eta t+\mathcal{G}(T(0))\big).
  \end{align}
  Set $G(t) =\mathcal{G}^{-1} \big(\eta t+\mathcal{G}(T(0))\big)$, we obtain the function $G(t)$ stated in this corollary.
\end{proof}
\begin{remark_}
  It is not difficult to verify that when $T(t)$ is a constant and $f(T)=T$, Corollary \ref{cor:QLD_convergence_time-dependent_T} is reduced to Theorem \ref{thm:QLD_convergence}.
\end{remark_}
\begin{remark_}
  By this corollary, we can actualize  Remark \ref{remark:QLD_convergence} by judiciously selecting a $T(t)$ such that $f(T(t))$ is minimized to the desired precision level $\varepsilon$.
\end{remark_}

\subsubsection{Three time-dependent parameters $\eta(t),\hbar(t),T(t)$ }
Finally, we consider the case where all three parameters $\eta(t),\hbar(t),T(t)$ are time-dependent. If \cref{cor:QLD_convergence_time-dependent_eta,cor:QLD_convergence_time-dependent_hbar,cor:QLD_convergence_time-dependent_T} are satisfied simultaneously, then we have
\begin{align}
  \lr{V}_{t} - V(x^*) \le \text{\note{$\zeta$}}k f(T(t)) + O(e^{-\int_0^t \eta(t^\prime)dt^\prime}),
\end{align}
whose proof, similar to the proof of \cref{cor:QLD_convergence_time-dependent_eta,cor:QLD_convergence_time-dependent_hbar,cor:QLD_convergence_time-dependent_T}, is omitted here. The proof for quasar-convex case is essentially the same, where the convergence still holds
\begin{align}
    \lr{V}_{t} - V(x^*) \le \frac{\text{\note{$\zeta$}}}{r}k f(T(t)) + O(e^{-r\int_0^t \eta(t^\prime)dt^\prime}).
\end{align}

\subsection{Objective functions} \label{sec:objective_functions}
In this part, we succinctly introduce the objective functions used in our nonconvex function experiments.
Many objective functions of global optimization are multidimensional \cite{layeb2022new, floudas2013handbook, al2015unconstrained, jamil2013literature, simulationlib, leng2023quantum}. However, we can only conduct one-dimensional experiments on classical computers due to the computational cost (each one-dimensional QLD experiment requires approximately $14$ hours). Consequently, we transform several multidimensional objective functions into one-dimensional ones, ensuring the preservation of their most significant characteristics within the landscapes. Finally, $7$ nonconvex functions are split into $5$ different categories by the characteristics of their landscapes \cite{leng2023quantum}. Each of them only has one global minimum $x^*=0$. By shifting the functions vertically, we set the global minimum value $V(x^*)=0$.

Ref.~\cite{leng2023quantum} also explains how to categorize different objective functions. The category  ``Ridges or Valleys'' is featured by small deep sections that divide the domain. ``Basin'' is featured by flatness near the global minimum of function values. ``Flat'' is featured by its flatness at elevated function values. ``Studded'' is identified by a foundational shape that exhibits high-frequency perturbations. The category ``Simple'' are functions that can be solved efficiently by standard gradient descent algorithms. The functions in each category are listed in Table~\ref{tab:table1}, whose precise formulations can be found in Appendix \ref{sec:objective_functions_appendix}.

\begin{table}[htbp]
    \centering
        \caption{Test functions classification.}
    \begin{tabular}{@{}ll@{}}
    \toprule
   \textrm{Features}          & \textrm{Function names}                    \\
     \midrule
      \textrm{Ridges or Valleys} & \textrm{Deflected Corrugated Spring (DCS)} \\
      \textrm{Basin}             & \textrm{Csendes}                           \\
      \textrm{Flat}              & \textrm{Michalewicz}                       \\
      \textrm{Studded}           & \textrm{Alpine 1, Bohachevsky 2, Griewank} \\
      \textrm{Simple}            & \textrm{Double well}                       \\
      \bottomrule
    \end{tabular}
   \label{tab:table1}  
\end{table}

\subsection{\label{sec:time-dependent-qld}Time-dependent Quantum Langevin Dynamics}
In the previous experiments, we analyze roles of $T,\eta, \hbar$ in QLD, which also lead to the question why QLD would have a better performance if these parameters are time-dependent, as different ranges of parameters have different effect in different phases of the optimization process. At the beginning of the evolution, we set a high temperature to prompt the system to escape from local minimum. Then, we reduce the temperature to cool the system, allowing it to reach a more optimal, lower energy state. The discussion concerning $\hbar$ is also similar.  At the beginning of the evolution, larger $\hbar$ can accelerate the rate of quantum tunneling, helping system escape from local minimum. After seeking the global minimum, smaller $\hbar$, weakening the quantum tunneling effect, can prevent system from tunneling back to local minimum. If $\hbar$ and $T$ decrease at the same time, $\hbar$ must decrease faster than $T$ as the effect of  tunneling back to local minimum is very significant at a mild temperature, termed as “quantum reverse tunneling”. This could be due to the large value of $\hbar$, which violates the approximation $\hbar\Omega \ll kT$ as stated in \cref{cor:QLD_convergence_time-dependent_eta,cor:QLD_convergence_time-dependent_hbar,cor:QLD_convergence_time-dependent_T}. This violation could trigger the ``quantum reverse tunneling'' phenomenon. Therefore, $\hbar(t)$ needs to decrease faster than $T(t)$ as \eq{eq:time_hbar} and \eq{eq:time_T}. This might ensure that the system can concentrate on the global minimum after escaping from local minimum.

From the above analysis, we decide to use inverted Sigmoid function to establish time-dependent $\hbar(t)$ and $T(t)$, which is also well known as the (inverted) Richards curve \cite{richards1959flexible}.  The inverted Sigmoid function $Y(t)$ is defined as
\begin{align}
  Y(t)=Y_l+\frac{Y_u-Y_l}{1+\exp({\tilde{k}(t-\tau)})},
  \label{eq:inverted_Sigmoid}
\end{align}
where $Y_l$ and $Y_u$ are the lower and upper bounds, respectively, $\tilde{k}$ is the growth (decline) rate, and $\tau$ is the time of maximum growth (decline).

\begin{figure}[htbp]
\centering
  \includegraphics[width=0.35\textwidth]{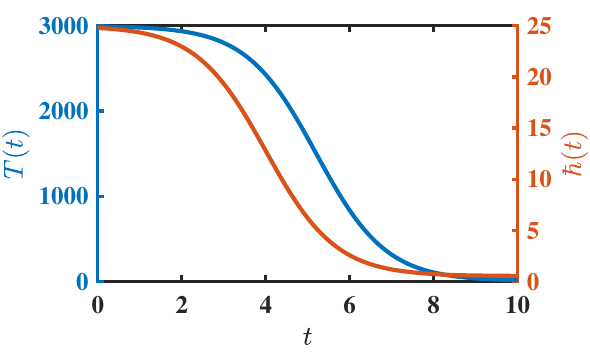}
  \caption{Time-dependent $\hbar(t)$ and $T(t)$ for $t_f=10$.}
  \label{fig:time_hbar_T}
\end{figure}

In all experiments of time-dependent QLD, $\hbar(t)$ is set to be
\begin{align}
  \hbar(t)=0.5+\frac{25-0.5}{1+\exp({1.2\cdot(t-0.4\cdot t_f)})},
  \label{eq:time_hbar}
\end{align}
and $T(t)$ is set to be
\begin{align}
  T(t)=2+\frac{3000-2}{1+\exp({1.2\cdot(t-0.52\cdot t_f)})},
  \label{eq:time_T}
\end{align}
where $t_f$ is the total evolution time. The curves of $\hbar(t)$ and $T(t)$ are shown in Fig.~\ref{fig:time_hbar_T}.

Same finite difference method and implicit Euler method as Section~\ref{sec:Implement_QLD} are used to numerically solve the evolution of system. The only difference is that time-dependent $\hbar(t)$ and $T(t)$ make $\hat{\mathcal{L}}$ of \eq{eq:Gao_L_matrix} be a time-dependent matrix $\hat{\mathcal{L}}(t)$. Dirichlet boundary condition and approximately uniform distribution are still kept here.

\Pra{Resolution $n=128$, domain $x\in [-1,1]$, space step $\Delta x=\frac{2}{n}$, time step $\Delta t=0.05\Delta x^2$, total evolution time $t_f=10$, $\hbar(t)$ and $T(t)$ are \eq{eq:time_hbar} and \eq{eq:time_T}, characteristic damping rate $\eta=10$, $\Omega=300$, $m=1$, initial state $\psi_0(x)=\sqrt{p(x)}$.}

\subsection{Quantum algorithms}\label{sec:quantum_al_lists}
We introduce another two quantum algorithms: Quantum Hamiltonian Descent (QHD) and Quantum Adiabatic Algorithm (QAA) as competitors of QLD.
\subsubsection{Quantum Hamiltonian Descent (QHD)}
We mainly follow Ref.~\cite{leng2023quantum} to implement QHD. The Hamiltonian of QHD is given by
\begin{align}
  H(t)=\left(\frac{2}{s_1+t^3}\right)\left(-\frac{1}{2} \nabla^2\right)+2\left({s_2+t^3}\right)V(x).
  \label{eq:QHD_Hamiltonian}
\end{align}
In the ideal case, $s_1=s_2=0$
\footnote{The original form of Ref.~\cite{leng2023quantum} set $s_2$ to zero, however, our symplectic leapfrog scheme is not as stable as shifted Fourier transform used in Ref.~\cite{leng2023quantum}. In order to maintain the stability of numerical calculation in symplectic leapfrog scheme, we also set $s_2$ to be a small positive number.}. We would like to note here that the damping parameter $\frac{1}{t^3}$ is just one possible choice, made for the purpose of comparing QHD with the classical Nesterov's accelerated gradient descent algorithm directly. It is possible for QHD to achieve better performance with alternative damping parameter.

Then, we discretize the space into grids similar to Section~\ref{sec:Implement_QLD} and use symplectic leapfrog scheme \cite{Mauger} to evolve this Hamiltonian system. 

\Pra{Resolution $n=128$, domain $x\in [-1,1]$, space step $\Delta x=\frac{2}{n}$, time step $\Delta t=0.05\Delta x^2$, total evolution time $t_f=10$, $s_1=s_2=0.01$, initial state being the uniform distribution.}

\subsubsection{Quantum Adiabatic Algorithm (QAA)}
Quantum Adiabatic Algorithm \cite{farhi2001quantum} is a famous quantum computing algorithm that takes advantage of the quantum adiabatic theorem to solve general optimization problems. Suppose the Hamiltonian changes gradually from initial $H_i$ to final $H_f$, quantum adiabatic theorem states that if the system is initially in the $n$-th eigenstate of $H_i$, it will remain in the $n$-th eigenstate of $H_f$ at final time $t_f$ if the evolution is slow enough \cite{griffiths2018introduction}. To be more precise, it follows the Schr\"{o}dinger equation
\begin{align}
  \im \frac{\partial \psi}{\partial t}= \left[(1-g(t))H_i+g(t)H_f \right]\psi(t).
  \label{eq:QAA}
\end{align}
where $g(t)$ is a time-dependent function that satisfies $g(0)=0$ and $g(t_f)=1$.
In order to solve optimization problems, we set $H_f$ to be the problem Hamiltonian that encodes the problem of interest. The final state of the system represents the solution to the optimization problem. By measuring the final state, we can obtain the solution to the problem. Following Refs.~\cite{leng2023quantum, cohen2020portfolio,date2021adiabatic}, we adopt Radix-2 representation to solve continuous optimization problems. The initial Hamiltonian $H_i$ is given by sum of Pauli-$X$ operators
\begin{align}
  H_i=-\sum_{j=1}^N X_j,
\end{align}
whose ground state is the uniform superposition of all basis states. The final Hamiltonian $H_f$ is given by diagonal matrix $V=\text{diag}(V(x_1),\ldots,V(x_{2^N}))$.

The evolution of \eq{eq:QAA} is calculated numerically by symplectic leapfrog scheme \cite{Mauger}.

\Pra{Resolution $n=2^N=128$, domain $x\in [-1,1]$, space step $\Delta x=\frac{2}{n}$, time step $\Delta t=0.05\Delta x^2$, total evolution time $t_f=10$, $g(t)=\frac{t}{t_f}$, initial state being the uniform distribution.}

\subsection{Classical algorithms}\label{sec:classical_al_lists}
This part describes Stochastic Gradient Descent (SGD) which is the classical counterpart of QLD, and another prestigious Nesterov's accelerated gradient descent (NAGD). Both of them are as competitors of QLD.

\subsubsection{Stochastic Gradient Descent (SGD)}\label{sec:SGD}
Stochastic Gradient Descent is widely used in  machine learning and deep learning for training models to minimize the loss function \cite{shi2020learning}. Starting from an initial point $x_0$, SGD updates the point by
\begin{align}
  x_{k+1}=x_k-s \nabla f(x_k)-s\xi_k,
\end{align}
where $s$ is the step size (learning rate), $\xi_k$ is a noise term at $k$-th iteration. The effective evolution time at the $k$-th step is $t_k=sk$. In our experiments, we set $\xi_k$ to be a Gaussian random noise with zero mean and non-zero variance \cite{leng2023quantum}.

\Pra{Domain $x\in [-1,1]$, learning rate $s=0.05\cdot (\frac{2}{128})^2$, total evolution time $t_f=10$, variance of Gaussian random noise $\sigma=1$, $20000$ initial points $x_0$ are uniformly randomly distributed on $[-1,1]$.}

\subsubsection{Nesterov's accelerated gradient descent (NAGD)}\label{sec:NAGD}
Nesterov's accelerated gradient descent \cite{nesterov1983method} is the classical counterpart of QHD, which iterates according to
\begin{align}
  \begin{split}
    x_{k}&=y_{k-1}-s \nabla f(y_{k-1}),\\
    y_{k}&=x_{k}+\frac{k-1}{k+2}(x_{k}-x_{k-1}),
  \end{split}
\end{align}
where $s$ is the step size. The initial points are chosen as $x_0=y_0$. In some cases, it will converge faster than the standard gradient descent. The effective evolution time at the $k$-th step is $t_k=sk$ \cite{leng2023quantum}.

\Pra{Domain $x\in [-1,1]$, learning rate $s=0.05\cdot (\frac{2}{128})^2$, total evolution time $t_f=10$, $20000$ initial points $x_0=y_0$ are uniformly randomly distributed on $[-1,1]$.}

\subsection{Results of the comparison}\label{sec:three_phases_theorem}
In above experiments, we fix the domain of $x$ to $[-1,1]$ and all algorithms share the same total evolution time $t_f=10$ and \note{the} same time step $\Delta t$ \note{and} $s=0.05\cdot (\frac{2}{128})^2$. Besides, all quantum algorithms share same resolution $n=128$. Seven objective functions in Table~\ref{tab:table1}  with different features are tested for each algorithm.

\begin{figure}[htbp]
\centering
  \includegraphics[width=0.37\textwidth]{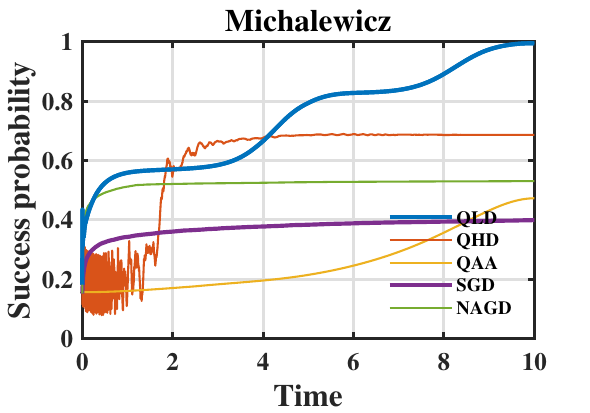}
  \caption{Success probabilities of Michalewicz function for different algorithms at different (effective) evolution times.}
  \label{fig:Michalewicz_success}
\end{figure}

Here, we first show the results of Michalewicz function in Fig.~\ref{fig:Michalewicz_success} and Fig.~\ref{fig:Michalewicz_time_evolution}, while the results of other objective functions are shown in Appendix~\ref{sec:numerical_exp_appendix} due to space constraints. In these numerical experiments, we demonstrate that time-dependent QLD leads to superior performance in many landscapes.
\begin{figure*}[htbp]
  \centering
  {\includegraphics[width=1\textwidth]{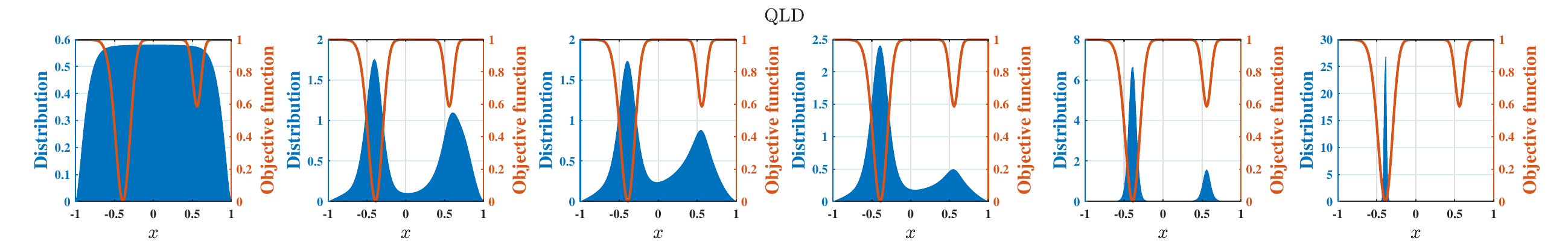}}  \\
  \vspace{-0.161cm}
  {\includegraphics[width=1\textwidth]{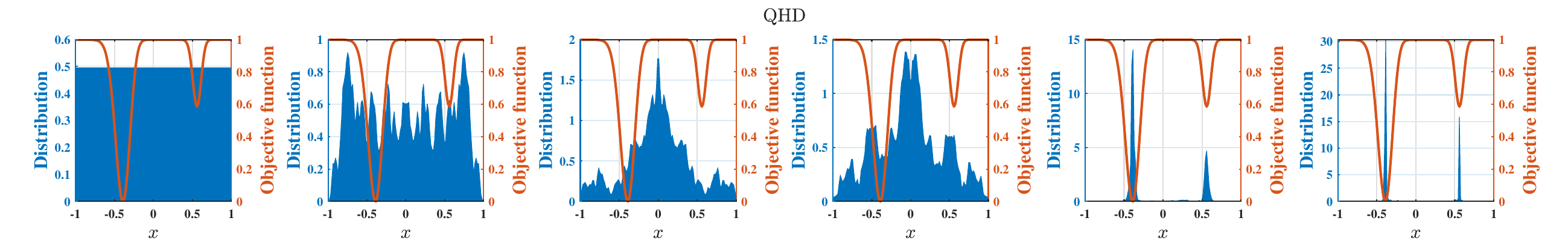}}\\
  \vspace{-0.161cm}
  {\includegraphics[width=1\textwidth]{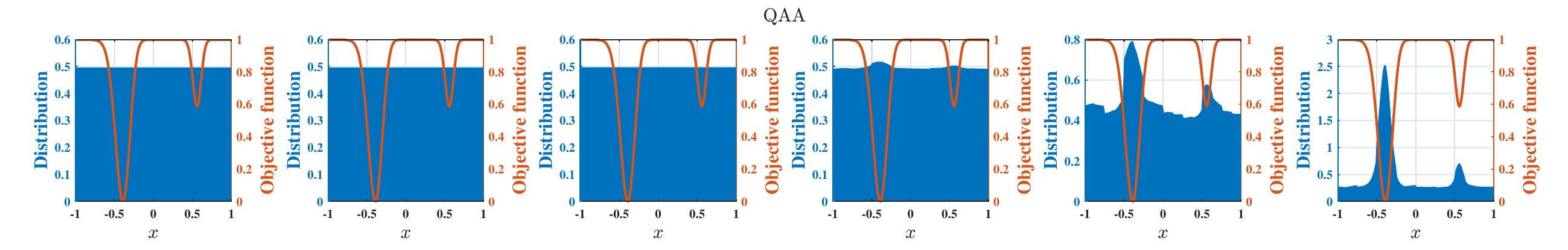}}\\
  \vspace{-0.161cm}
  {\includegraphics[width=1\textwidth]{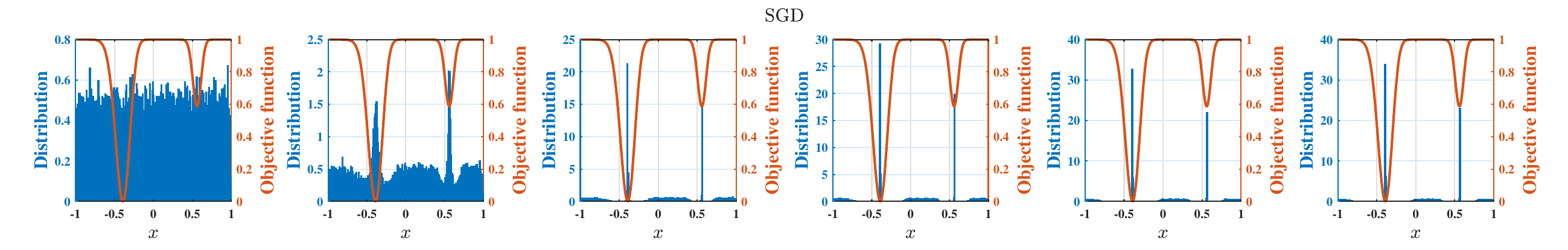}}\\
  \vspace{-0.16cm}
  {\includegraphics[width=1\textwidth]{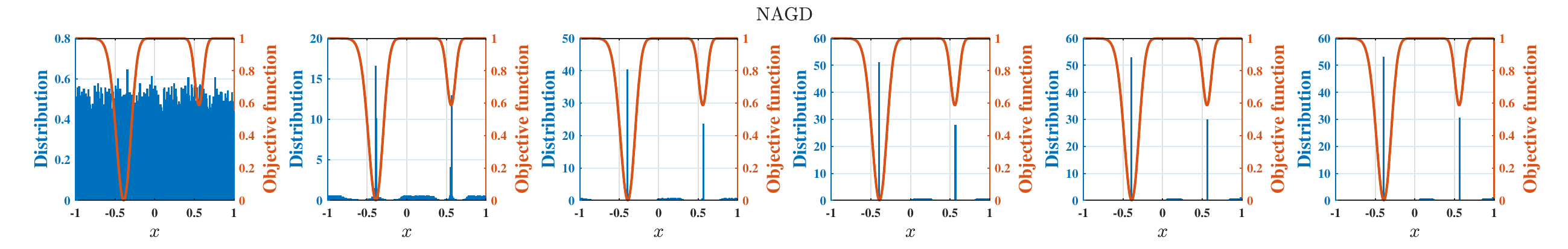}}\\
  \vspace{-0.17cm}
  {\includegraphics[width=1\textwidth]{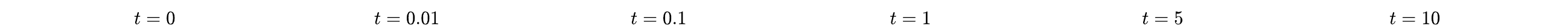}}
  \caption{Probability distributions of Michalewicz function $(x^*=-0.3911)$ for different algorithms at different (effective) evolution times $t=0,0.01,0.1,1,5,10$.}
  \label{fig:Michalewicz_time_evolution}
\end{figure*}

Finally, based on the observations from Fig.~\ref{fig:Michalewicz_success}, Fig.~\ref{fig:Michalewicz_time_evolution}, and Appendix~\ref{sec:numerical_exp_appendix}, we categorize the entire evolution process into three distinct phases. This classification not only provides a comprehensive interpretation but also parallels the three phases of QHD \cite{leng2023quantum}. In essence, we identify three key phases in the QLD evolution: the \textbf{thermal} phase, the \textbf{cooling} phase, and the \textbf{descent} phase.

\textbf{Thermal phase.  }We set a high temperature and a large $\hbar$ at first to leverage its tunneling and thermal capability. At this stage, the wave function rapidly narrows the search domain. It serves as the global search phase in QHD \cite{leng2023quantum}.

\textbf{Cooling phase.  }We extract energy from the heat bath and weaken the quantum tunneling effect artificially (low down the temperature and $\hbar$), to prepare our algorithm for concentrating to the global minimum in the low-energy subspace.

\textbf{Descent phase.  }In the last phase, both the thermal and tunneling effect are insignificant. The wave function becomes more concentrated in a slow convergence rate.

\section{Comparison of query complexity for different algorithms}\label{app:query_com}
In this section, we delve into the comparison of query complexity among various algorithms, i.e., the classical query complexity to local information and the quantum query complexity to the evaluation oracle \cite{liu2023quantum,childs2022quantum}. We assume that the quantum evaluation oracle $O_V$ allows us to query the function value of $V$ in a coherent way
\begin{align}
  O_V \ket{x}\ket{z} = \ket{x}\ket{z + V(x)}, \quad \forall x \in  \mathbb{R}^d, z \in \mathbb{R}~.
  \label{eq:oracle_OV}
\end{align}
Correspondingly, there is a classical evaluation oracle $O_{V}^{cl}$ that allows us to query the function value of $V$ in a local way. In the following, we will discuss the query complexity of QLD, QHD, QAA, SGD, and NAGD.

\subsection{Quantum query complexity for QLD}
The quantum query complexity associated with QLD can be inferred from the complexity involved in the simulation of the Lindblad system, given that QLD is a Lindblad system itself. Here, we mainly follow Proposition \ref{prop:complexity_Lindblad} to simulate QLD. Definition \ref{def:block_encoding}, Definition \ref{def:state_pre} and Proposition \ref{prop:linear_combination} are stated for the references in constructing block-encoded matrices used in Proposition \ref{prop:complexity_Lindblad}.
\begin{theorem_}\label{thm:QLD_query_complexity}
  Given a Lindblad equation with form \eq{eq:Gao_Lindblad},
  \note{$${O}\left(t\lrb{\hbar^2 n^2+\|V\|_{\max}+\frac{\eta m kT}{\hbar^2}+\eta n} \frac{\log(t(\hbar^2 n^2+\|V\|_{\max}+\frac{\eta m kT}{\hbar^2}+\eta n)/\epsilon)}{\log\log(t(\hbar^2 n^2+\|V\|_{\max}+\frac{\eta m kT}{\hbar^2}+\eta n)/\epsilon)}\right)$$}
  queries to $O_V$ are needed to simulate \eq{eq:Gao_Lindblad} for time $t$ with error at most $\epsilon$ in terms of the diamond norm, where $n$ is the number of discretization grids along  each axis of domain and $\|V\|_{\max}$ is the max-norm of $V$.
\end{theorem_}
\begin{proof}
  We can divide Hamiltonian into two parts $H=E_k+V$. Then by finite difference method mentioned in Section~\ref{sec:Implement_QLD}, it is very easy to present $E_k$ and $V$ in the matrix forms. Consider discretizing the domain into $n$ grids along each axis, $E_k=-\frac{\hbar^2 \mathbf{L} }{2m \Delta x^2}$, where $\mathbf{L}$ is the Laplacian matrix associated to the central difference method and $\Delta x$ is the width of discretizing grid \cite{liu2023quantum}, and $V=\text{diag}(V(x_1),\ldots,V(x_n))$ is a diagonal matrix by discretizing $V(x)$ in the grid, which is given by the oracle $O_V$ \cite{leng2023quantum}.

  By Definition \ref{def:block_encoding}, it is not difficult to give an $(\alpha_{E_k},a,\xi_{E_k})$-block-encoded matrix $U_{E_k}$ of $E_k$ and an $(\alpha_{V},a,\xi_V)$-block-encoded matrix $U_V$ of $V$, where $\alpha_{E_k}$ is \note{${O}(\frac{\hbar^2}{\Delta x^2})={O}(\hbar^2n^2)$} and $\alpha_{V}$ is ${O}(\|V\|_{\max})$. \note{Here, $m$ is set to $1$ as in the previous numerical experiments.}
  Then, Proposition \ref{prop:linear_combination} derives a $(\beta,a+b,\xi_{E_k}+\xi_V)$-block-encoded matrix $\tilde{W}_H$ of $H$ with $\beta\geq |\alpha_{E_k}|+|\alpha_{V}|$. In addition, Proposition~\ref{prop:linear_combination} also tells one query to $\tilde{W}_H$ can be implemented by one query to $U_{E_k}$ and one query to $U_V$.

  Now, we have gotten a $(\beta,a+b,\xi_{E_k}+\xi_V)$-block-encoded matrix $\tilde{W}_H$ of $H$. In our Lindblad equation \eq{eq:Gao_Lindblad}, the unique $L$ is given by $L=\mu x+\im \nu p$. Presenting $L$ in the matrix form  by finite difference method, we have $L=\mu \cdot\text{diag} (x_1,\ldots,x_n)+\nu\hbar\cdot \frac{\mathbf{D}}{\Delta x}$, where $\mathbf{D}$ is the differential matrix associated to the central difference method. Then, we can easily give an $(\alpha_{L},a,\xi_{L})$-block-encoded matrix $U_L$ of $L$, where $\alpha_{L}$ is at most \note{${O}(\mu + \frac{\nu \hbar}{\Delta x})={O}(\mu + {\nu \hbar}{n})$}.

  Using Proposition \ref{prop:complexity_Lindblad}, $\|\mathcal{L}_{}\|_{\text{be}}=\beta+\alpha_L^2$, where the smallest $\beta$ we can achieve is \note{ $|\alpha_{E_k}|+|\alpha_{V}|={O}(\hbar^2 n^2)+{O}(\|V\|_{\max})$}, which gives 
  \note{\begin{align}
      \|\mathcal{L}_{}\|_{\text{be}}&=O\lrb{\hbar^2 n^2+\|V\|_{\max}+\mu^2 +\hbar^2 \nu^2 n^2+ 2\hbar \mu \nu n}\\
      &=O(\hbar^2 n^2 +\|V\|_{\max}+\mu^2 +\frac{\eta^2 n^2}{4 \mu^2} +\eta n ).
  \end{align}
  Following \eq{eq:mu_approx_1} and \eq{eq:mu_approx}, $$\|\mathcal{L}_{}\|_{\text{be}}=O\lrb{\hbar^2 n^2+\|V\|_{\max}+\frac{\eta m kT}{\hbar^2}+\eta n}.$$}
  
  So, $ O\left(t(\|\mathcal{L}_{}\|_{\text{be}}) \frac{\log(t(\|\mathcal{L}_{}\|_{\text{be}})/\epsilon)}{\log\log(t(\|\mathcal{L}_{}\|_{\text{be}})/\epsilon)}\right)$ queries to $\tilde{W}_H$ and $U_L$ are needed to simulate $e^{\mathcal{L}t}$ with error at most $\epsilon$. Recall that each query to $\tilde{W}_H$ can be implemented by one query to $U_{E_k}$ and one query to $U_V$, and one query to $U_V$ can be applied by one query to oracle $O_V$. So, $O\left(t(\|\mathcal{L}_{}\|_{\text{be}}) \frac{\log(t(\|\mathcal{L}_{}\|_{\text{be}})/\epsilon)}{\log\log(t(\|\mathcal{L}_{}\|_{\text{be}})/\epsilon)}\right)$ queries to $O_V$ are needed to simulate $e^{\mathcal{L}t}$ with error at most $\epsilon$.
\end{proof}

\begin{definition_}[Block-encoding, Definition 43 of \cite{gilyen2019quantum}]\label{def:block_encoding}
  Let $A\in\mathbb{C}^{2^q\times2^q}$ be a $q$-qubit operator. Then a $(q+a)$-qubit unitary $U_A$ is an  $(\alpha,a,\xi)$ block-encoding of $A$ if
  \begin{align*}
    \|\alpha(\bra{0^a}\otimes I_q)U(\ket{0^a}\otimes I_q)-A\|\leq\xi~.
  \end{align*}
\end{definition_}

\begin{definition_}[State preparation pair, Definition 51 of \cite{gilyen2019quantum}]\label{def:state_pre}
  Let $y=(y_0,\ldots,y_{m-1})\in\mathbb{C}^m$ and $\|y\|_1\leq\beta$. The pair of unitaries $(P_L,P_R)$ is called a $(\beta,q,\xi)$- state-preparation pair for a nonzero vector $y$ if $P_L\ket{0^q}=\sum_{j=0}^{2^q-1} c_j \ket{j}$ and $P_R \ket{0^q}=\sum_{j=0}^{2^q-1} d_j \ket{j}$ such that $\sum_{j=0}^{m-1} \beta (c_j^* d_j-y_j)\leq \xi$ and for all $j=m,m+1,\ldots, 2^q-1$ we have $c_j^*d_j=0$.
\end{definition_}

\begin{proposition_}[Linear combination of block-encoded matrices, Proposition 5 of \cite{takahira2021quantum}]\label{prop:linear_combination}
  Let $A=\sum_{j=0}^{m-1}y_j A^{(j)}$ be a $q$-qubit operator, where $A^{(j)}\in\mathbb{C}^{2^q\times2^q}$ and $y=(y_0,\ldots,y_{m-1})\in\mathbb{C}^m$. If $U_j$ is an $(\alpha_j,a_j,\xi_j)$-block-encoding of $A^{(j)}$ and $(P_R,P_L)$ is a $(\beta,b,0)$-state-preparation pair for $(y_0\alpha_0,y_1\alpha_1,\ldots, y_{m-1}\alpha_{m-1})$, then
  \begin{eqnarray*}
    \tilde{W}_A=&\big(P_L^{\dagger} \otimes I_{a+q} \big)\left(\sum_{j=0}^{m-1}\ket{j}\bra{j}\otimes (I_{a-a_j}\otimes U_j)+\sum_{j=m}^{2^b-1}\ket{j}\bra{j}\otimes I_{a+q}\right) \big(P_R\otimes I_{a+q}\big)
  \end{eqnarray*}
  is a $(\beta, a+b, \sum_{j=0}^{m-1} |y_j|\xi_j)$-block-encoding of $A=\sum_{j=0}^{m-1}y_j A^{(j)}$, where $a=\max_j a_j$.
\end{proposition_}

\begin{proposition_}[Complexity for Lindblad simulation, Theorem 11 of \cite{li2022simulating}]\label{prop:complexity_Lindblad}
  Suppose we are given an $(\alpha_0,a,\epsilon^{\prime})$-block-encoding $U_H$ of $H$ and an $(\alpha_i,a,\epsilon^{\prime})$-block-encoding $U_{L_j}$ for each $L_j$ $(j=1,\ldots,m)$. For all $t,\epsilon\geq0$ with $\epsilon^{\prime}\leq \epsilon/(t\|\mathcal{L}_{}\|_{\text{be}})$, there exists a quantum algorithm for simulating $e^{\mathcal{L}t}$ using
  \begin{align*}
    O\left(t\|\mathcal{L}_{}\|_{\text{be}} \frac{\log(t\|\mathcal{L}_{}\|_{\text{be}}/\epsilon)}{\log\log(t\|\mathcal{L}_{}\|_{\text{be}}/\epsilon)}\right)
  \end{align*}
  queries to $U_H$ and $U_{L_j}$, and some additional $1$- or $2$- qubit gates, where $\|\mathcal{L}_{}\|_{\text{be}}=\alpha_0+\sum_{j=1}^{m} \alpha_j^2$.
\end{proposition_}

\subsection{Query complexity of other algorithms}
\subsubsection{Quantum query complexity for QHD}
Because QHD simulates a time-dependent Hamiltonian system, its quantum query complexity  can be inferred from the complexity involved in the simulation of the time-dependent Schr\"{o}dinger equation. Here, we mainly follow Refs.~\cite{childs2022quantumreal,leng2023quantumclassical} and use quantum simulation in the real space to calculate the quantum query complexity for QHD in terms of the quantum oracle $O_V$ (\eq{eq:oracle_OV}).

Given a  Nesterov’s time-dependent $H(\tau )=\frac{2}{s_1+\tau^3}\left(-\frac{1}{2} \nabla^2\right)+2{\tau^3}V(x)$ with $s_1\ll 1$ \cite{leng2023quantum}, we can get the number of queries to $O_V$ are $O\left(\|\tilde{V}\|_{\max,1}  \frac{\log (\|\tilde{V}\|_{\max, 1} t  /\epsilon)}{\log\log (\|\tilde{V}\|_{\max, 1} t /\epsilon)}\right)$ (by the Theorem 8 of Ref.~\cite{childs2022quantumreal}, and the Theorem 4 and Lemma 9 of Ref.~\cite{leng2023quantumclassical}), where $\tilde{V}=\tilde{V}(x,\tau)=2\tau ^3 V(x)$. Therefore, the total number of queries to  $O_V$ is
\begin{align}
  O\left(\left(t^4\|V\|_{\max}\right)\frac{\log \left(\left(t^4\|V\|_{\max}\right)\big/\epsilon\right)}{\log\log \left(\left(t^4\|V\|_{\max}\right)\big/\epsilon\right)}\right),
\end{align}
where the factor $t^4$ comes from the Nesterov’s damping parameter $\tau^3$ in $\tilde{V}(x,\tau)$.

\subsubsection{Quantum query complexity for QAA}
QAA is a quantum algorithm which evolves according to the time-dependent Schr\"{o}dinger equation, so we mainly follow the Theorem 4 of Ref.~\cite{berry2020time} to calculate its query complexity. In the case of QAA, the total number of queries to $O_V$ is
\begin{align}
  O\left(t\|V\|_{\max}\frac{\log (t\|V\|_{\max}/\epsilon)}{\log\log (t\|V\|_{\max}/\epsilon)}\right).
\end{align}

\subsubsection{Classical query complexity for SGD}
For SGD, we assume that the classical evaluation oracle $O_{V}^{cl}$ allows us to query the function value of $V$ in a local way, so the total number of queries to $O_{V}^{cl}$ is at least the number of iterations $t/s$ timing the number of samples $N_s$, which derives $\Theta(N_st/s)$ \cite{liu2023quantum}. In the numerical experiment setting of Section \ref{sec:SGD}, $s=\Delta t=O(\Delta x^2)=O(1/n^2)$, the query complexity becomes $\Theta(N_stn^2)$.

\subsubsection{Classical query complexity for NAGD}
The classical query complexity for NAGD is same as SGD, which is $\Theta(N_st/s)$. In the numerical experiment setting of Section \ref{sec:NAGD}, it becomes $\Theta(N_stn^2)$.

\subsection{Comparison of query complexity for different algorithms}
From the above discussion, when $t$ is significant, loosely speaking, QLD and QAA have similar quantum query complexity $\tilde{O}(t)$ which is much better than $\tilde{O}(t^4)$ of QHD.
Furthermore, when the total evolution time is kept consistent across all quantum and classical algorithms, the quantum query complexity of QLD becomes comparable to those of SGD and NAGD, which are also $\tilde{O}(t)$.
This also indicates the superiority of QLD in scenario where the evolution time is significant.

\note{We also need to note that the roles of $t$ in QLD and QAA are different. Indeed, the interpretation of $t$ as mixing time in QLD, influenced by both the Hamiltonian and dissipative components, differs  from the adiabatic scaling time in QAA, which relates to the gap of the time-dependent Hamiltonian. As a result, the time scales required for each algorithm to achieve optimal performance can differ significantly. In our numerical experiments, we only try to compare the performance of each algorithm at the same time scale $t$.}

\section{Discussion and Conclusions}
In conclusion, we demonstrate the possibility of solving continuous optimization problems through the application of a specific class of open dissipative quantum dynamics. We have theoretical guarantee of the convergence of QLD; meanwhile, we verify our intuition and provide a feasible cooling strategy through numerical experiments. 
In particular, although in some landscapes other algorithms have already performed well, by controlling the parameters of QLD, it can outperform all other algorithms in  other landscapes.

From a high-level perspective, this work can be seen as an attempt to construct optimization algorithms inspired by quantum physical systems (open quantum systems), which includes guiding the configuration of parameters based on their inherent physical significance. Through this approach, it is possible to maximally leverage the advantages offered by quantum mechanics in the process of addressing optimization problems.

However, there are several issues that necessitate further investigation.  Theoretically, we believe that both the advantages and shortcomings of QLD can be interpreted more thoroughly.

In particular, due to the complex formula of the quantum Langevin dynamics, we only provide the convergence of QLD under the approximations of $\hbar \Omega \ll kT$ and $T\to 0$. Since the limit $\hbar \to 0$ clearly gives a transition to the \note{semi-}classical Langevin equation, where all commutators vanish in this limit, in a sense, we might just prove the convergence of QLD in the \note{semi-}classical limit, and use its quantum effect in the process of convergence. Whether the convergence still exists without these approximations remains further discussion with more meticulous mathematical and physical analysis. Numerically, since the simulation in the one-dimensional landscape alone has consumed a considerable amount of our computational resources, numerical experiments in high dimension requires more efficient numerical methods and more powerful classical computers. This also demonstrates the necessity of analog (or digital) quantum computers.

\note{Moreover, we only consider the jump operator as a linear combination of the position and momentum operator in this work, which follows from Ref.~\cite{gao1997dissipative}. However, the efficacy of alternative jump operators and the potential enhancement of optimizing performance through their adoption remain open questions. In conclusion, many intriguing questions remain for further discussion, and we believe that our work will inspire more investigation into optimization using open quantum systems.}


\section*{Acknowledgements}
We thank Zhiyan Ding, Jiaqi Leng, Lin Lin, Hao Wu, Yukai Wu, Shuohan Zhang, and Zhennan Zhou for helpful discussions.

\section*{Declarations}

\noindent
\textbf{Funding. }Z. C. and Y. L. acknowledge research funding from Tsien Excellence in Engineering Program of Tsinghua University and Tsinghua University Initiative Scientific Research Program. T. L. was supported by the National Natural Science Foundation of China (Grant Numbers 62372006 and 92365117), and a startup fund from Peking University.

\vspace{0.2cm}

\noindent
\textbf{Financial interests. }The authors declare that they have no financial interests.

\vspace{0.2cm}
\noindent
\textbf{Data availability. }The data that support the findings of this study are available from the corresponding author, upon reasonable request.

\bibliographystyle{arxiv_submit}
\bibliography{QLD_arxiv_submit}


\clearpage
\appendix
\section{Numerical results of different objective functions}

In this appendix, we elucidate the specifics of numerical implementation in Appendix \ref{app:Implement_QLD}, articulate the exact forms of all objective functions in Appendix \ref{sec:objective_functions_appendix}, and analyze numerical results of all algorithms across different objective functions in Appendix \ref{sec:numerical_exp_appendix}.

\subsection{\label{app:Implement_QLD} Implementation details of numerical experiments for QLD}

QLD is simulated by numerically solving \eq{eq:Gao_Lindblad_xy}. At first, we discretize each axis of  domain into $n$ grids \cite{press2007numerical} and  use central difference scheme in space, such that all operators in the  \eq{eq:Gao_Lindblad_xy} can be approximated by matrices, making  \eq{eq:Gao_Lindblad_xy} become
\begin{align}
  \frac{\partial \hat{\rho}}{\partial t}= \hat{\mathcal{L}} \hat \rho,
  \label{eq:Gao_L_matrix}
\end{align}
where $\hat \rho(t)$ is column vector of length $n^2\times1$ transformed from $\rho(x,y,t)$, $\hat{\mathcal{L}}$ is an $n^2\times n^2$ matrix representing the sum of all operators in central difference scheme. In order to overcome the instability of numerical calculation, we employ the implicit Euler method (backward Euler method) in the time scheme, which is given by
\begin{align}
  \hat{\rho}(t+\Delta t)=\left(I-\Delta t \hat{\mathcal{L}}\right)^{-1} \hat{\rho}(t),
\end{align}
where $\Delta t$ is time step.
Although implicit method costs a lot in computational time, it is more stable than explicit method.

\begin{figure}[h]
  \centering
  \includegraphics[width=0.4\textwidth]{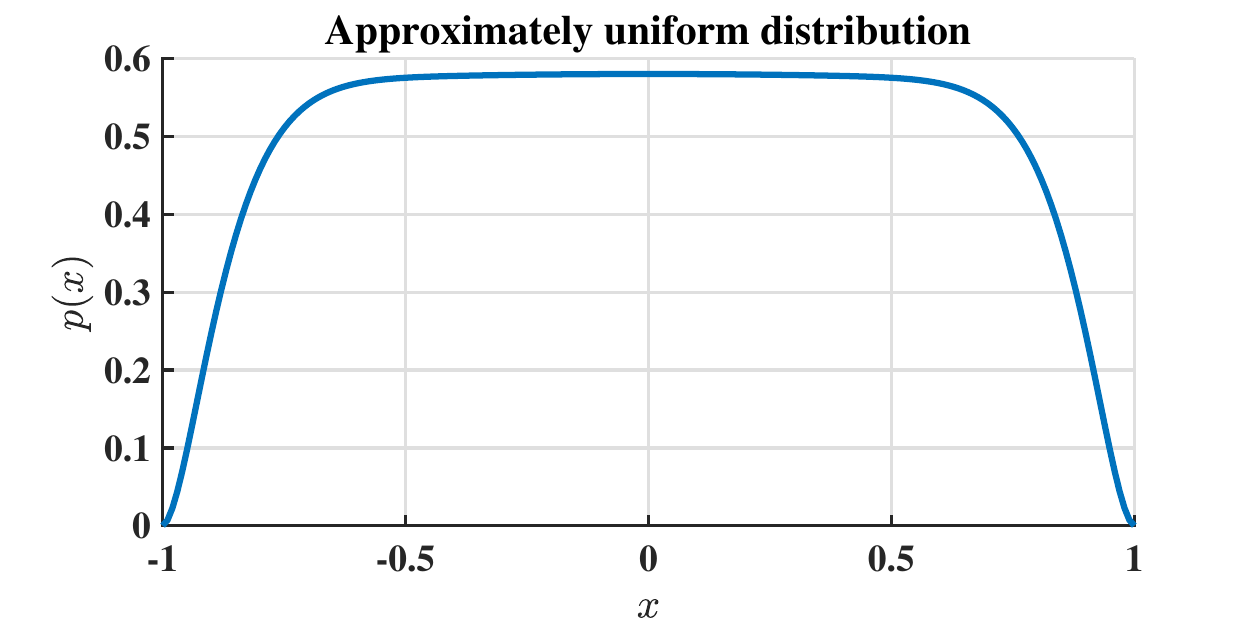}
  \caption{The approximately uniform distribution in the interval $[-1,1]$.}
  \label{fig:approx_uniform}
\end{figure}

Because the objective function reaches positive infinity at the boundary of the domain in most cases, it is reasonable to use the Dirichlet boundary condition in our numerical experiments. However,  other efforts are still needed to avoid the instability of the numerical calculation. In order to maintain the stability at the boundary, we set initial probability distribution to be an approximately uniform distribution with form
\begin{align}
  p(x)=(-0.4742x^{10}-0.0063x^2+0.4903)^2\big/Z_p,
\end{align}
where $Z_p$ is the normalization factor. (See Fig.~\ref{fig:approx_uniform}.) It is a good approximation of the uniform distribution in the interval $[-1,1]$ and also approximates zero at the boundary to achieve the numerical stability.
In the following part, we will use this approximately uniform distribution to initialize our wave function $\psi_0(x)$ at time $t=0$, which is given by
\begin{align}
  \psi_0(x)=\sqrt{p(x)}.
\end{align}

We also establish the time step size $\Delta t=0.05\Delta x^2 $, which is small compared to the width of the discretization grid $\Delta x$ to ensure the stability of the numerical computation. In all numerical experiments, the domain of $x$ is $[-1,1]$. We only consider the one-dimensional case because each one-dimensional QLD experiment requires approximately $14$ hours.

Nonetheless, a crucial challenge persists: determining the values of each parameter in \eq{eq:Gao_Lindblad_xy}. The system's behavior is highly sensitive to the quantities chosen for all parameters in \eq{eq:Gao_Lindblad_xy}. Refs.~\cite{gao1997dissipative, gao1998lindblad, gao1995femtosecond} study the dynamics of vibrational relaxation of a Morse oscillator, mimic the O$_2$-Pt(111) bond, induced by equilibrium electrons. We have nondimensionalized this real physical system to obtain the baseline values for all parameters.

\subsection{The precise formulations of objective functions}\label{sec:objective_functions_appendix}
We choose seven objective functions with different features in Table~\ref{tab:table1} to test the performance of different algorithms, as discussed in Section \ref{sec:objective_functions}. The precise formulations and global minimum $x^*$ of these objective functions defined in $x\in [-1,1]$  are listed below,
\begin{itemize}
  \item Deflected Corrugated Spring ($x^*=0$):
        \begin{align}
          V(x)=0.1\left((6.5x)^2+25\right)-\cos\left (5\sqrt{ (6.5x)^2+25}\right).
        \end{align}
  \item Csendes ($x^*=0$):
        \begin{align}
          V(x)=x^6(2+\sin(1/x)).
        \end{align}
  \item Michalewicz ($x^*=-0.3911$):
          \begin{align}
            V(x)=-\sin(1.2(x+1.7))\sin^2\left(\frac{2(1.2(x+1.7))^2}{\pi}\right)^{20}.
          \end{align}
  \item Alpine 1 ($x^*=0$):
        \begin{align}
          V(x)= |8x\sin(8x)+0.8x|.
        \end{align}
  \item Bohachevsky 2 ($x^*=0$):
        \begin{align}
          V(x)=9x^2-0.3\cos(9\pi x)+0.3.
        \end{align}
  \item Griewank ($x^*=0$):
        \begin{align}
          V(x)=\frac{1}{4000}(15x)^2-\cos(15x)+1.
          \label{eq:Griewank}
        \end{align}
  \item Double well ($x^*=-0.6472$):
        \begin{align}
          V(x)=83.9808x^4-64.8x^2+7.2x+17.0680.
        \end{align}
\end{itemize}

\subsection{\label{sec:numerical_exp_appendix}Analysis of numerical results}
\begin{figure*}[htbp]
  \centering
  {\includegraphics[width=0.3\textwidth]{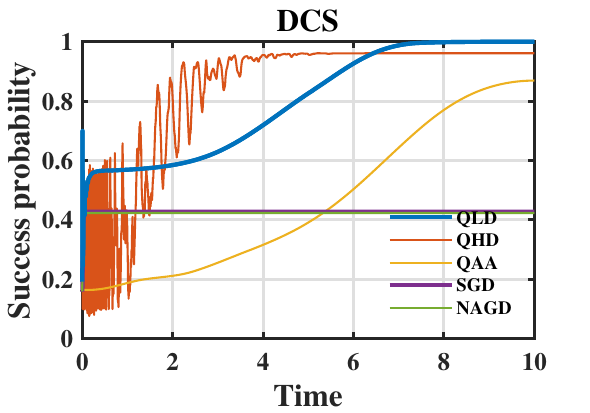}}
  {\includegraphics[width=0.3\textwidth]{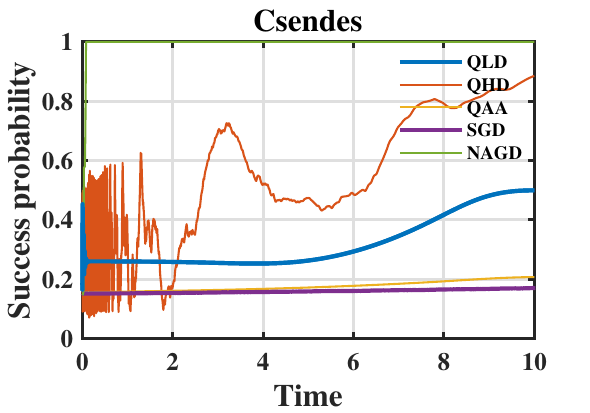}}
  {\includegraphics[width=0.3\textwidth]{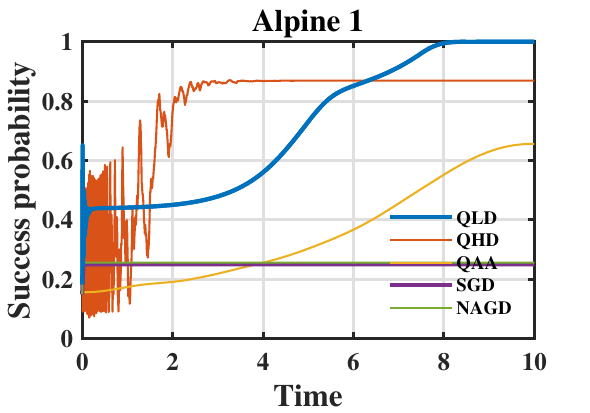}}

  {\includegraphics[width=0.3\textwidth]{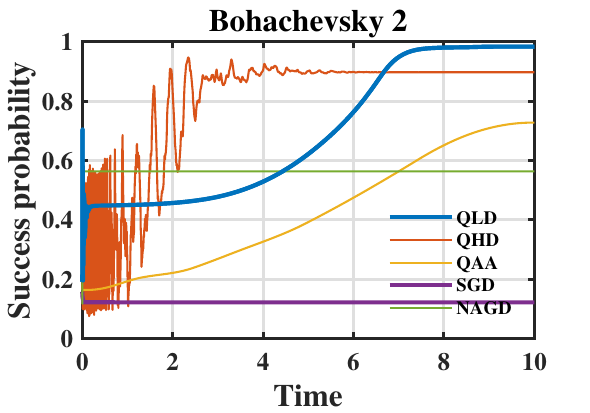}}
  {\includegraphics[width=0.3\textwidth]{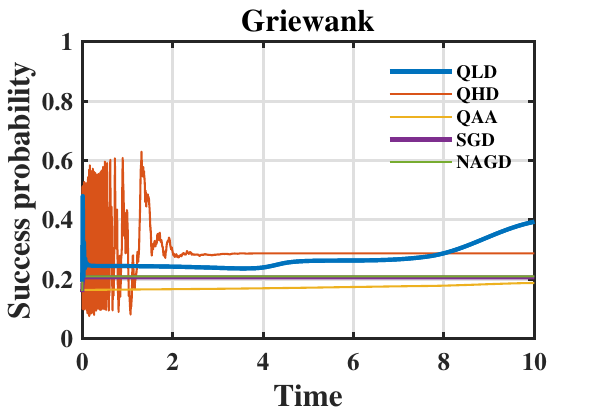}}
  {\includegraphics[width=0.3\textwidth]{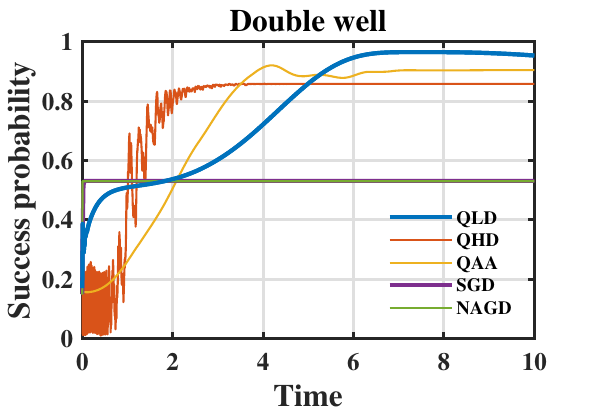}}

  \caption{Success probabilities of six objective functions for different algorithms at different (effective) evolution times.}
  \label{fig:rest_success}
\end{figure*}

\begin{figure*}[p]
  \centering
  {\includegraphics[width=1\textwidth]{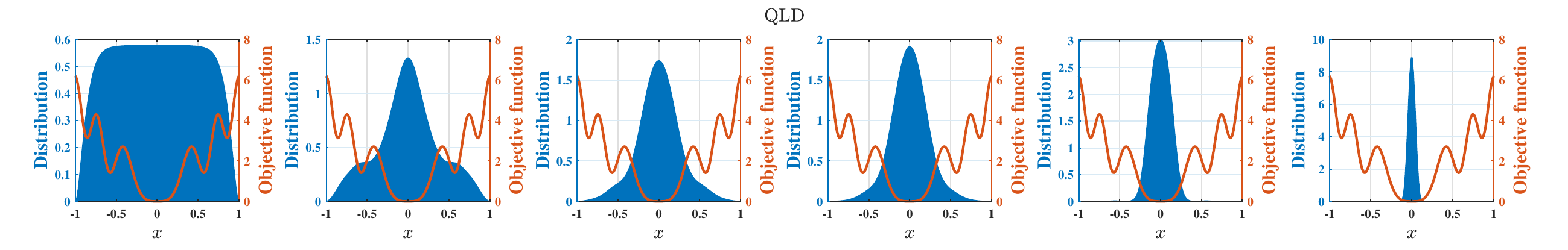}}  \\
  \vspace{-0.16cm}
  {\includegraphics[width=1\textwidth]{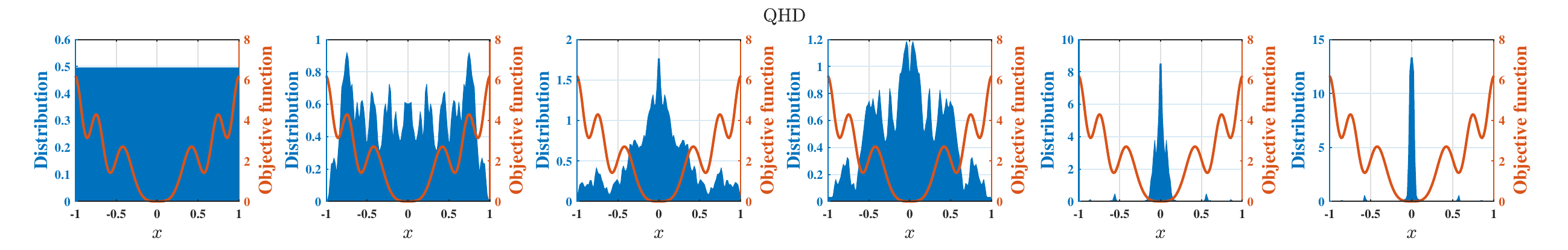}}\\
  \vspace{-0.16cm}
  {\includegraphics[width=1\textwidth]{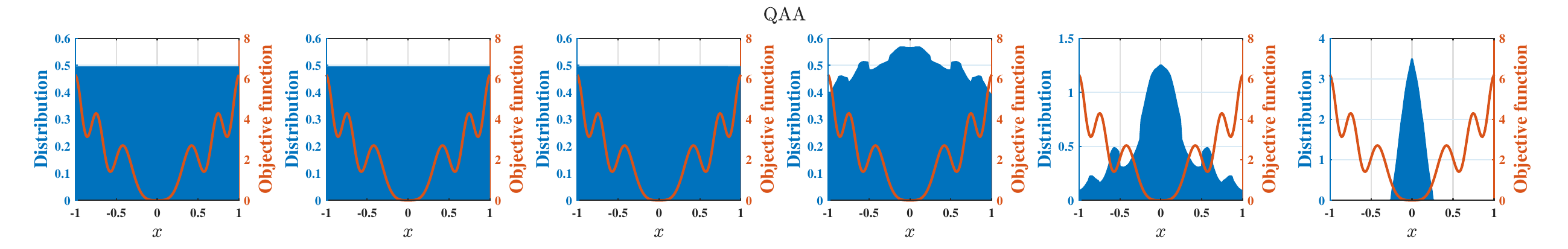}}\\
  \vspace{-0.16cm}
  {\includegraphics[width=1\textwidth]{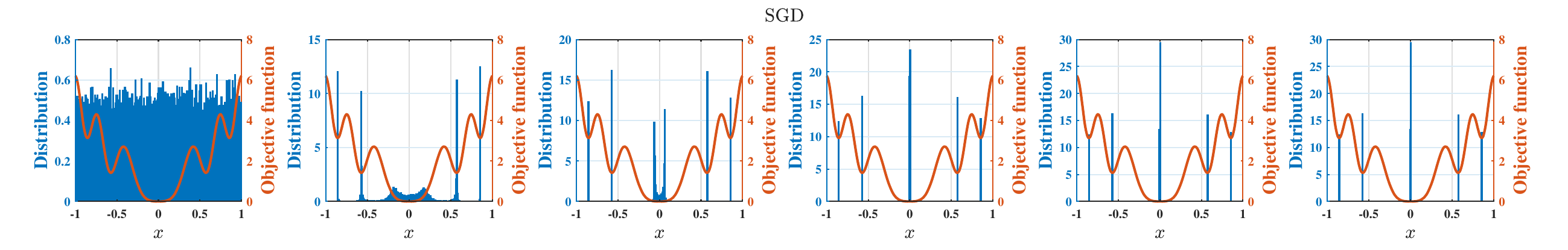}}\\
  \vspace{-0.16cm}
  {\includegraphics[width=1\textwidth]{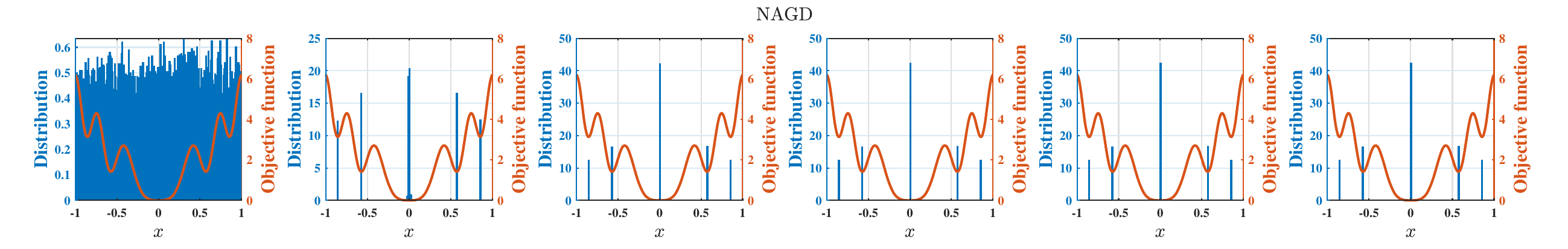}}\\
  \vspace{-0.17cm}
  {\includegraphics[width=1\textwidth]{figs/time_plot.pdf}}
  \caption{Probability distributions of DCS function $(x^*=0)$ for different algorithms at different (effective) evolution times $t=0,0.01,0.1,1,5,10$.}
  \label{fig:rest_DCS}
\end{figure*}

\begin{figure*}
  \centering
  {\includegraphics[width=1\textwidth]{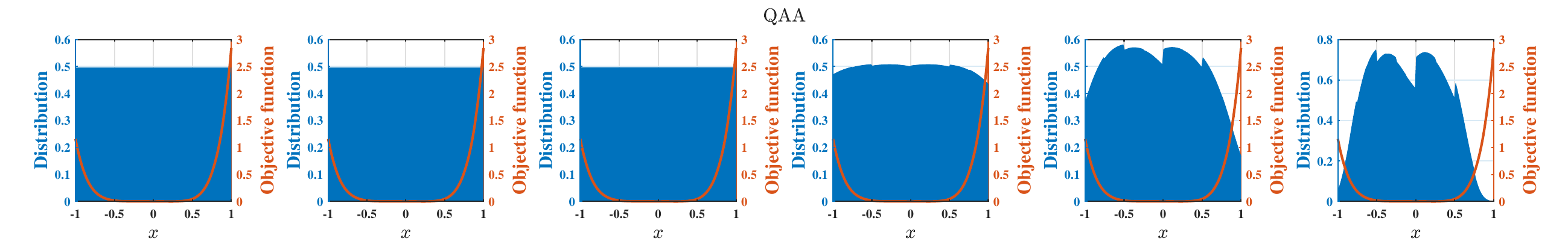}}\\
  \vspace{-0.16cm}
  {\includegraphics[width=1\textwidth]{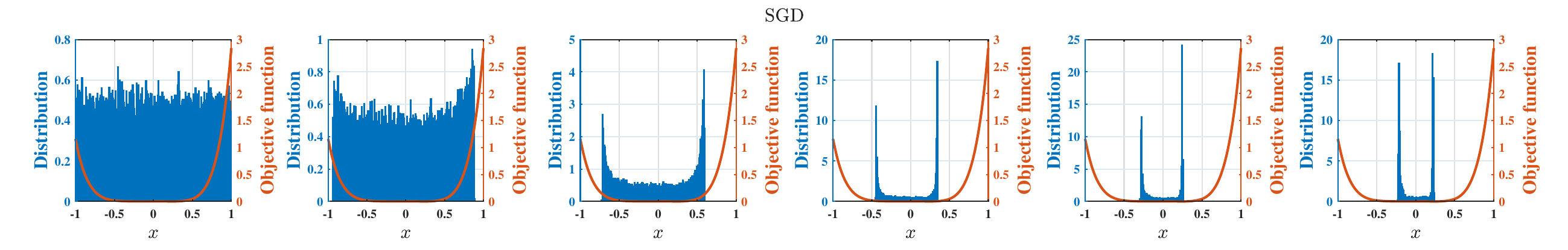}}\\
  \vspace{-0.16cm}
  {\includegraphics[width=1\textwidth]{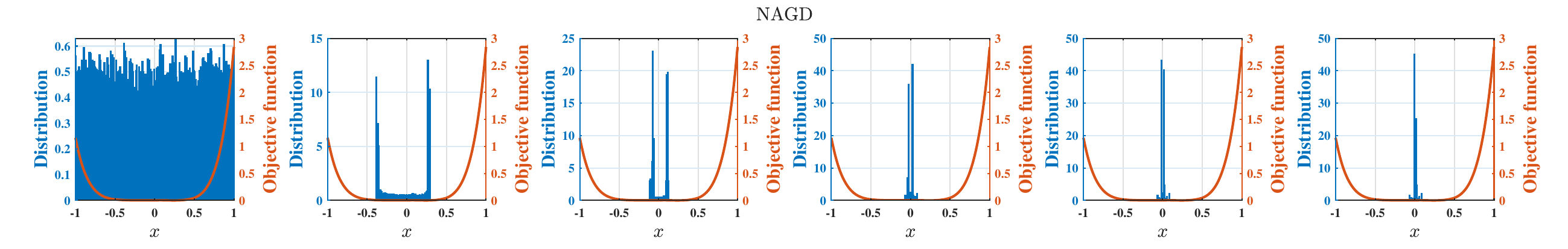}}\\
  \vspace{-0.17cm}
  {\includegraphics[width=1\textwidth]{figs/time_plot.pdf}}
  \caption{Probability distributions of Csendes  function $(x^*=0)$  for different algorithms  at different (effective) evolution times $t=0,0.01,0.1,1,5,10$.}
  \label{fig:rest_Csendes}
\end{figure*}

\begin{figure*}
  \centering
  {\includegraphics[width=1\textwidth]{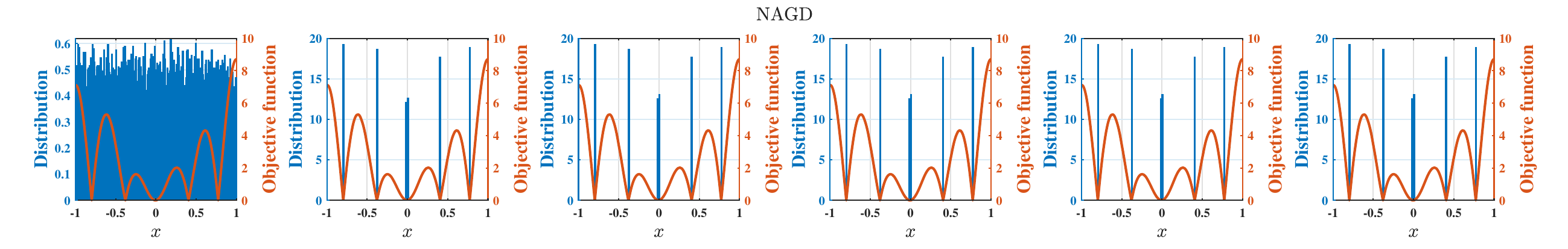}}\\
  \vspace{-0.17cm}
  {\includegraphics[width=1\textwidth]{figs/time_plot.pdf}}
  \caption{Probability distributions of Alpine 1 function $(x^*=0)$  for different algorithms  at different (effective) evolution times $t=0,0.01,0.1,1,5,10$.}
  \label{fig:rest_alpine_1}
\end{figure*}

\begin{figure*}
  \centering
  {\includegraphics[width=1\textwidth]{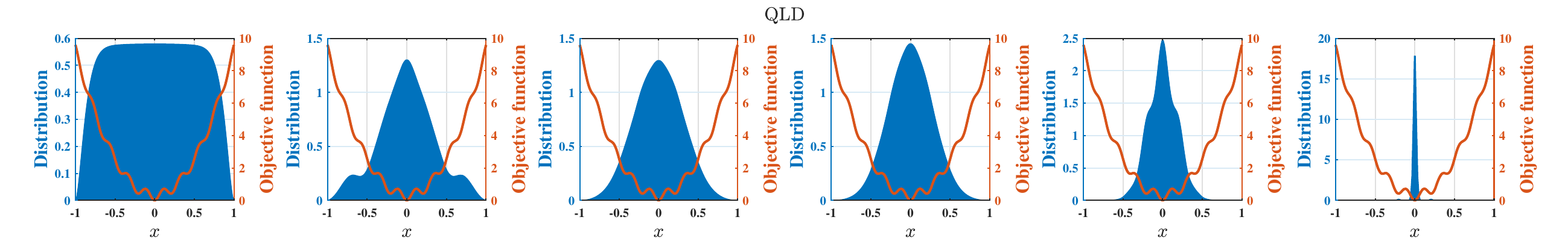}}  \\
  \vspace{-0.16cm}
  {\includegraphics[width=1\textwidth]{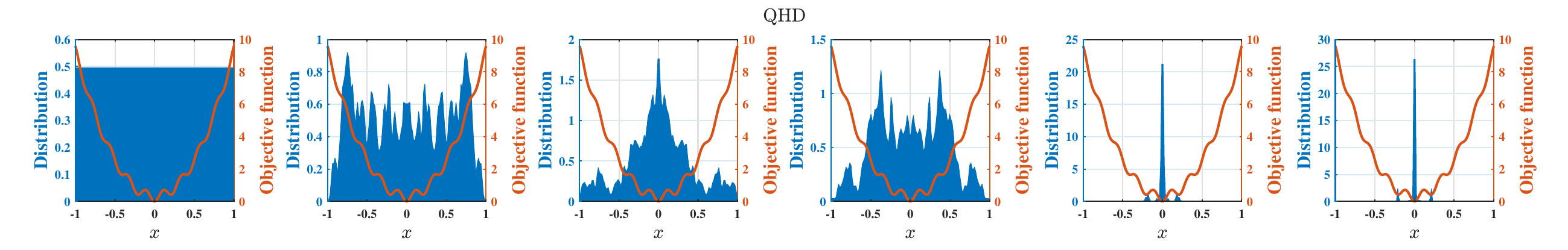}}\\
  \vspace{-0.16cm}
  {\includegraphics[width=1\textwidth]{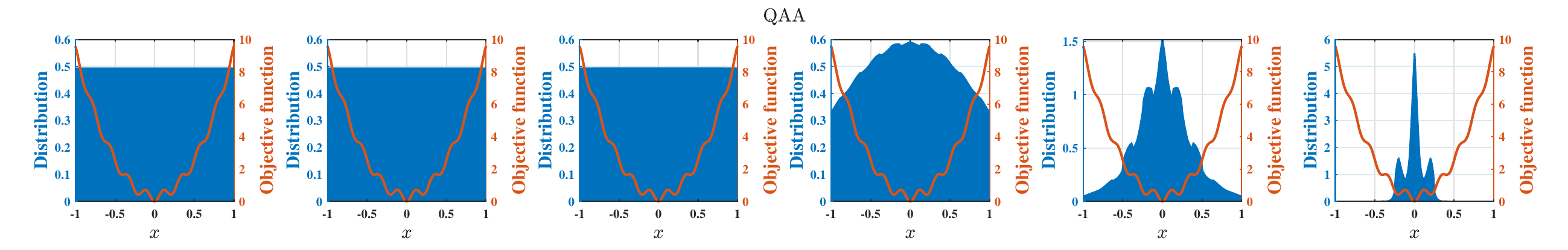}}\\
  \vspace{-0.16cm}
  {\includegraphics[width=1\textwidth]{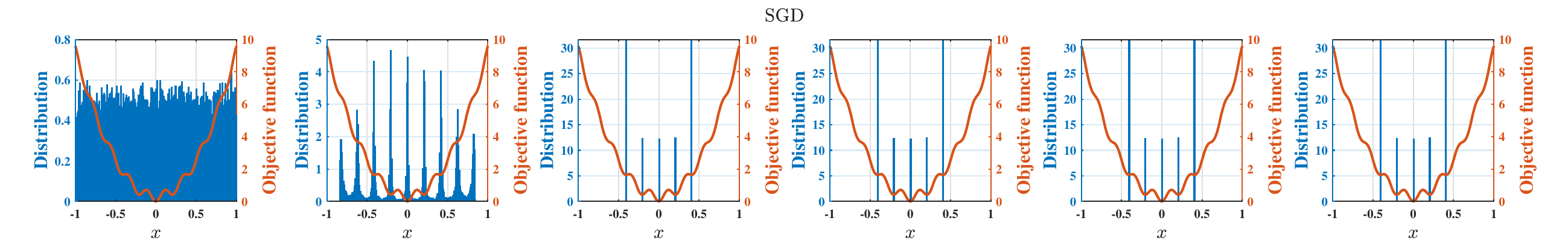}}\\
  \vspace{-0.16cm}
  {\includegraphics[width=1\textwidth]{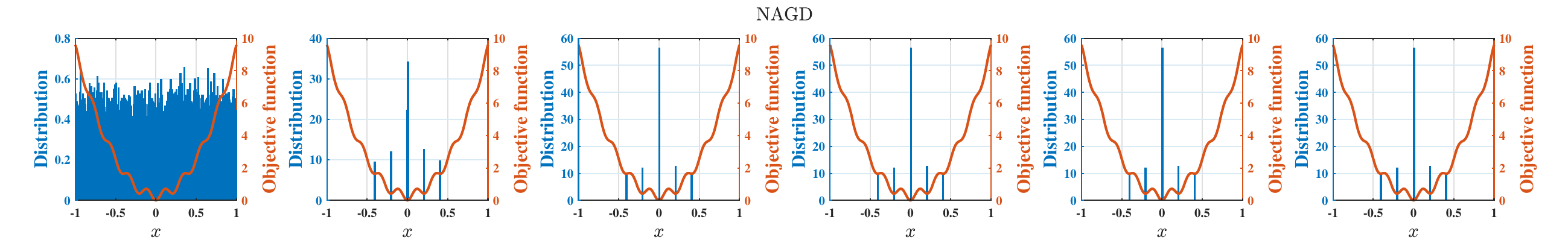}}\\
  \vspace{-0.17cm}
  {\includegraphics[width=1\textwidth]{figs/time_plot.pdf}}
  \caption{Probability distributions of Bohachevsky 2 function $(x^*=0)$  for different algorithms  at different (effective) evolution times $t=0,0.01,0.1,1,5,10$.}
  \label{fig:rest_Bohachevsky_2}
\end{figure*}

\begin{figure*}
  \centering
  {\includegraphics[width=1\textwidth]{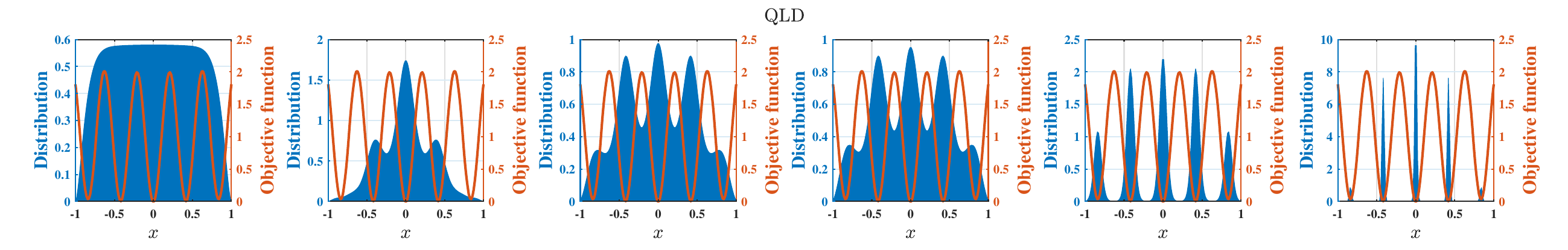}}  \\
  \vspace{-0.161cm}
  {\includegraphics[width=1\textwidth]{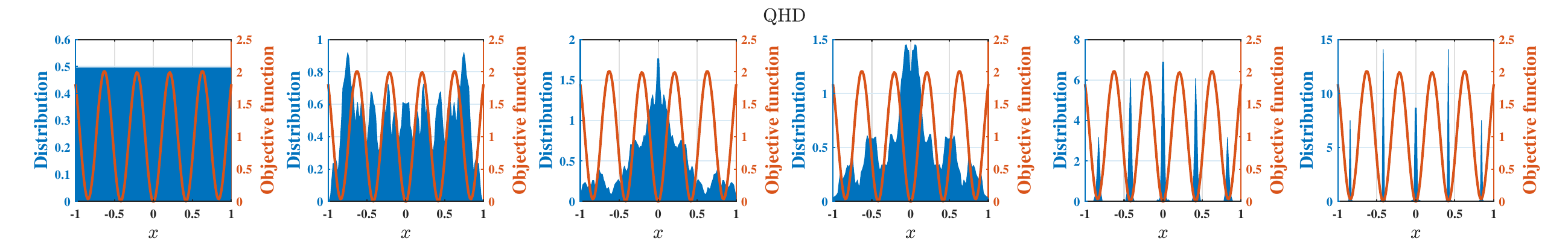}}\\
  \vspace{-0.161cm}
  {\includegraphics[width=1\textwidth]{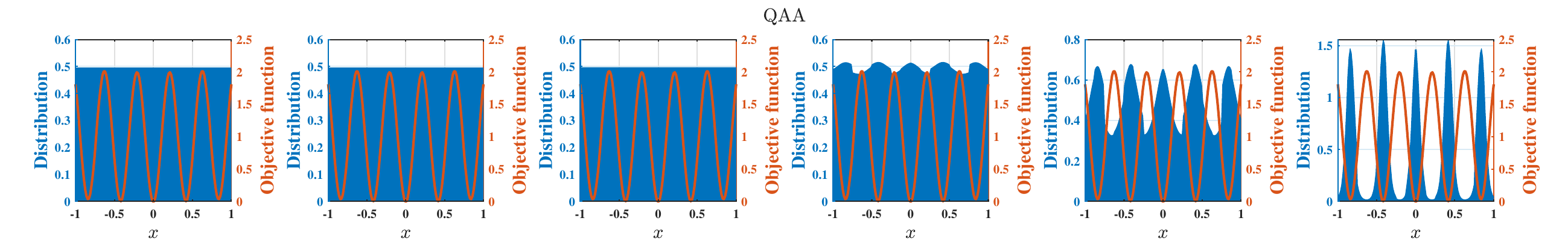}}\\
  \vspace{-0.161cm}
  {\includegraphics[width=1\textwidth]{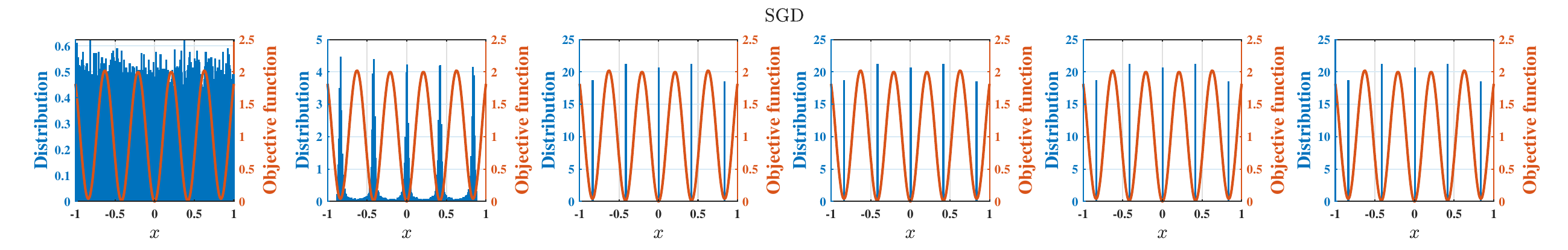}}\\
  \vspace{-0.165cm}
  {\includegraphics[width=1\textwidth]{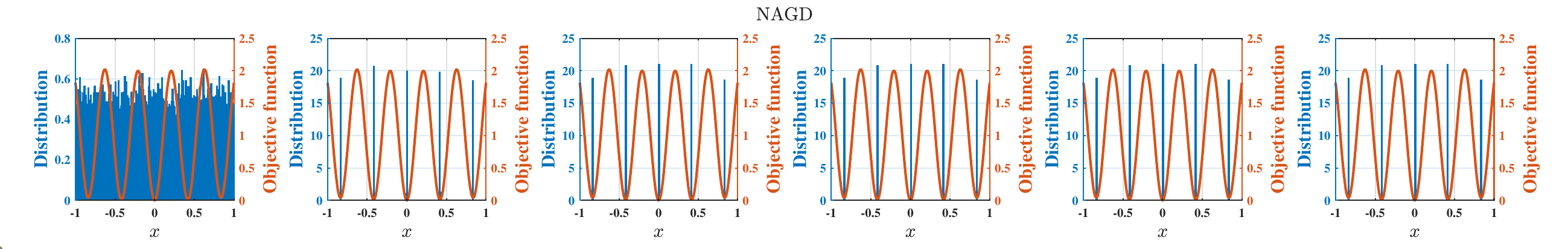}}\\
  \vspace{-0.17cm}
  {\includegraphics[width=1\textwidth]{figs/time_plot.pdf}}
  \caption{Probability distributions of Griewank function $(x^*=0)$  for different algorithms at different (effective) evolution times $t=0,0.01,0.1,1,5,10$.}
  \label{fig:rest_Griewank}
\end{figure*}

\begin{figure*}
  \centering
  {\includegraphics[width=1\textwidth]{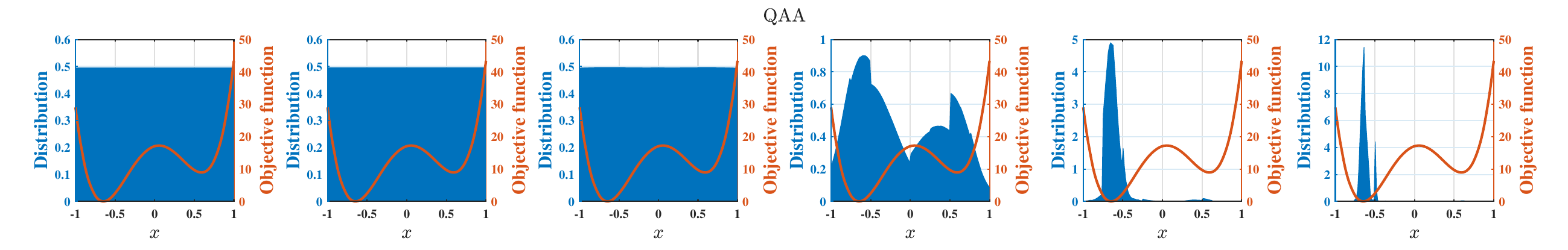}}\\
  \vspace{-0.161cm}
  {\includegraphics[width=1\textwidth]{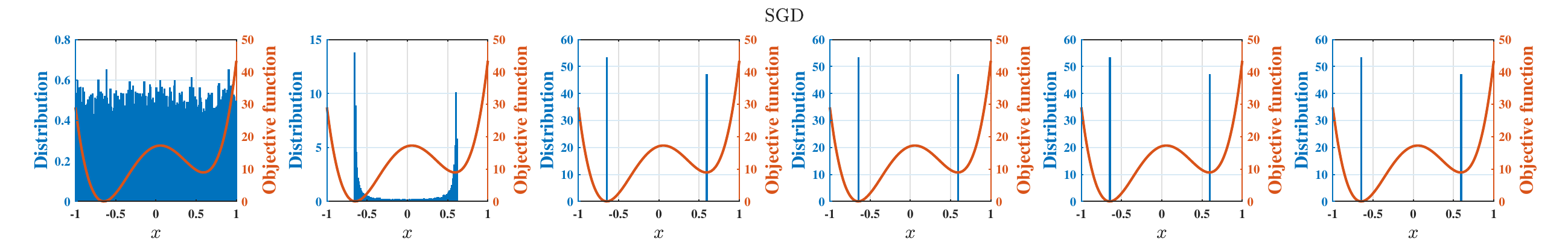}}\\
  \vspace{-0.165cm}
  {\includegraphics[width=1\textwidth]{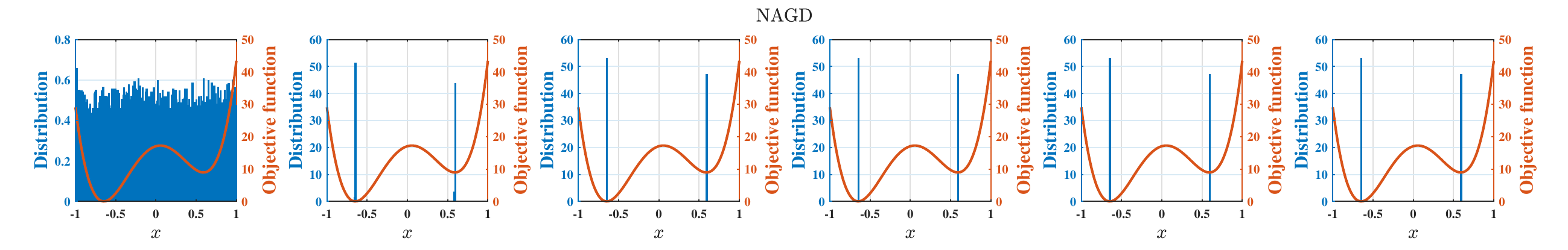}}\\
  \vspace{-0.161cm}
  {\includegraphics[width=1\textwidth]{figs/time_plot.pdf}}
  \caption{Probability distributions of Double well function $(x^*= -0.6472)$  for different algorithms at different (effective) evolution times $t=0,0.01,0.1,1,5,10$.}
  \label{fig:rest_double_well}
\end{figure*}

The results of Michalewicz function are shown in Section~\ref{sec:three_phases_theorem}. Due to space constraints, the numerical results of rest six objective functions are shown here.
The success probabilities of six objective functions for different algorithms at different (effective) evolution times are shown in Fig.~\ref{fig:rest_success}. The probability distributions of six objective functions for different algorithms at different (effective) evolution times are shown in  \cref{fig:rest_DCS,fig:rest_Csendes,fig:rest_alpine_1,fig:rest_Bohachevsky_2,fig:rest_Griewank,fig:rest_double_well}.

Based on the observations of Fig.~\ref{fig:rest_success}, QLD exhibits much higher success probability across most objective functions than other algorithms. Specifically, the phenomena of strong quantum tunneling and thermal effects equip QLD with superior capabilities to escape from local minima encircled by numerous shallow wells at the beginning of evolution. This advantage enables QLD to achieve nearly 100\% success probability in the Michalewicz, Alpine 1, Bohachevsky 2, and Deflected Corrugated Spring (DCS) functions which own several not-so-deep wells.  However, the persistence of quantum noise and thermal effects towards the end of the evolution process hinder QLD's ability to focus on the global minimum. Consequently, QLD performs poorly on the ``Basin'' function (Csendes), characterized by a flat region near the global minimum.
The observations of dynamical processes in these experiments further confirms the three-phase explanation of QLD, which is shown at the end of  Section~\ref{sec:three_phases_theorem}.

\begin{figure*}
  \centering
  {\includegraphics[width=1\textwidth]{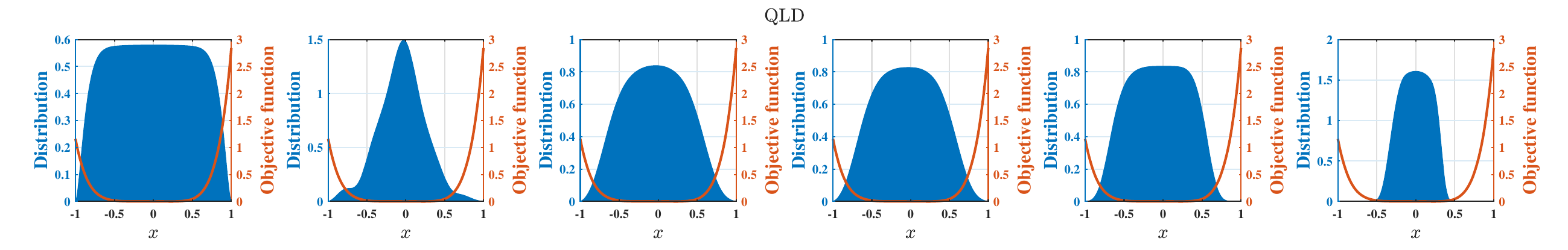}}  \\
  \vspace{-0.16cm}
  {\includegraphics[width=1\textwidth]{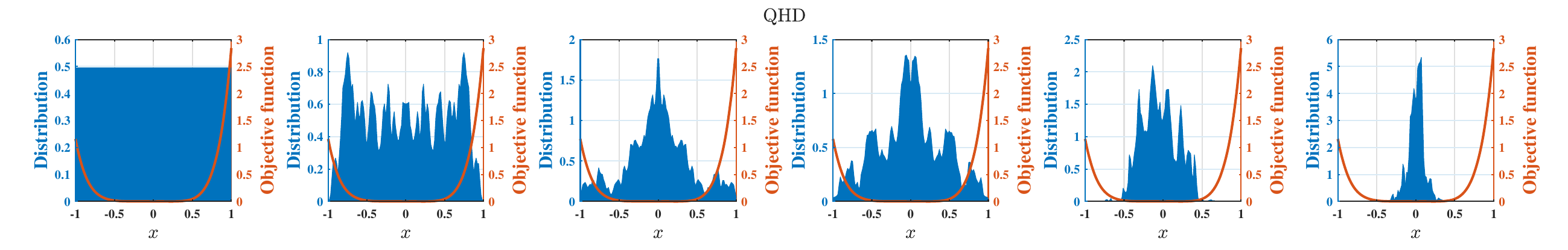}}
\end{figure*}

\begin{figure*}
  \centering
  {\includegraphics[width=1\textwidth]{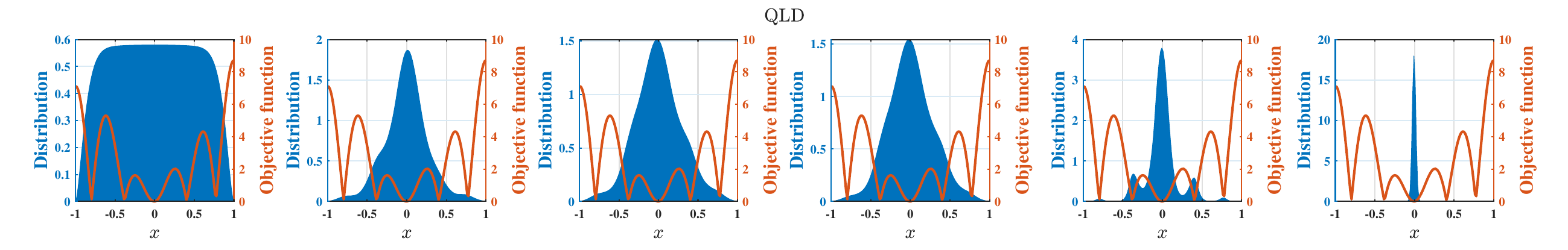}}  \\
  \vspace{-0.16cm}
  {\includegraphics[width=1\textwidth]{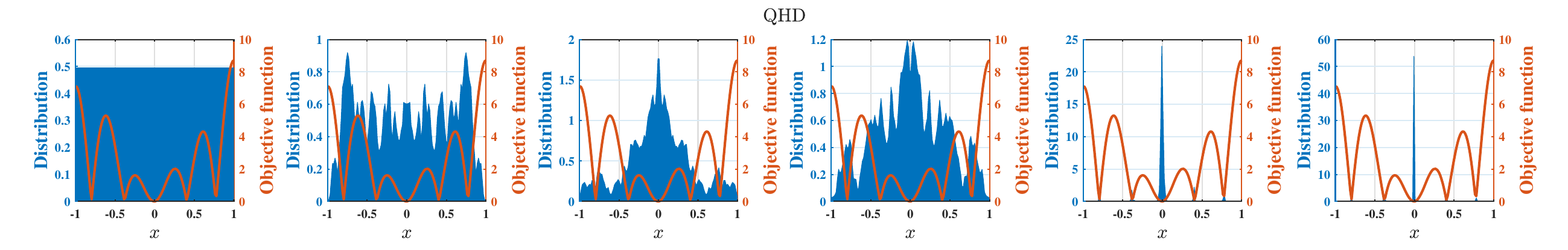}}\\
  \vspace{-0.16cm}
  {\includegraphics[width=1\textwidth]{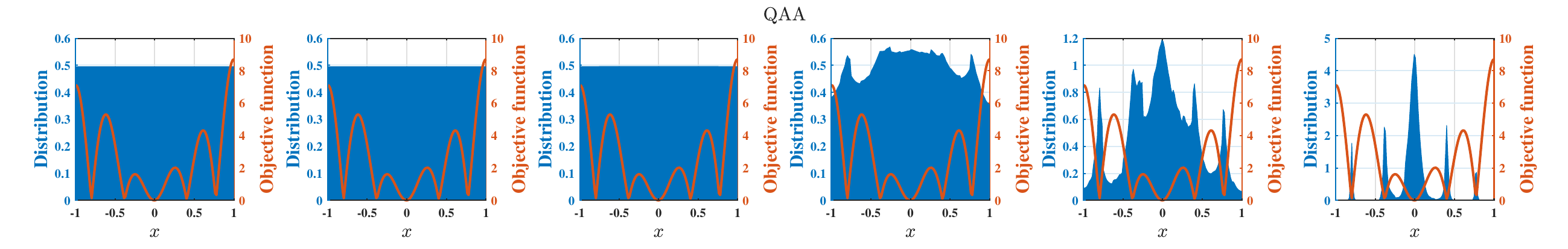}}\\
  \vspace{-0.16cm}
  {\includegraphics[width=1\textwidth]{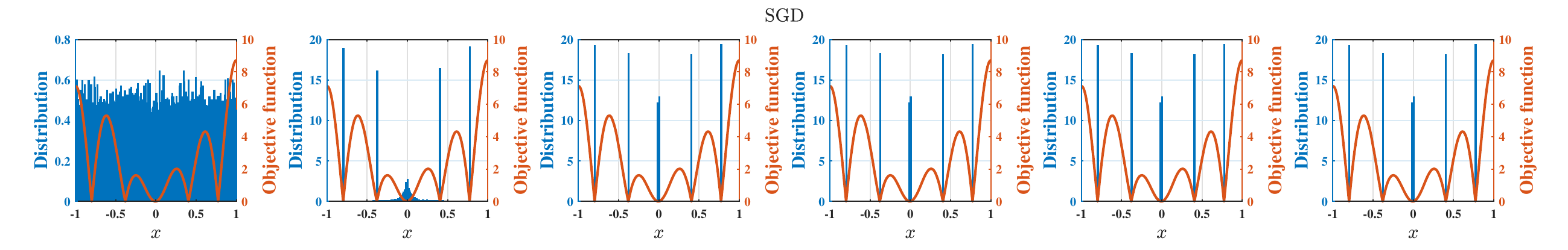}}
\end{figure*}

\begin{figure*}
  \centering
  {\includegraphics[width=1\textwidth]{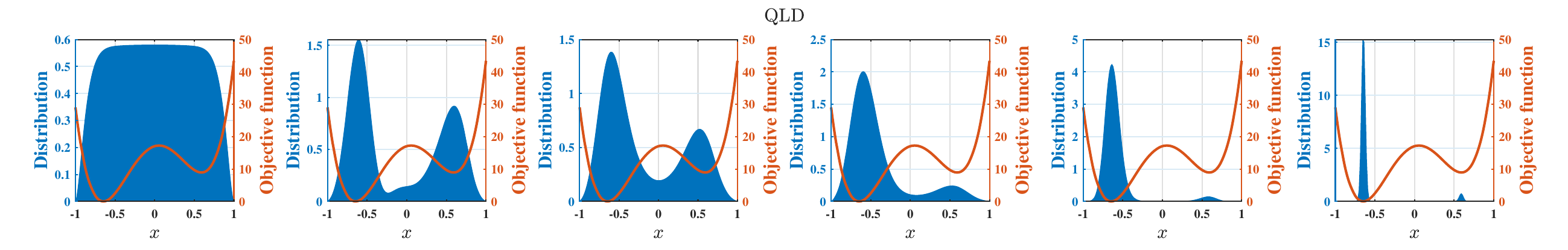}}  \\
  \vspace{-0.161cm}
  {\includegraphics[width=1\textwidth]{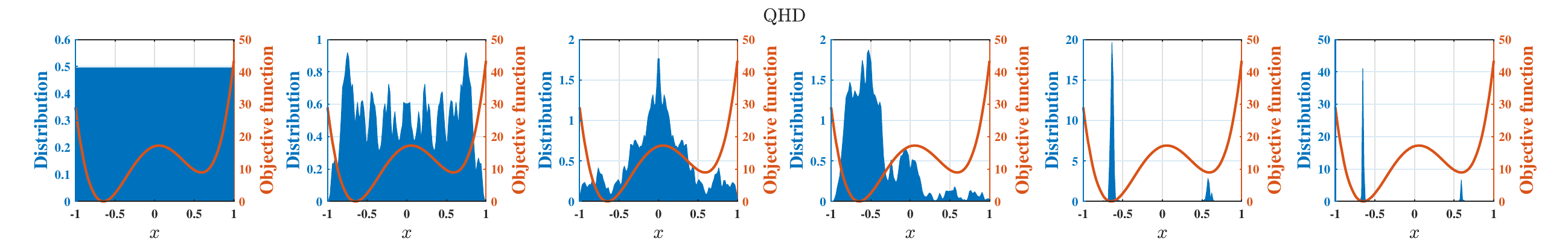}}
\end{figure*}

\end{document}